\documentclass{CUP-JNL-DCE}

\usepackage{latexsym}
\usepackage{graphicx}
\usepackage{multicol,multirow}
\usepackage{amsmath,amssymb,amsfonts}
\usepackage{mathrsfs}
\usepackage{amsthm}
\usepackage{apacite}
\usepackage{rotating}
\usepackage{appendix}
\usepackage{subcaption}
\usepackage{placeins}
\usepackage[authoryear]{natbib}

\begin{document}

\begin{Frontmatter}

\title[Enhancing Industrial X-ray Tomography by Data-Centric Statistical Methods]
{Enhancing Industrial X-ray Tomography by Data-Centric Statistical Methods}

\author[1]{Jarkko Suuronen}
\author[2]{Muhammad Emzir}
\author[3]{Sari Lasanen}
\author[2]{Simo S\"arkk\"a}
\author*[1]{Lassi Roininen}\email{lassi.roininen@lut.fi}

\authormark{Jarkko Suuronen \textit{et al.}}

\address[1]{\orgdiv{School of Engineering Science}, \orgname{Lappeenranta-Lahti University of Technology}, \orgaddress{\street{PO Box 20}, \postcode{FI-53851 Lappeenranta}, \country{Finland}}}

\address[2]{\orgdiv{Department of Electrical Engineering and Automation}, \orgname{Aalto University}, \orgaddress{\street{PO Box 11000}, \postcode{FI-00076 Aalto}, \country{Finland}}}

\address[3]{\orgdiv{Sodankyl\"a Geophysical Observatory}, \orgname{University of Oulu}, \orgaddress{\street{PO Box 8000}, \postcode{FI-90014 University of Oulu}, \country{Finland}}}



\abstract{
X-ray tomography has applications in various industrial fields such as sawmill industry, oil and gas industry, chemical engineering, and geotechnical engineering.
In this article, we study Bayesian methods for the X-ray tomography reconstruction.
In Bayesian methods, the inverse problem of tomographic reconstruction is solved with help of a statistical prior distribution which encodes the possible internal structures by assigning probabilities for smoothness and edge distribution of the object.
We compare Gaussian random field priors, that favour smoothness, to non-Gaussian total variation, Besov, and Cauchy priors which promote sharp edges and  high-contrast and low-contrast areas in the object.
We also present computational schemes for solving the resulting high-dimensional Bayesian inverse problem with 100,000-1,000,000 unknowns.
In particular, we study the applicability of a no-U-turn variant of Hamiltonian Monte Carlo methods and of a more classical adaptive Metropolis-within-Gibbs algorithm for this purpose.
These methods also enable full uncertainty quantification of the reconstructions.
For faster computations, we use maximum a posteriori estimates with limited-memory BFGS optimisation algorithm. 
As the first industrial application, we consider sawmill industry  X-ray  log tomography.
The logs have knots, rotten parts, and even possibly metallic pieces, making them good examples for non-Gaussian priors.
Secondly, we study drill-core rock sample tomography, an example from oil and gas industry.
We show that Cauchy priors produce smaller number of artefacts than other choices, especially with sparse high-noise measurements, and choosing Hamiltonian Monte Carlo enables systematic uncertainty quantification.
}

\begin{policy}[Impact Statement]
Industrial X-ray tomography reconstruction accuracy depends on various factors, like the equipment, measurement geometry and constraints of the target.
For example dynamical systems are harder targets than static ones.
The harder and noisier the setting becomes, the more emphasis goes on mathematical modelling of the targets.
Bayesian statistical inversion is a common choice for difficult measurement settings, and its limitations mainly come from the choice of the a priori models. 
 Gaussian models are widely studied, but they provide smooth reconstructions. Total variation priors are not invariant under mesh changes, so doing systematic uncertainty quantification, like data-centric sensor optimisation, cannot be done with them.  Besov and Cauchy priors however provide systematic non-Gaussian random field models, which can be used for contrast-boosting tomography. The drawback is higher computational cost.
Hence, the techniques developed here are useful for non-time-critical applications with difficult measurement settings.
In these cases, the methods developed may provide significantly better reconstructions than the traditional  methods, like filtered back-projection.

\end{policy}

\end{Frontmatter}

\section{Introduction}
\label{sec1}

X-ray tomography has applications in various industrial fields such as sawmill industry, where it can be used for detecting knots, rotten parts and foreign objects in sawmills \citep{Shustrov:2019,Zolotarev:2019}. 
In oil and gas industry, X-ray tomography can be used to analyze drill-core rock samples to identifying pore structures \citep{mendoza:2019} and other structural properties of rock. 
In chemical engineering, X-ray microtomography can be used to measure the internal structure of substances  at the micrometer level \citep{Ou2017}. In manufacturing it  can be used in  nondestructive testing \citep{Garcea2018,Rotella2018},  endurance testing \citep{Piao2019}, and   dimensional metrology \citep{Kruth2011,VillarrageGomez2019}.  In geotechnical engineering, X-ray microtomography  can be used to measure soil properties in laboratories, while a closely related  travel-time tomography  can be used to measure  the  structure of soil and rocks from cross-borehole measurements \citep{Ernst2007,Huai2016}.  

The principle of X-ray tomography is that we transmit X-ray radiation to the object, which penetrates the object of interest over a collection of propagation paths, and the attenuated X-ray radiation is measured in a detector system \citep{peterinvaitos}.
This allows us to estimate internal properties of an unknown object given the noise-perturbed indirect measurements.
However, typically, we can transmit and measure the X-rays only from a limited number of angles around the object which makes it harder to reconstruct the internal structure of the object from the measurements.
The reconstruction problem is thus an inverse problem where the measurements give only a limited amount of information on the object of interest \citep{statistical}.

In order to successfully reconstruct the internals of the object from the limited number of measurements, we need to introduce additional information to the reconstruction process.
In this article, we study so-called Bayesian methods \citep{kaipio}, where we introduce a statistical prior model for the possible internal structures in form of a probability distribution.
This prior model encodes the information on what kind of structures are more likely and which are less likely than others.
For example, a Gaussian random field prior puts higher probability for smoother structures whereas total variation (TV), Besov, and Cauchy priors favor structures that can have sharper edges between the substructures \citep{gerardonvaitos}.
The advantage of the Bayesian formulation is that it provides uncertainty quantification mechanism to the reconstruction problem, as it not only provides single reconstruction, but also the error bars for the reconstruction in form of a probability distribution \citep{mcmcvar}.

The Bayesian formulation of the X-ray tomography problem transforms the solution to the associated inverse problem to a Bayesian inference problem, where we need to use computational methods from Bayesian statistics to solve it.
In this article, we use Markov chain Monte Carlo (MCMC) methods, and in particular, Metropolis-within Gibbs (MwG) sampling and Hamiltonian Monte Carlo (HMC) which are powerful tools for this purpose.
We also experiment simpler solution methods which find the maximum posterior probability reconstructions (maximum a posteriori, MAP, estimates) by using numerical optimization.

In the experiments we concentrate on two industrial problems.
The first one is sawmill industry application, doing log tomography for detecting knots, rotten parts and foreign objects in sawmills \citep{Shustrov:2019,Zolotarev:2019}.
The second application is drill-core rock sample tomography, in oil and gas industry, for identifying pore structures  \citep{mendoza:2019}. 

%




  
\subsection{X-ray tomography as a Bayesian statistical inverse problem}

In general,  reducing the number of measurements tends to add artefacts to the tomographic reconstruction. This means that we need to carefully evaluate the accuracy of the reconstruction for different levels of sparsity. For this kind of problems, a typical approach is to deploy Bayesian statistical inversion techniques in the sense of \citet{kaipio}.
Bayesian  inversion is the theory and practical data analysis of  noisy indirect measurements within the Bayesian estimation framework. 
Following \citet{peterinvaitos}, X-ray radiation measured at single detector pixel over a given propagation path is
\begin{equation} \label{eqn:direct}
    y_{\theta,s} = \iint  \mathcal{X}\left(x_1,x_2\right) \delta\left(s- x_1 \cos \theta - x_2 \sin \theta\right) dx_1 dx_2 + e_{\theta,s}=: \mathcal{A}_{\theta,s}\mathcal{X}+e_{\theta,s},
\end{equation}
where $y_{\theta,s}$ is measured value at angle $\theta$ with translation $s$ and
$\mathcal X$ is varying X-ray attenuation  coefficient field inside the object of interest. 
For practical computations, we discretise the operation $\mathcal{A}_{\theta,s}\mathcal{X}$ 
by approximating $\mathcal{X}$ with a  piecewise defined function with constant values on square pixels, and obtain a  finite-dimensional presentation $y = AX +e$, where $y\in \mathbb{R}^n$, $A\in\mathbb{R}^{n\times m}$, $e\in\mathbb{R}^n$, and $X\in\mathbb{R}^m$ represent the approximated values of $\mathcal X$ on pixels.     

In X-ray tomography, the noise process $e$ is often modelled as a compound Poisson distributed noise \citep{Whiting2002,poisson}.
We make a  simplification, and  assume that $e$ is zero-mean Gaussian white noise with covariance $C=\sigma^2 I$.
As our main aim is in contrast-boosting priors, so we assume that this simplification does not affect the generality of the results.
The solution of the tomography problem is then an a posteriori probability density  via the Bayes formula 
\begin{equation} \label{eqn:bayes}
    \pi \left(X \vert y\right)  = \frac{\pi\left(X\right)\pi\left(y\vert X\right)}{\pi\left(y\right)} \propto \pi\left(X\right)\pi\left(y\vert X\right) =  \pi\left(X\right) \exp\left(-\frac{1}{2} \left(y-AX\right)^TC^{-1}\left(y-AX\right)\right),
\end{equation}
where $\pi(X)$ is the a priori density, that is, the probabilistic description of the unknown we know before any measurements are taken, $\pi(y\vert X)$ is the likelihood density,  and $\pi(y)$ is a norming constant, which we shall omit,  as we carry out  computations by using unnormed  densities.

\subsection{Literature review}
Sparse-angle tomography is an ill-posed inverse problem \citep{Natterer2001}, and thus in order to have stable solutions, we need to set the prior $\pi(X)$. 
As our objective is in reconstructing unknown objects with sharp interfaces, which are often refereed as edges, we need to choose a prior density that can model the edges. 
A standard choice  is to use Gaussian random field priors \citep{bardsley}.
They are easy to use through their analytic properties, that is, they are fully defined by their means and covariances.
The drawback of Gaussian priors is that they are locally smoothing, and thus cause contrast artefacts near  sharp edges. 
Conversely, any contrast-boosting prior needs to be non-Gaussian.

When computing posterior estimates, the unavoidable non-Gaussianity of any contrast-boosting  prior,  leads to computational complexity and high computational cost.
In two-dimensional industrial tomography, with even small or moderate mesh sizes, a typical problem has around 100,000-1,000,000 unknown parameters. 
High-dimensionality in industry poses problems, especially in time-critical applications, where speed is valued.
Optimisation methods are natural starting points as they are computationally less expensive. 
For example, maximum a posteriori (MAP) estimates, expectation-maximisation algorithm, and variational Bayesian methods are flavours of posterior analysis \citep{variational}.  
They provide computational efficiency, but they do not provide full  uncertainty quantification.  

To fully quantify uncertainty, we need to explore the whole posterior, that is,   use Markov chain Monte Carlo (MCMC) techniques. A straightforward choice is to use Metropolis-within-Gibbs (MwG), as the one given in \citet{markkanen:2019}. 
There is a growing number of MCMC algorithms designed especially for high-dimensional problems \citep{Cotter2013,Law2014,Beskos2017}.
Unfortunately, non-Gaussianity of the contrast-boosting priors breaks many of the assumptions of the function-space MCMC methods. For example, there is currently no preconditioned Crank-Nicolson algorithm for the pixel-based Cauchy prior constructions.
MwG is a standard choice, but it typically has some single, more problematic, pixels requiring significantly long MCMC chains.
As an alternative, we shall use Hamiltonian Monte Carlo \citep[HMC,][]{neal2012mcmc}. It has been used for a wide range of applications and it is well-suited for large-dimensional problems \citep{Beskos2011hilbert, Beskos_2013}. In particular, we use the HMC variant with no-U-turn sampling  \citep[HMC-NUTS,][]{hoffman2011nouturn}.

There are several priors, which promote  
contrast-boosting or edge-preserving inversion. 
A standard choice for edge-preserving inversion are the total variation (TV) priors, which are based on using $L^1$-norms. 
In statistics, these methods are called LASSO. 
This method, however, has a drawback, which rises from the finiteness of the     prior  moments. When we make the discretisation denser and denser, the TV-priors converge to Gaussian priors in the discretisation limit. 
This behaviour was studied by \citet{lassas_siltanen:2004}, and they showed that the estimators are not consistent under mesh refinement for Bayesian statistical inverse problems. 
This, naturally, means that doing uncertainty quantification under TV-prior assumption is not consistent with respect to the change of the mesh.

Recently, several  hierarchical models, which promote more versatile behaviours, have been developed. These include, for example, deep Gaussian processes \citep{dunlop_et_al:2018,Emzir2019a},  level-set methods  \citep{Dunlop2017}, mixtures of compound Poisson processes and Gaussians \citep{Hosseini2017}, and stacked Mat\'ern fields via stochastic partial differential equations \citep{roininen_et_al:2019}. 
The problem with hierarchical priors is  that in the posteriors the parameters and hyperparameters  may become  strongly coupled, which means that  vanilla MCMC methods become problematic and, for example, re-parameterisations are needed for sampling the posterior efficiently \citep{chada_et_al:2019,Karla2018}. 
In level-set methods, the number of levels  is  usually low, because experiments have shown that the method  deteriorates when the number of levels is increased.

Here, we utilise a different approach, and make the prior non-Gaussian by construction without hierarchical modelling or  compromising on uncertainty quantification. 
The first choice is to utilise priors based on Besov $\mathcal{B}_{p,q}^{s}$ norms \citep{lassas_saksman_siltanen:2009}.
These priors are constructed on wavelet basis, typically Haar wavelets, and they have well-defined non-Gaussian discretisation limit behaviour.
Besov priors have been utilised in a number of studies for Bayesian inversion, see, e.g., \citet{katinvaitos}. 
The problem with Besov priors is the structure of  wavelets and the truncation of the wavelet series.  The coefficients of the wavelets are  random but  actually not interchangeable,  since their decay is necessary for convergence of the series.
That typically means that we make an unnecessary and strong prior assumption for edge locations  based on wavelet properties, for example, consider a one-dimensional inverse problem on domain (0,1), we prefer an edge at 1/4 over an edge at 1/3 for Haar wavelets. 

General $\alpha$-stable random field priors can be constructed with Karhunen-Lo\`eve (KL) expansions. For KL expansions, see \citet{Berlinet}, and for stable field expansions, see \citet{Samorodnitsky,Sullivan}.
Third option  is to use pixel-based approaches \citep{markkanen:2019, Bolin2014,mendoza:2019}, which cover also some
$\alpha$-stable processes.
As $\alpha$-stable processes, both KL-expansions and pixel-based approaches  are valid, but one should note that the different approaches lead to different statistical objects. 
We stress that both approaches lead to well-posedness of the inverse problem.
In this paper, we shall utilise  certain pixel-based approaches, and limit the discussion to the Cauchy difference priors, which is a special case $\alpha = 1$.

\subsection{Contribution and organisation of this work}

The contributions of this paper are two-fold, first of all we make a large-scale numerical comparison for the X-ray tomography problem with different Gaussian and non-Gaussian (TV, Besov, Cauchy) prior assumptions. To the best of the authors' knowledge, this kind of comparison has not yet been published. Secondly, we apply a carefully designed HMC algorithm to the high-dimensional non-Gaussian inverse problem. In particular, we show its applicability for the sparse-angle tomography problem.

The rest of this paper is organised as follows: In Section \ref{section:priors}, we review Gaussian, TV, Besov and Cauchy priors. In Section \ref{section:samplers}, we introduce the necessary MwG and HMC tools. 
In Section \ref{section:numericalexamples}, we have two synthetic case studies: 3D-imaging of logs in sawmills, and drill-core  tomography problem.
Finally, in Section \ref{section:Conclusion}, we conclude the study, and make some notes on future developments.

\section{Random field priors}
\label{section:priors}

For Gaussian, TV and Besov priors, we write the priors as $
    \pi(X) \propto \exp\left(-G(X)\right)$,
where $G(X)$ is $L^2$-norm of differences of $X$ for the Gaussian case,  $L^1$-norm of differences for TV, and 
Besov  $\mathcal B^s_{p,q}$-norm  for a  wavelet expansion with  random coefficients $X_1,\dots,X_m$ for Besov case. 
We will describe these distributions below and their respective $G(X)$. In the following, we will describe separately the  Cauchy prior, as it is based on another type of construction.

\subsection{Gaussian prior}

Let us assume a zero-mean Gaussian prior, which is fully defined with one of the following choices,
\begin{equation*}
    G(X) = \frac{1}{2}X^T\Sigma^{-1}X = \frac{1}{2}X^TQX = \frac{1}{2}X^TL^TLX = \frac{1}{2}(LX)^TLX,
\end{equation*}
where $\Sigma$ is the covariance matrix, $Q$ is the precision matrix and $L$ is a square-root (e.g., Cholesky factor) of the precision matrix.
Common choices  are exponential and squared exponential covariances, Mat\'ern covariances, and Brownian motion covariance. 
Different choices lead to different kinds of presentations, for example, squared exponential leads to full matrices for all $\Sigma,Q,L$, but with certain choices, the Mat\'ern covariances have sparse $Q,L$ matrices due to the Markov property \citep{roininen_et_al:2019}. 

Two-dimensional difference priors can be obtained by choosing $L$ to be a discretisation of the following operator equation $ \nabla \mathcal{X} = \mathcal{W}$,
where 
$\mathcal W=(\mathcal W_1,\mathcal W_2)^T$ with two statistically independent white noise random fields $\mathcal W_1$ and $\mathcal W_2$. Thus we have $LX=W$ for the discrete presentation, which we will solve in the least-squares sense. 
We note that the precision matrix $Q=L^TL$ is not invertible, thus this is an improper prior, but if we impose zero-boundary conditions, we have a proper prior. For X-ray tomography zero-boundary conditions are natural choices, as typically the domain of interest is larger than the object of interest, and thus putting zero-boundary is justified. 
Now, let us denote by $X_{i,j}$ the unknown at pixel $(i,j)$.
As a summation formula with zero-boundary conditions, we can write the prior then as 
\begin{equation*}
    G_{\mathrm{Gauss}}(X)     = G_{\mathrm{pr}}(X) + G_{\mathrm{boundary}}(X)
    = \sum_{i=1}^I\sum_{j=1}^J \left(\frac{\left(X_{i,j}-X_{i-1,j}\right)^2}{\sigma^2_{\mathrm{pr}}}+ \frac{\left(X_{i,j}-X_{i,j-1}\right)^2}{\sigma^2_{\mathrm{pr}}} \right)+
    \sum_{i,j\in \partial\Omega} \frac{X_{i,j}^2}{\sigma^2_{\mathrm{boundary}}},
\end{equation*}
where by $\partial\Omega$ we mean the boundary indices of the discretised domain $\Omega$, $\sigma^2_{\mathrm{pr}}$ is regularisation parameter, and $\sigma^2_{\mathrm{boundary}}$ zero-boundary parameter.

\subsection{Total variation prior}

Similarly to the Gaussian prior, we can write a two-dimensional TV prior with 
\begin{equation*}
    G_{\mathrm{TV}}(X)     = G_{\mathrm{pr}}(X) + G_{\mathrm{boundary}}(X)
    = \sum_{i=1}^I\sum_{j=1}^J \alpha \left( \left\vert X_{i,j}-X_{i-1,j}\right\vert+  \left\vert X_{i,j}-X_{i,j-1}\right\vert \right)+ \sum_{i,j\in \partial\Omega} \alpha_{\mathrm{boundary}}
     \left\vert X_{i,j}\right\vert,
\end{equation*}
where $\alpha, \alpha_{\mathrm{boundary}} $ are regularisation parameters. 
We note that in statistics, these distribution are often called Laplace distributions.
This form of TV prior is anisotropic. According to \citet{isotropic}, it can be shown that  an isotropic form is obtained when we choose 
\begin{equation*}
    G_{\mathrm{pr}}(X) = \sum_{i=1}^I\sum_{j=1}^J \alpha \left( \left( X_{i,j}-X_{i-1,j}\right)^2+  \left( X_{i,j}-X_{i,j-1}\right)^2 \right)^{1/2}.
\end{equation*}
Using higher-order TV priors would be possible \citep{htv}, but often the basic first order TV prior is selected because it preserves sharp edges within the reconstruction much better. 
The major drawback of TV prior is that due to its finite moments, the prior is not discretisation-invariant and it resembles Gaussian difference prior on dense enough meshes. It is notable that  MAP estimate and the Conditional Mean (CM) estimate differ drastically of each other in the limit \citep{lassas_siltanen:2004}.  
While the inconsistency is not desirable, the prior is still used in many practical applications. 
In the numerical experiments, we use anisotropic TV prior.

\subsection{Besov prior}

Gaussian and TV priors are based on Gaussian and Laplace distributions.
Besov priors are based on wavelet coefficients of the reconstruction, and in practise calculated  by discrete wavelet transform (DWT) \citep{lassas_saksman_siltanen:2009}. 
The continuous  wavelet transform decomposes the reconstruction into approximation and detail coefficients. 
For detailed treatise on Besov priors, see  \citep{katinvaitos}. 
Besov prior  for  $X$ derives from  wavelet expansion for  function $\mathcal X$
\begin{equation}
\label{expansion}
     \mathcal{X} = \sum_{d_1=0}^{2^{k}-1} \sum_{d_2=0}^{2^{k}-1}   \left\langle\,  \mathcal{X}  , \phi_{k,d_1,d_2}\right\rangle \phi_{k,d_1,d_2} + \sum_{r=k}^{\infty} \sum_{d_1=0}^{2^r-1}\sum_{d_2=0}^{2^r-1}\sum_{t=1}^{3}  \left\langle\,  \mathcal{X}  , \psi_{r,d_1,d_2,t}\right\rangle \psi_{r,d_1,d_2,t},
\end{equation}
and its Besov norm is
\begin{equation*} 
    \left\Vert\mathcal X\right\Vert_{\mathcal B^s_{p,q}}  = 
    \left(\sum_{d_1=0}^{2^{k}-1} \sum_{d_2=0}^{2^{k}-1}  \left\vert \left\langle\,  \mathcal{X}  , \phi_{k,d_1,d_2}\right\rangle
    \right\vert ^p   +   \sum_{r=k}^{\infty} 2^{rq(s+1-2/p)} \sum_{d_1=0}^{2^r-1}\sum_{d_2=0}^{2^r-1}\sum_{t=1}^{3}
    \left\vert \left\langle\,  \mathcal{X}  , \psi_{r,d_1,d_2,t}\right\rangle
    \right\vert^p \right)^{1/p}.
\end{equation*}
In Equation \eqref{expansion}, $\phi$ is the father wavelet and $\psi$ is the mother wavelet. 
The subindices refer to different scales and translations of the functions \citep{Meyer}. 
We utilise the prior for Besov space $\mathcal B^1_{1, 1}$, which reduces the norm into absolute sum of wavelet coefficients. Other Besov spaces  than $\mathcal B^1_{1, 1}$ would be doable as well, but in those cases the wavelet coefficients would have different weights or they would be raised to other powers than one.  
We use Haar wavelets as our father and mother wavelets. 

In practise, we compute the two-dimensional DWT of  $X$, which is now represented as a reconstruction of size of $2^n \times 2^n,\, n\in \mathbb{N}$ square pixels. 
The Besov prior is defined by
 \begin{equation*}
     G_{\text{Besov}}(X) =  \sum_{d_1=0}^{2^{k}-1} \sum_{d_2=0}^{2^{k}-1}  \left\vert a_{k,d_1,d_2}\right\vert + \sum_{r=k}^{k_{\max}} \sum_{d_1=0}^{2^r-1}\sum_{d_2=0}^{2^r-1}\sum_{t=1}^{3} \left\vert b_{r,d_1,d_2,t}\right\vert.
\end{equation*}
Terms $a_{k,d_1,d_2}$ refer to the approximation coefficients of the DWT of $X$ and terms $b_{r,d_1,d_2,t}$ to the corresponding detail coefficients.   As DWT algorithm, we choose  a matrix operator method described by \citet{matriisi}. A more common and faster way to calculate the DWT would be to use fast wavelet transform, but we want to keep track of how changing the value of an individual pixel alters the wavelet coefficients and therefore the prior, too. This property is needed in the MwG algorithm, because we propose new samples for each component of $X$ in a sequential manner and calculating the whole DWT every time for all the components  would be too time-consuming.  

Besov  priors are discretisation-invariant and therefore  should remain consistent if the resolution is increased \citep{lassas_saksman_siltanen:2009}. Their drawback is the tendency to produce block artefacts into the point estimates: the wavelet coefficients at certain low scale are different, even though they are adjacent to each other. This makes Besov  priors location-dependent, that is, they are not translation-invariant.

\subsection{Cauchy  prior}

While the previous three priors were constructed based on different types of norms, Cauchy priors are constructed based on Cauchy walk, or more profoundly  general $\alpha$-stable random walk. 
These one-dimensional objects have well-defined limits, and the generalisation in \citep{markkanen:2019} to two-dimensional setting is given as a prior density 
\begin{equation*}
    \pi (X) \propto \prod_{i=1}^{I}\prod_{j=1}^{J} \frac{h\lambda}{\left(h\lambda\right)^2 + \left(X_{i,j} - X_{i-1,j}\right)^2}\frac{h\lambda}{(h\lambda)^2 + \left(X_{i,j} - X_{i,j-1}\right)^2},
\end{equation*}
where $h$ is discretisation step in both coordinate directions, and $\lambda$ is the regularisation parameter.
These priors have been used for X-ray tomography in oil and gas industry \citep{mendoza:2019} and subsurface imaging \citep{muhumuza2019bayesianbased}. 

A second formulation of the Cauchy difference prior was derived in \citep{chada_et_al:2019}. 
In that, the starting point was $\alpha$-stable sheets, which lead to difference approximations of the form
\begin{equation*}
    \pi (X) \propto \prod_{i=1}^{I}\prod_{j=1}^{J} \frac{h^2\lambda}{\left(h^2\lambda\right)^2 + \left(X_{i,j} - X_{i-1,j}-X_{i,j-1}+X_{i-1,j-1}\right)^2}.
\end{equation*}
We note that Cauchy difference prior is theoretically justified in the discretisation limit, which is not the case with TV prior. 
Unlike  Gaussian difference prior, Cauchy difference prior favours small increments and steep transitions relatively much more. This means that Cauchy difference prior preserves  edges within the lattice and does not smooth them out. A disadvantage of Cauchy difference prior is its anisotropic nature.

\section{Posterior estimates}
\label{section:samplers}

The two most common estimators drawn from the posterior density \eqref{eqn:bayes} are the MAP and CM estimates.
For MAP estimation, we will use Limited-memory BFGS (L-BFGS) optimisation algorithm, which belongs to the family of quasi-Newtonian methods.
For CM estimation, we shall use MCMC methods, as they also enable uncertainty quantification of the estimators.

\subsection{Adaptive Metropolis-within-Gibbs}

We shall use single-component adaptive Metropolis-Hastings (SCAM) \citep{haario2005}  to generate samples from the posterior distributions. It is an adaptive variation of the MwG algorithm, which  is a generalisation of Gibbs sampling. 
In Gibbs sampling, the samples are generated by generating new values for each component of the target distribution from the component-wise conditional distributions. However, if it is not possible to obtain samples from such distributions analytically, MwG can be used instead, since it utilises Metropolis-Hastings acceptance/rejection step for proposing new values for the components. In theory, both methods are applicable also in very high-dimensional distributions.  

The adaptive MwG algorithm  adapts a Gaussian transition kernel for each component during the burn-in period. This means that algorithm calculates the variance of each component in the chain and sets  new proposal variances for them. By this way, the algorithm does not require manual tuning, unlike the non-adaptive MwG algorithm does. 
We note that self-adaptation does not fix the fundamental problem of component-wise sampling -- the algorithm is unable to sample from distributions, which support consists of diagonally separated regions, since the algorithm cannot jump from one region to another without transforming the coordinate system of the distribution. If the distribution is heavy-tailed, the variance-based adaptation might not converge to a reasonable proposal variance at all because the conditional distributions have infinite or undefined variances. Single component robust adaptive Metropolis-Hastings (RAM) \citep{Vihola_2011} tunes the proposal variance according to the acceptance ratio of each component  so far in the chain and hence should work better with heavy-tailed distributions.  Furthermore, strong correlations in high-dimensional distributions might impair the sampling efficiency of component-wise samplers.

Another feature of MwG is that since the candidate generating kernel is centred at the previous sample and no matter what the proposal covariance is, there will be significant correlation between the samples in the chain. This decreases the effective sample size of the generated samples, since they are not actually independently drawn  from the target distribution.  
%

\subsection{Hamiltonian Monte Carlo}

Here, we will summarise the Hamiltonian Monte Carlo \citep[HMC,][]{neal2012mcmc} method. For an expanded conceptual descriptions of the HMC, the reader is directed to \citet{betancourt2017}.
In HMC, the geometric information of the typical set is used to guess of a new sample location. The Hamiltonian dynamic is used in to determine the path between the current sample position and the next sample. Specifically, one considers the negative logarithm of the probability density function of the target distribution as a potential energy function \citep{neal2012mcmc,barp}. Then one introduces an auxiliary momentum   $P$ and kinetic energy function $K(P)$ with suitable mass matrix $M$, which are used along with the potential energy function to form the Hamiltonian function:
\begin{equation*}
H(X,P) = U(X) + K(P) = -\log\left(P\right) + K(P) = -\log\left(P\right) + \frac{P^T M^{-1} P}{2}.
\end{equation*}
In order to sample from the target distribution, one starts following a Hamiltonian trajectory from the previous sample in the chain by setting $X^*= X_{k-1}$ and by sampling a  momentum vector from its proposal distribution. Then, one approximates the following Hamiltonian equations with a symplectic integrator, like St\"ormer-Verlet: 
\begin{equation*}
\frac{\partial H(X^*,P)}{\partial P} = \frac{dX^*}{dt}, \quad
\frac{\partial H(X^*,P)}{\partial X^*} =-\frac{dP}{dt}.
\end{equation*}
As Hamiltonian is time-reversible and the trajectories preserve their volume in the phase space, the new proposal sample can be selected to be the last value of $X^*$ at the end of the trajectory. The final step is to apply a Metropolis acceptance/rejection step to correct for the numerical approximation error caused by time discretisation. The computational complexity of generating the samples scales with $m^\frac{5}{4}$ \citep{Beskos_2013}, whereas random walk Metropolis-Hastings algorithms scale with $m^2$  \citep{neal2012mcmc}.

Selecting appropriate step size of the symplectic integrator and the corresponding trajectory length is apparent problems of the basic HMC. If the step size is too big, the value of Hamiltonian is not preserved so precisely and most of the proposals are rejected. If it is too small, the extra computational burden makes the algorithm impractically slow. Also, too long overall trajectory length both wastes time and actually increases the change that the trajectory comes close to the initial state, whereas too short trajectory evidently means that the trajectory always ends near the starting point. A   self-adaptive  HMC algorithm No-U-Turn Sampler (NUTS) adapts the integration step size by changing its value with respect to acceptance ratio so far in the chain by routine called Dual Averaging \citep{hoffman2011nouturn}. NUTS does not use a fixed trajectory length, since it calculates the trajectory randomly in the opposite directions from the starting point. Furthermore, it uses a set of heuristics  to remain the  reversibility and to stop   simulating trajectories until  they start to turn back. These facts render NUTS a promising method for sampling from high-dimensional distributions with possibly severe correlations. However, even NUTS might need setting an appropriate mass  matrix to sample efficiently from  the target distribution. One way to select a mass matrix is to run preliminary MCMC chains to calculate an empirical covariance for the distribution and use it as the mass matrix. If the distribution is very high-dimensional, the mass matrix might be just diagonal. More sophisticated approach is to utilise  mass matrices, which are not constant. That is the case in  Riemannian Manifold HMC algorithms \citep{girolami2009riemannian}.

\section{Numerical experiments}
\label{section:numericalexamples}

We have two numerical experiments, the log tomography and the drill-core tomography.
The measurement setting is the same for both test cases: We shall use X-ray tomography with parallel beam geometry. The computational grid is $512\times 512$, so we have  262,144-dimensional posterior distributions. We have additive white noise models, according to Equation \eqref{eqn:direct}, we fix the standard deviation of the noise  to  1.5\% of the maximum value of the  line integrals.
The parameters of each prior distribution at each measurement scenario are set using a grid search. We select the parameter value, which gives approximately a minimum $L^2$-error between the MAP estimate given by the parameter value and the ground truth. Furthermore, in the log tomography case, we use a different log slice without a high density object to select the parameters by the grid search.  
For Besov  prior, we use 8 levels in the DWT.

For log tomography, we have a foreign object inside, thus it is a high-contrast target. The second experiment is a lower-contrast target with lots of distinct regions with different attenuation coefficient.
In simulations, we shall use the four priors, L-BFGS, adaptive MwG, and HMC-NUTS. 
The initial points of MCMC chains are  set to be the MAP estimate obtained with L-BFGS.
As MCMC chain lengths, we use 500,000 samples in MwG to adapt a proposal variance for each component and we use those variances to generate  400,000 samples to calculate CM estimates. For NUTS, we let the algorithm run 100 times, at which time  the dual averaging algorithm adapts a step length for the St\"ormer-Verlet symplectic integrator. Then we use that step length to generate 4,000 samples for CM estimates.

\subsection{Log tomography}

\begin{figure}
    \centering
    \begin{subfigure}[b]{3.3cm}
    \centering
    \includegraphics[height=3.0cm]{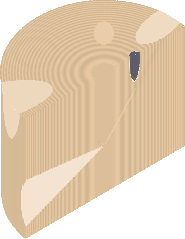}
    \caption{3D ground truth.}
    \label{gt3d}
    \end{subfigure} \hspace{1cm}
    \begin{subfigure}[b]{3.3cm}
    \centering
    \includegraphics[height=3.0cm]{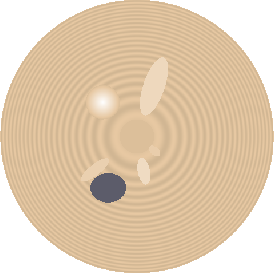}
    \caption{2D slice.}
    \label{gt2d}
    \end{subfigure}
    \hspace{1cm}
    \begin{subfigure}[b]{0.3cm}
    \includegraphics[height=4cm]{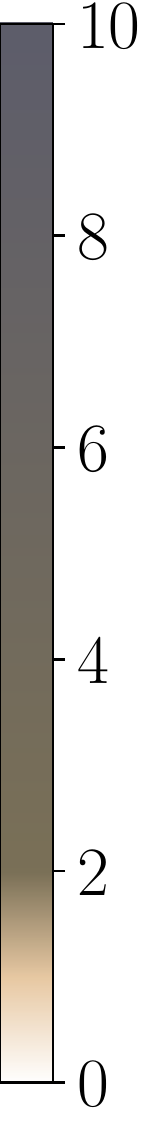} 
    \end{subfigure}

    \begin{subfigure}[b]{3.3cm}
    \includegraphics[height=3.3cm]{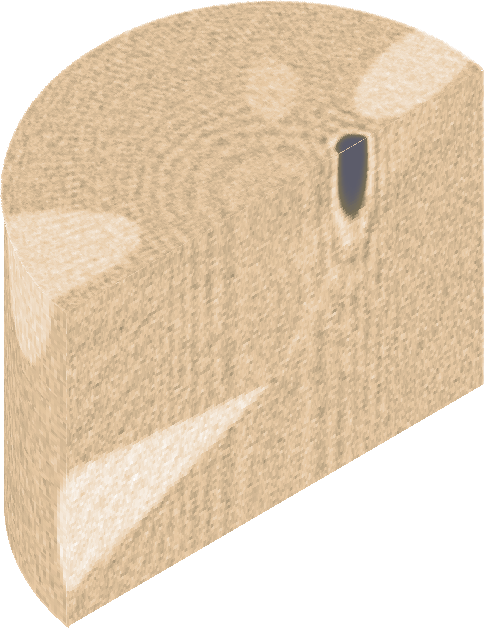}
    \caption{Gaussian, 90}\label{tikhonov3d90}\end{subfigure}   
    \begin{subfigure}[b]{3.3cm}
    \includegraphics[height=3.3cm]{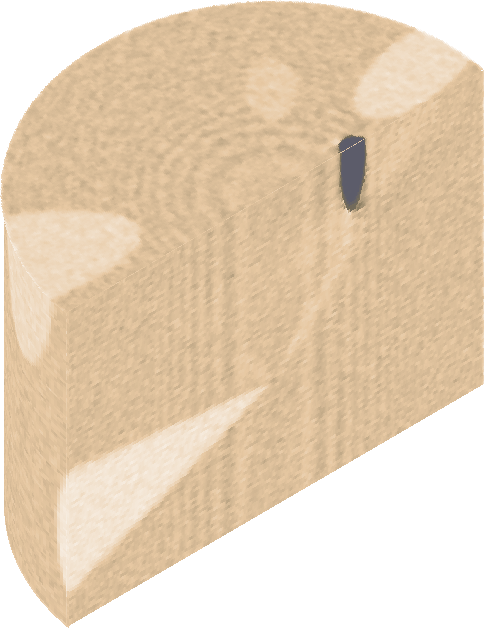}
    \caption{TV, 90}\label{tv3d90}\end{subfigure}   
    \begin{subfigure}[b]{3.3cm}
    \includegraphics[height=3.3cm]{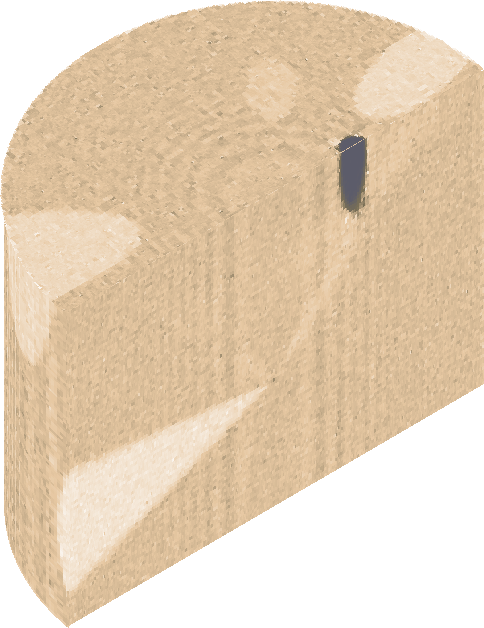}
    \caption{Besov, 90}\label{haar3d90}\end{subfigure}   
    \begin{subfigure}[b]{3.3cm}
    \includegraphics[height=3.3cm]{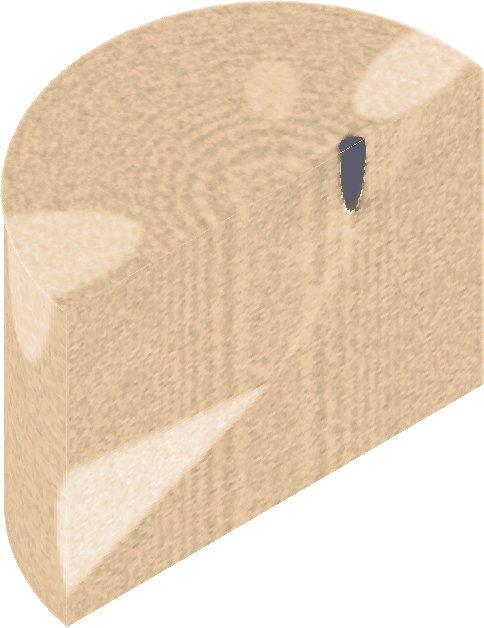}
    \caption{Cauchy, 90}\label{cauchy3d90}\end{subfigure}

    \centering
    \begin{subfigure}[b]{3.3cm}
    \includegraphics[height=3.3cm]{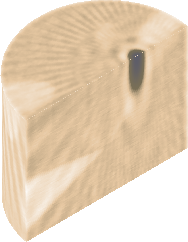}
    \caption{Gaussian, 30}\label{tikhonov3d30}\end{subfigure}   
    \begin{subfigure}[b]{3.3cm}
    \includegraphics[height=3.3cm]{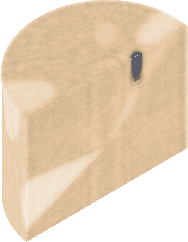}
    \caption{TV, 30}\label{tv3d30}\end{subfigure}   
    \begin{subfigure}[b]{3.3cm}
    \includegraphics[height=3.3cm]{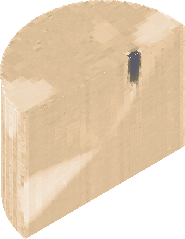}
    \caption{Besov, 30}\label{haar3d30}\end{subfigure}   
    \begin{subfigure}[b]{3.3cm}
    \includegraphics[height=3.3cm]{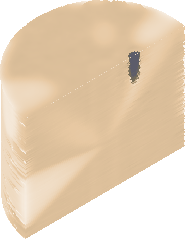}
    \caption{Cauchy, 30}\label{cauchy3d30}\end{subfigure}   

    \centering
    \begin{subfigure}[b]{3.3cm}
    \includegraphics[height=3.3cm]{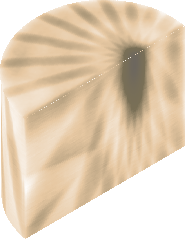}
    \caption{Gaussian, 10}\label{tikhonov3d10}\end{subfigure}   
    \begin{subfigure}[b]{3.3cm}
    \includegraphics[height=3.3cm]{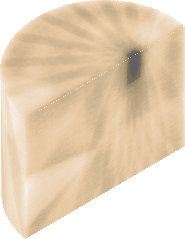}
    \caption{TV, 10}\label{tv3d10}\end{subfigure}   
    \begin{subfigure}[b]{3.3cm}
    \includegraphics[height=3.3cm]{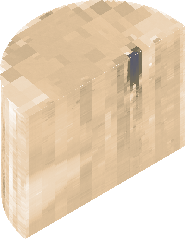}
    \caption{Besov, 10}\label{haar3d10}\end{subfigure}   
    \begin{subfigure}[b]{3.3cm}
    \includegraphics[height=3.3cm]{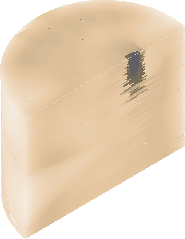}
    \caption{Cauchy, 10}\label{cauchy3d10}\end{subfigure}   
    
   \caption{Three-dimensional ground truth and two-dimensional ground truth with knots (light material) and metallic piece (black). Three-dimensional reconstructions obtained by stacking two-dimensional MAP estimates with different angles and priors} 
    \label{slices}
\end{figure}

Our test case is visualised in Figure \ref{slices}, with metallic piece in black and knots in light color, and a small rotten part in near white with more smooth behavior.
Besides the ground truth, we have also plotted the three-dimensional estimates, obtained by stacking two-dimensional MAP estimate slices computed with L-BFGS.
We have all the four different priors, and the number of equispaced measurement angles is 10, 30 or 90.
While all the methods seem to capture the main features in densest-data case, reducing the number of measurement leads to severe artefacts with Gaussian and TV priors, and the Besov prior promotes blocky estimate. Cauchy prior seems to capture the metallic piece even in the most difficult case, but it struggles  with knots.

\begin{figure}
    \begin{subfigure}[b]{3.3cm}
    \includegraphics[height=3.3cm]{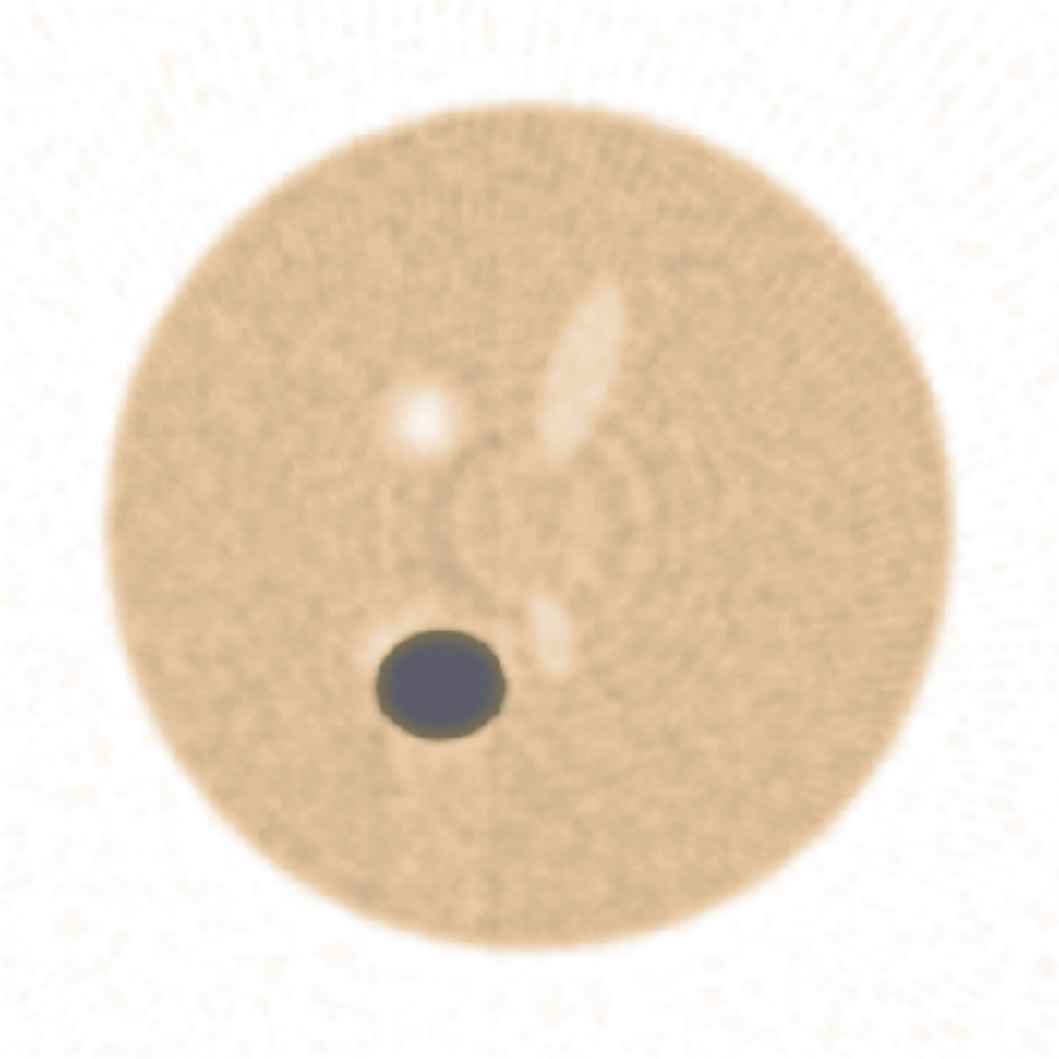}
    \caption{Gaussian, 90}\label{tikhonov18090}
    \end{subfigure}
    \begin{subfigure}[b]{3.3cm}
    \includegraphics[height=3.3cm]{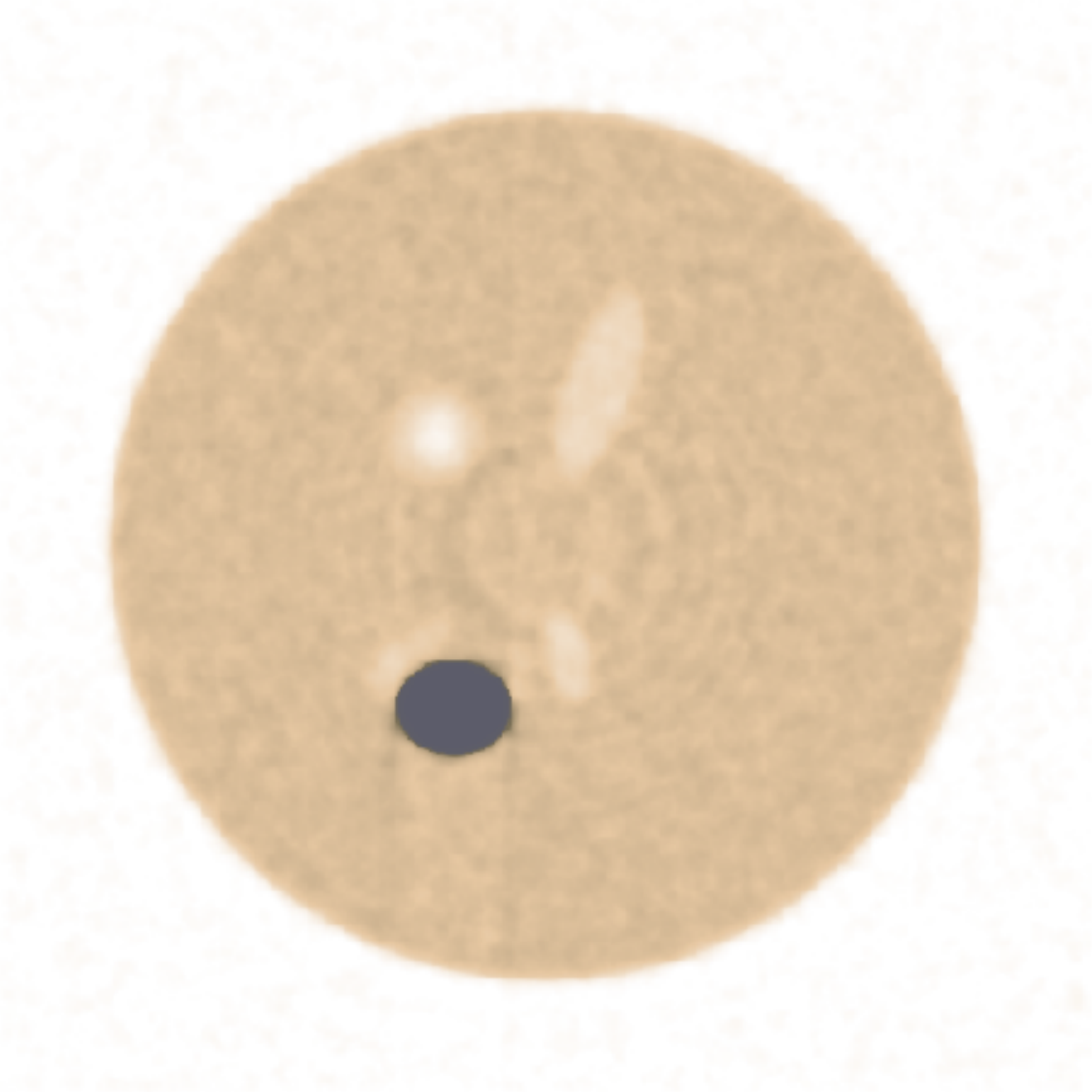}
    \caption{TV, 90 }\label{tv18090}\end{subfigure}
    \begin{subfigure}[b]{3.3cm}
    \includegraphics[height=3.3cm]{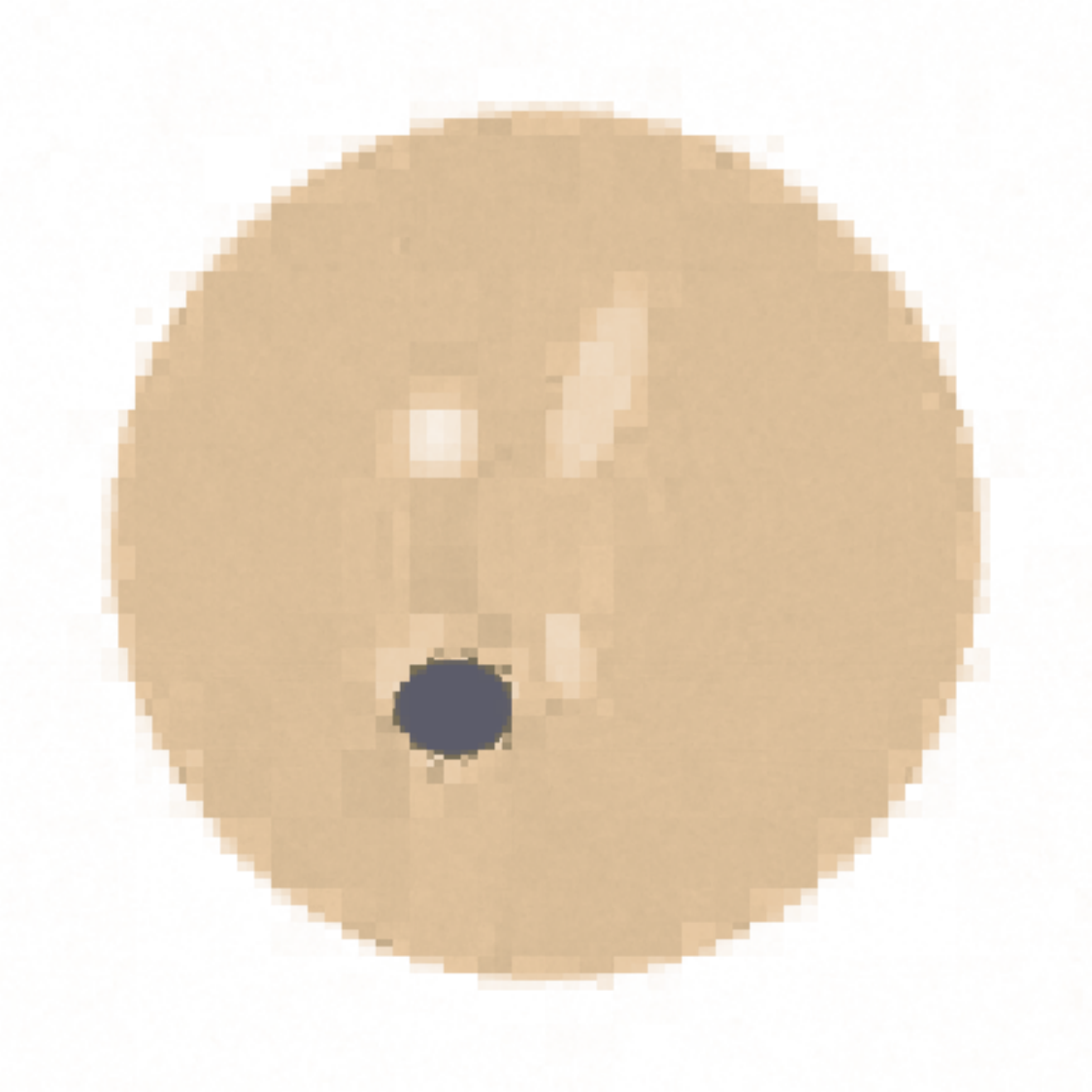}
    \caption{Besov, 90}\label{haar18090}
    \end{subfigure}
    \begin{subfigure}[b]{3.3cm}
    \includegraphics[height=3.3cm]{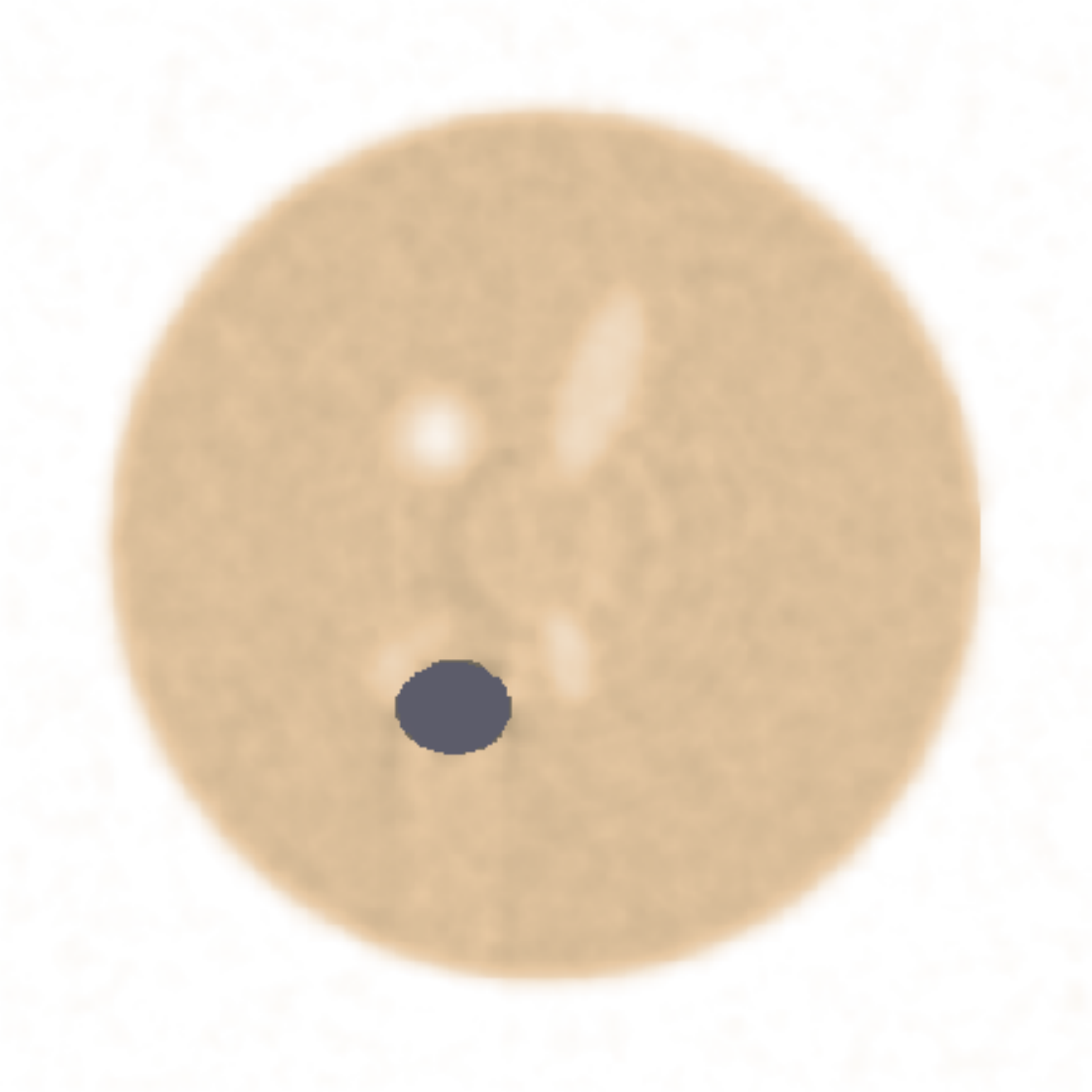}
    \caption{Cauchy, 90 }\label{cauchy18090}
    \end{subfigure}
    
     \begin{subfigure}[b]{3.3cm}
    \includegraphics[height=3.3cm]{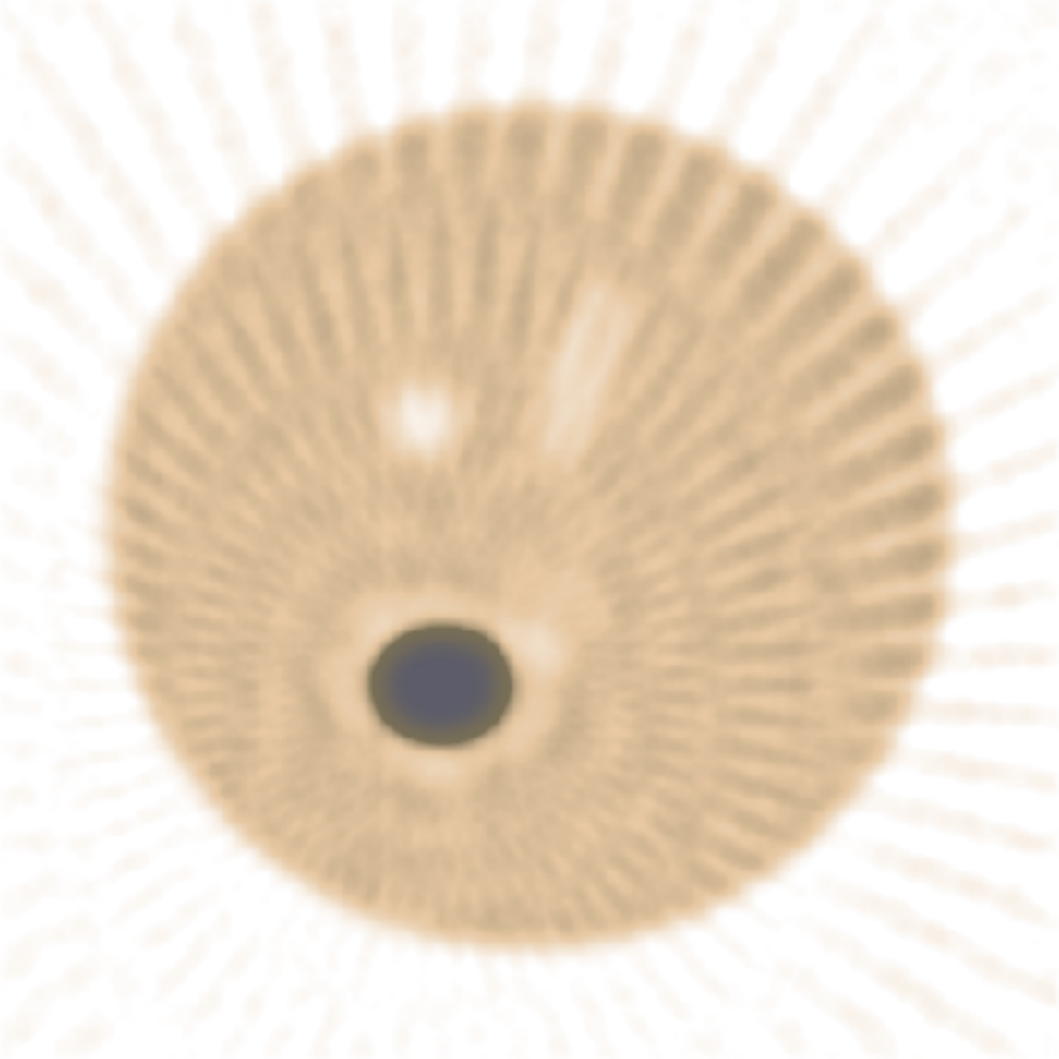}
    \caption{Gaussian, 30}\label{tikhonov18030}
    \end{subfigure}
    \begin{subfigure}[b]{3.3cm}
    \includegraphics[height=3.3cm]{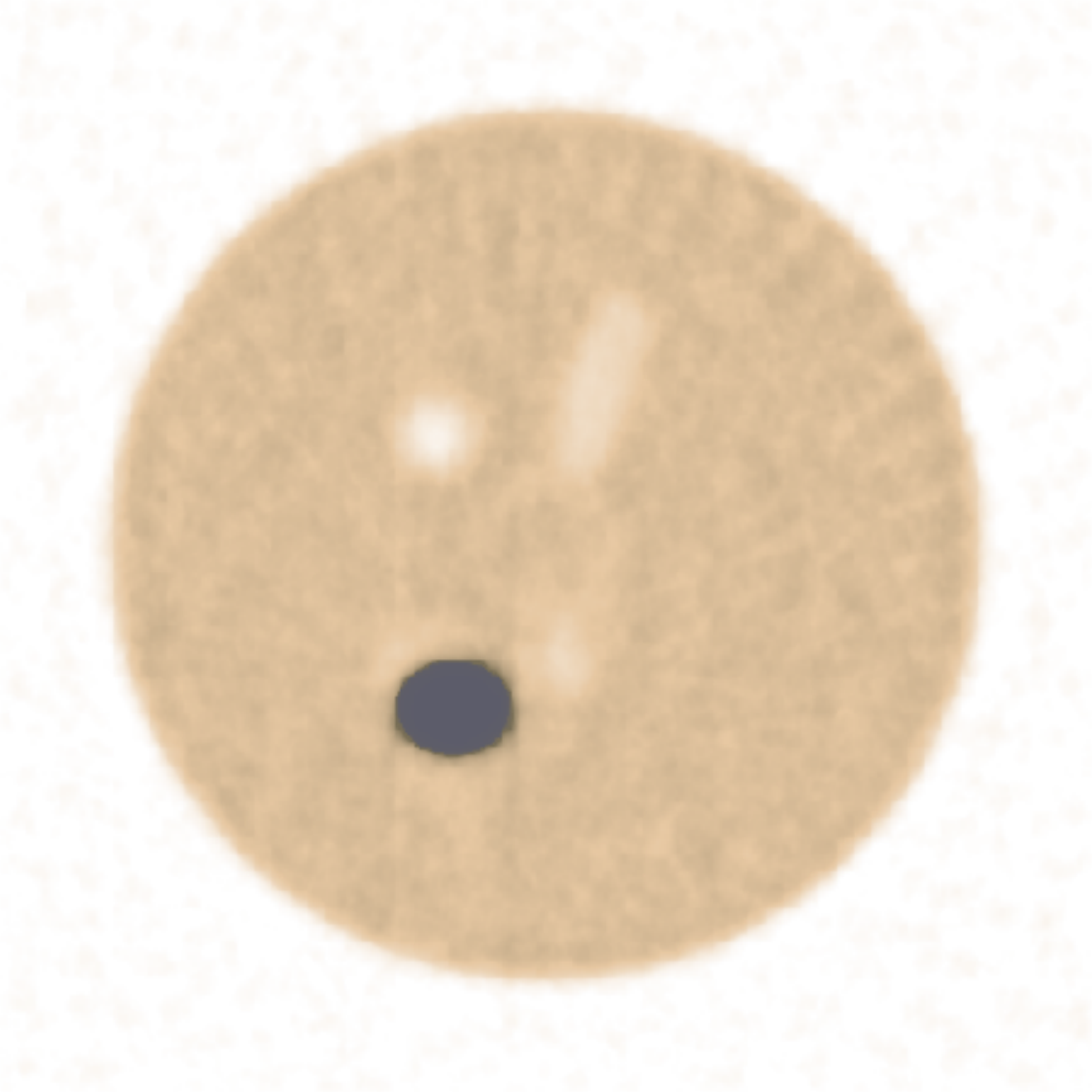}
    \caption{TV, 30}\label{tv18030}
    \end{subfigure}
     \begin{subfigure}[b]{3.3cm}
    \includegraphics[height=3.3cm]{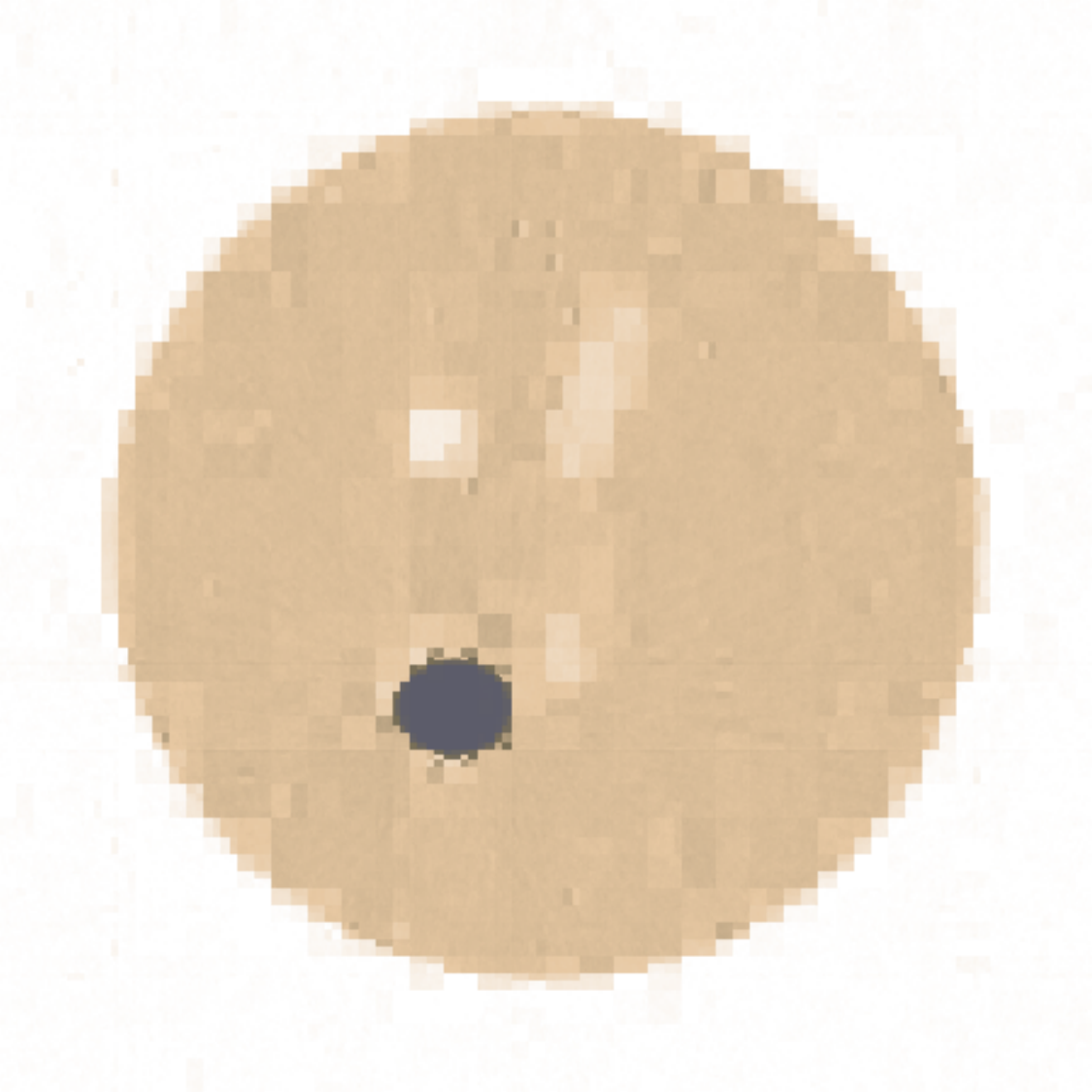}
    \caption{Besov, 30}\label{haar18030}
    \end{subfigure}
     \begin{subfigure}[b]{3.3cm}
    \includegraphics[height=3.3cm]{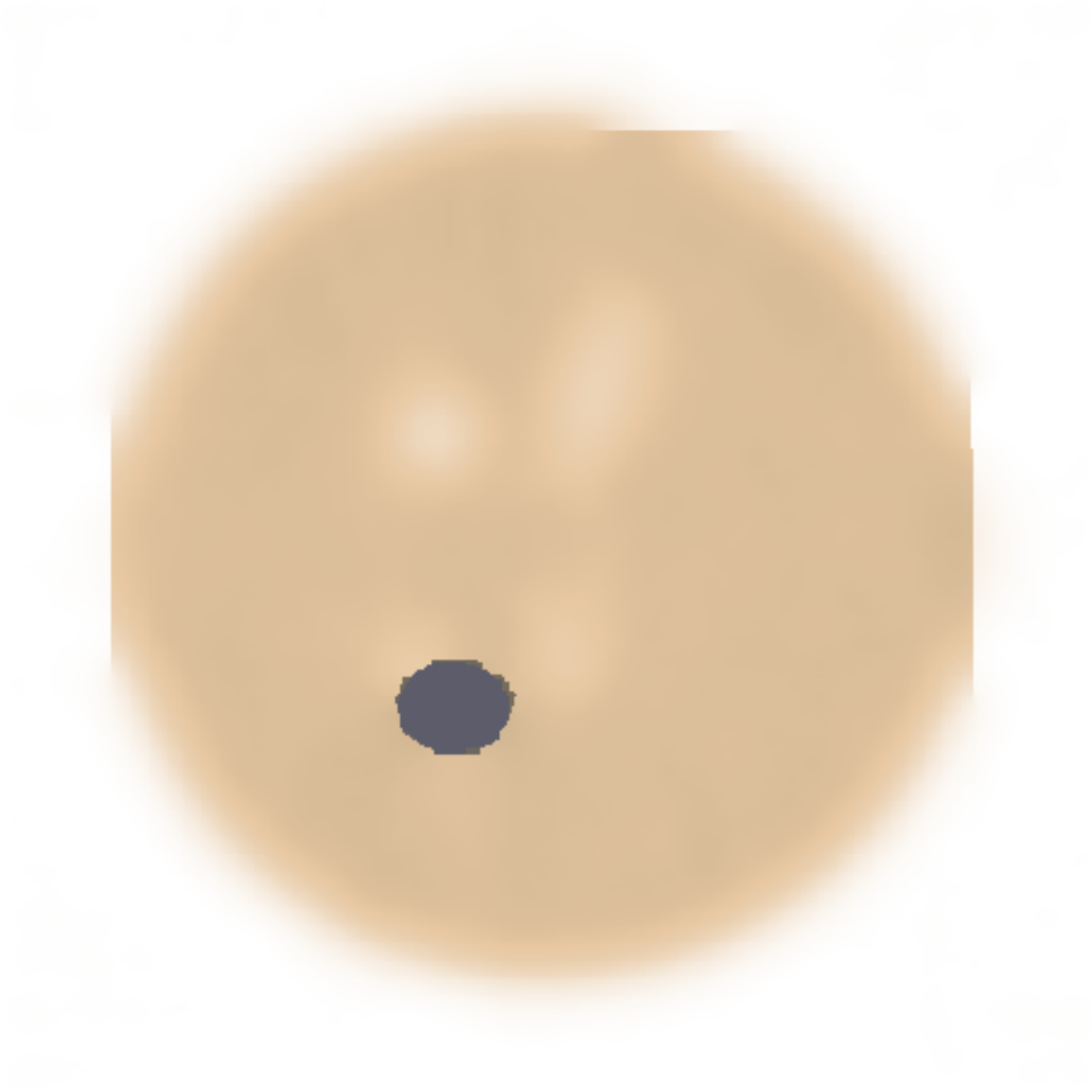}
    \caption{Cauchy, 30}\label{cauchy18030}
    \end{subfigure}
    \begin{subfigure}[b]{0.3cm}
    \includegraphics[height=4cm]{figs/wood/colorbar.pdf} 
    \end{subfigure}

    \begin{subfigure}[b]{3.3cm}
    \includegraphics[height=3.3cm]{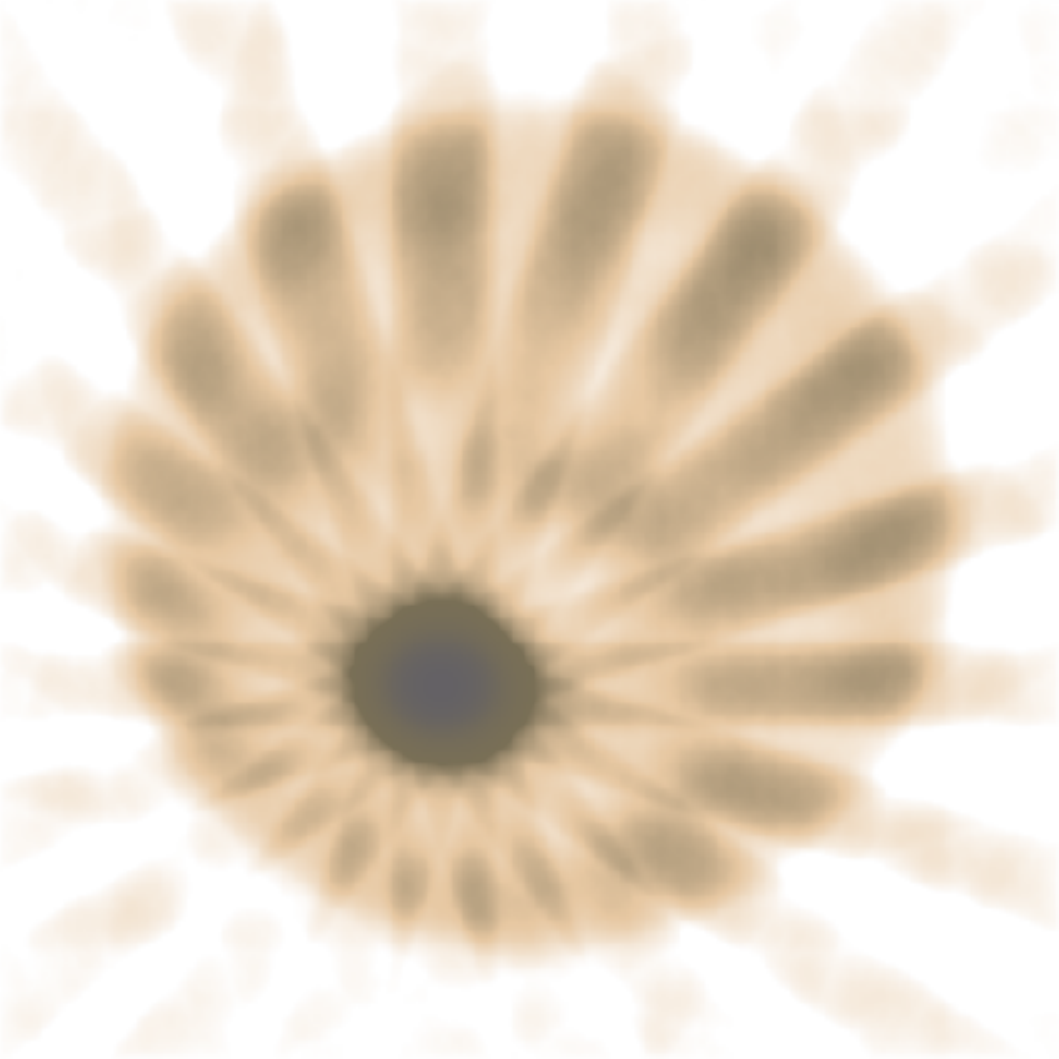}
    \caption{Gaussian, 10}\label{tikhonov18010}
    \end{subfigure}  
    \begin{subfigure}[b]{3.3cm}
    \includegraphics[height=3.3cm]{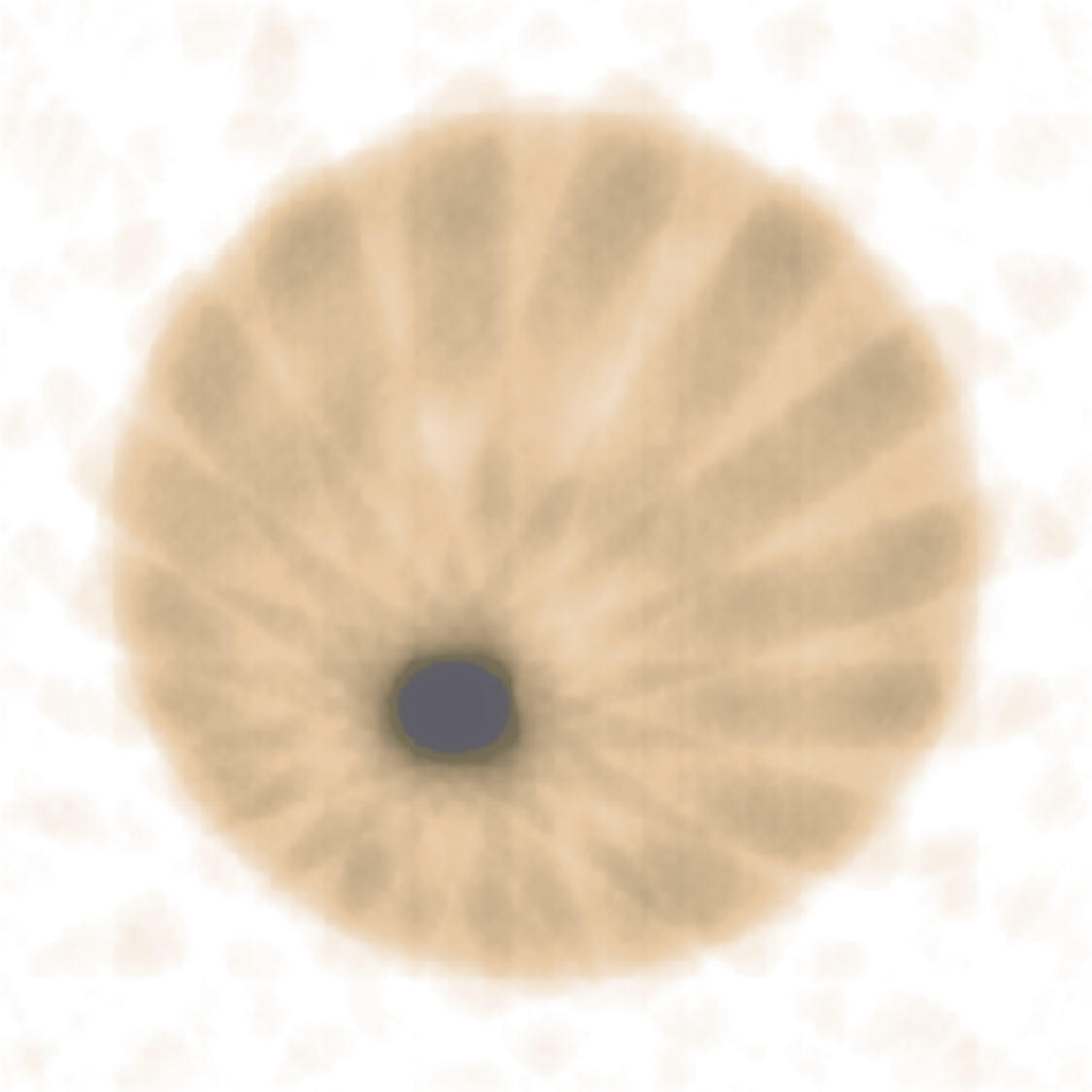}
    \caption{TV, 10}\label{tv18010}
    \end{subfigure}
     \begin{subfigure}[b]{3.3cm}
    \includegraphics[height=3.3cm]{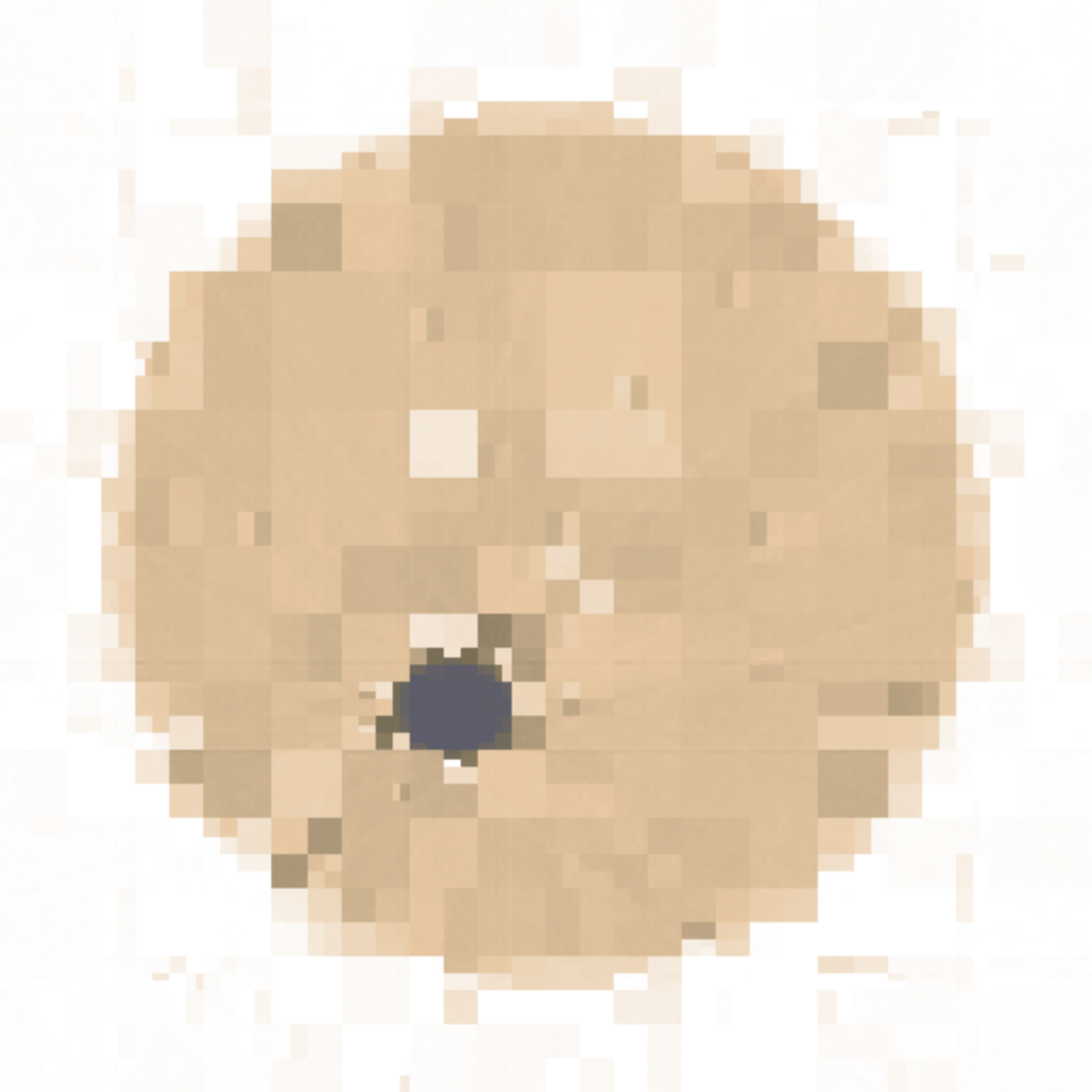}
    \caption{Besov, 10}\label{haar18010}
    \end{subfigure}
     \begin{subfigure}[b]{3.3cm}
    \includegraphics[height=3.3cm]{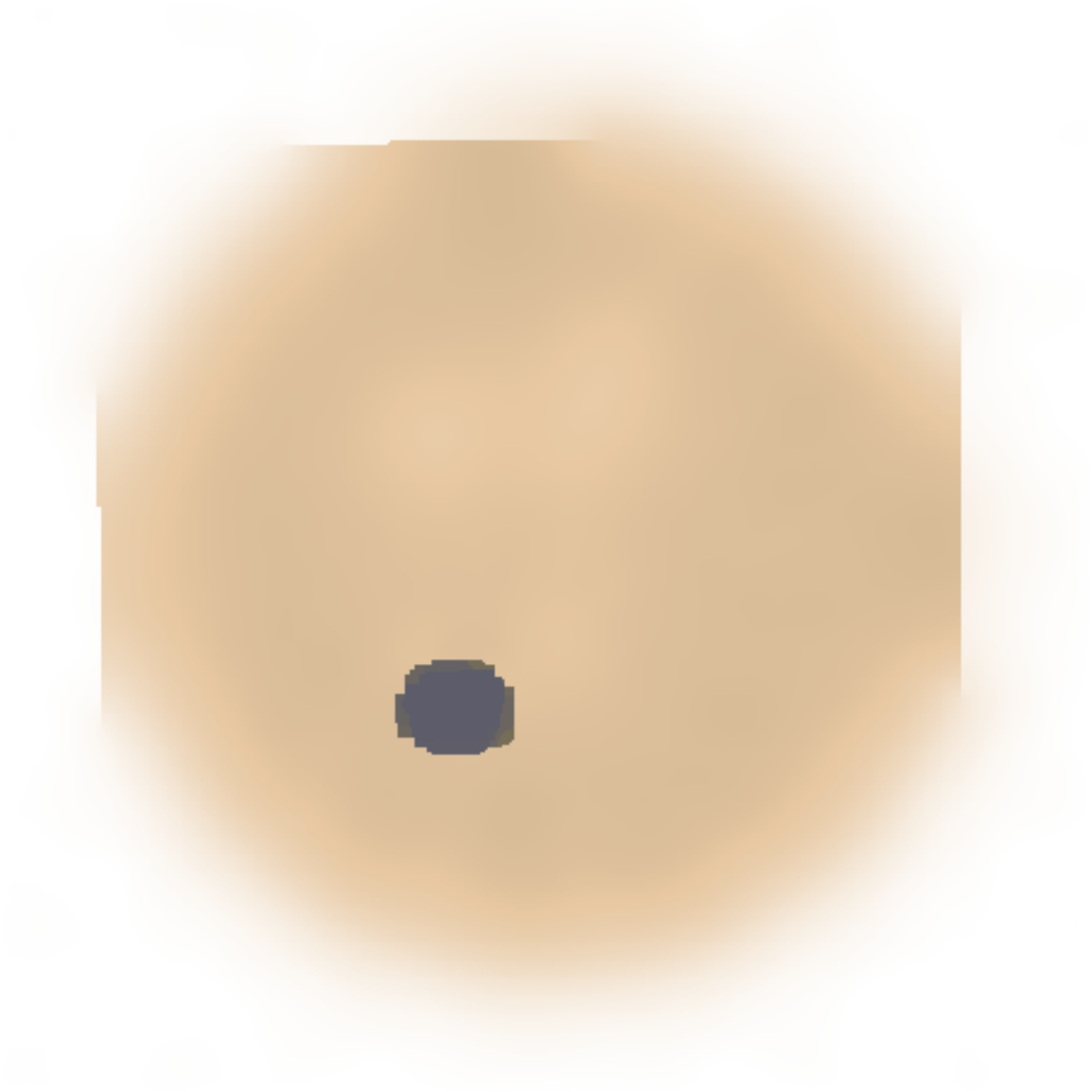}
    \caption{Cauchy, 10}\label{cauchy18010}
    \end{subfigure}

    \caption{Two-dimensional MAP estimates with different measurement angles and prior assumptions} 
    \label{logwide_map}
\end{figure}

It is easier to see the effects of different priors in the two-dimensional slices in Figure \ref{logwide_map}.
The densest scenario, with 90 angles, TV and Cauchy  priors produce very good MAP estimates recovering the edges. 
MAP estimate with Besov  prior looks blocky.
In the MAP estimate with Gaussian  prior, the edges of the metallic object are smoothed. 
Reducing the number of measurement angles causes visible artefacts into MAP-estimates with TV and Gaussian priors, but TV prior preserves the edges much better. 
Meanwhile, the rotation-dependency of Cauchy difference prior becomes apparent with coordinate-axis aligned sharp artefacts. 
The Besov prior MAP estimate remains more or less consistent even if the number of angles is decreased. 
However, in practice the block artefacts will spoil the MAP estimates and that is not useful in practical applications. 
We could decrease the number of wavelet transform levels in order to reduce blockiness, or use another wavelet family.

\begin{figure}
    \begin{subfigure}[b]{3.3cm}
    \includegraphics[height=3.3cm]{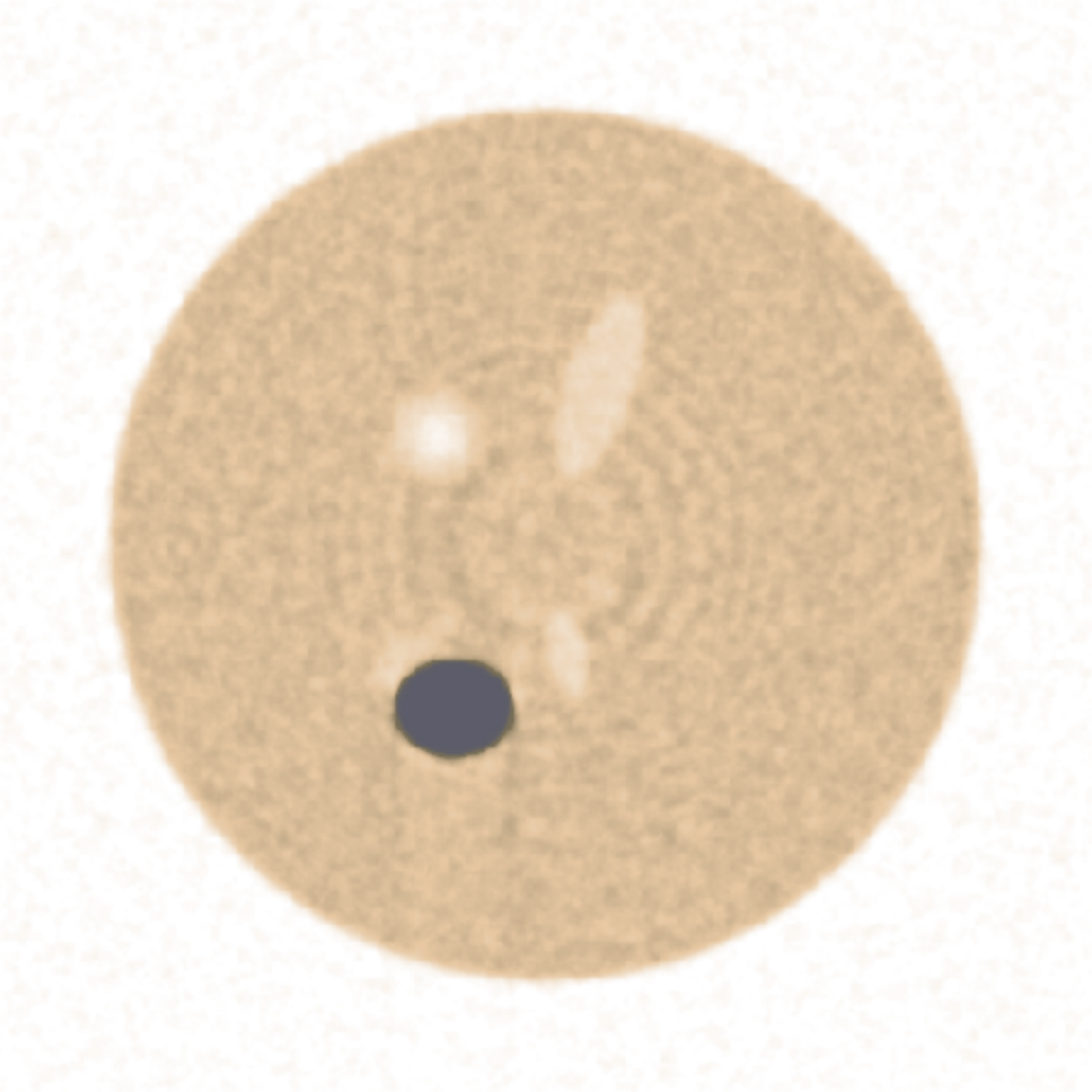}
    \caption{MwG, TV,  90  }\label{mwgtv18090}\end{subfigure}
    \begin{subfigure}[b]{3.3cm}
    \includegraphics[height=3.3cm]{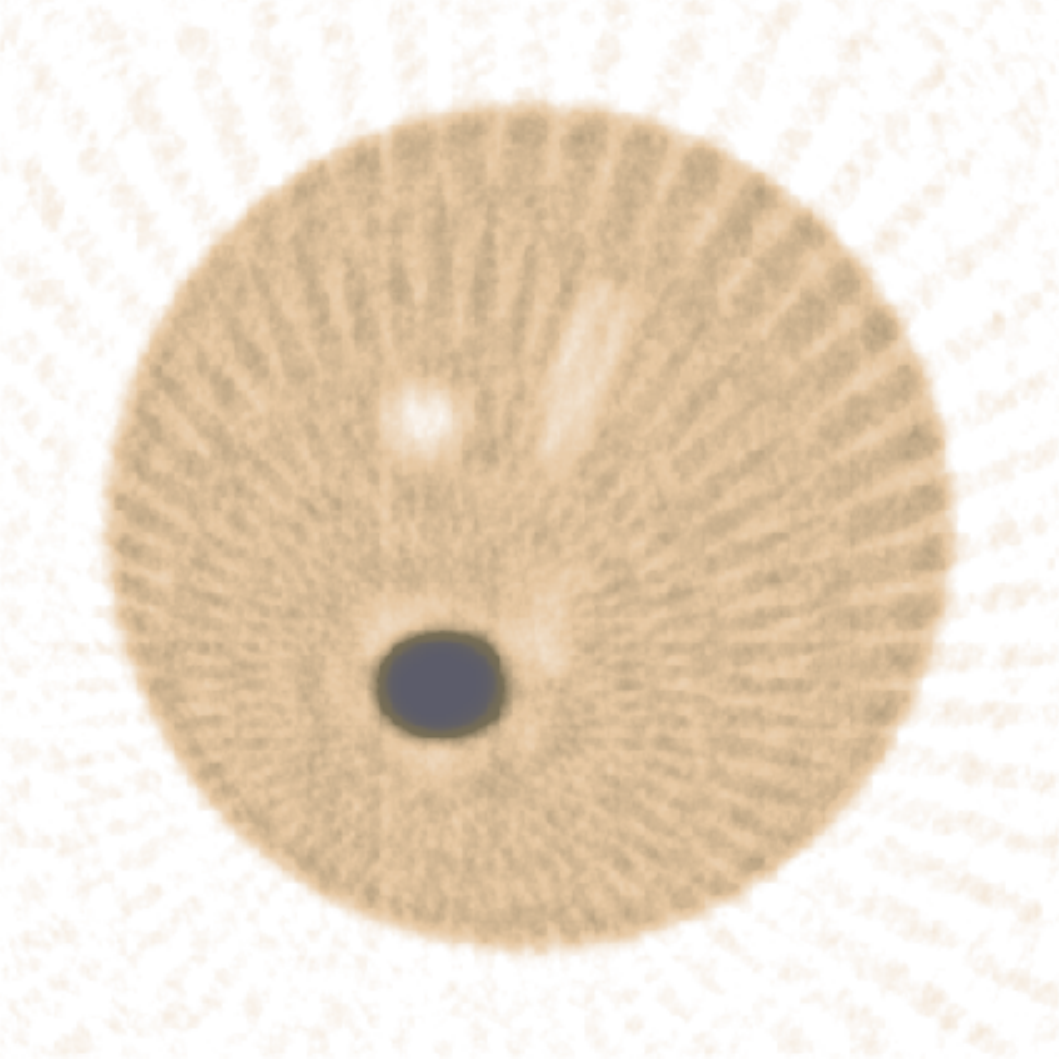}
    \caption{MwG, TV, 30  }\label{mwgtv18030}\end{subfigure}
    \begin{subfigure}[b]{3.3cm}
    \includegraphics[height=3.3cm]{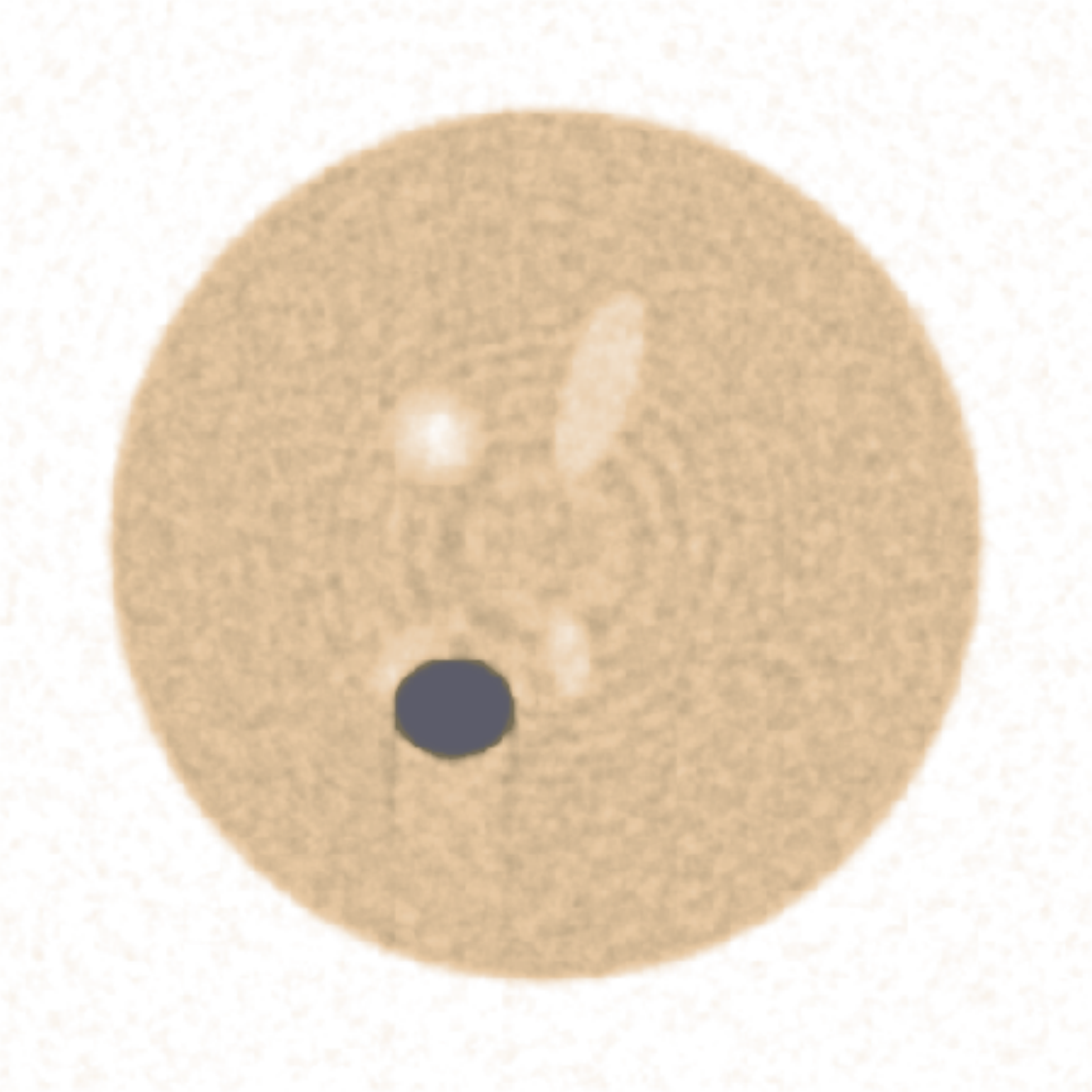}
    \caption{HMC, TV, 90}\label{hmctv18090}\end{subfigure}
    \begin{subfigure}[b]{3.3cm}
    \includegraphics[height=3.3cm]{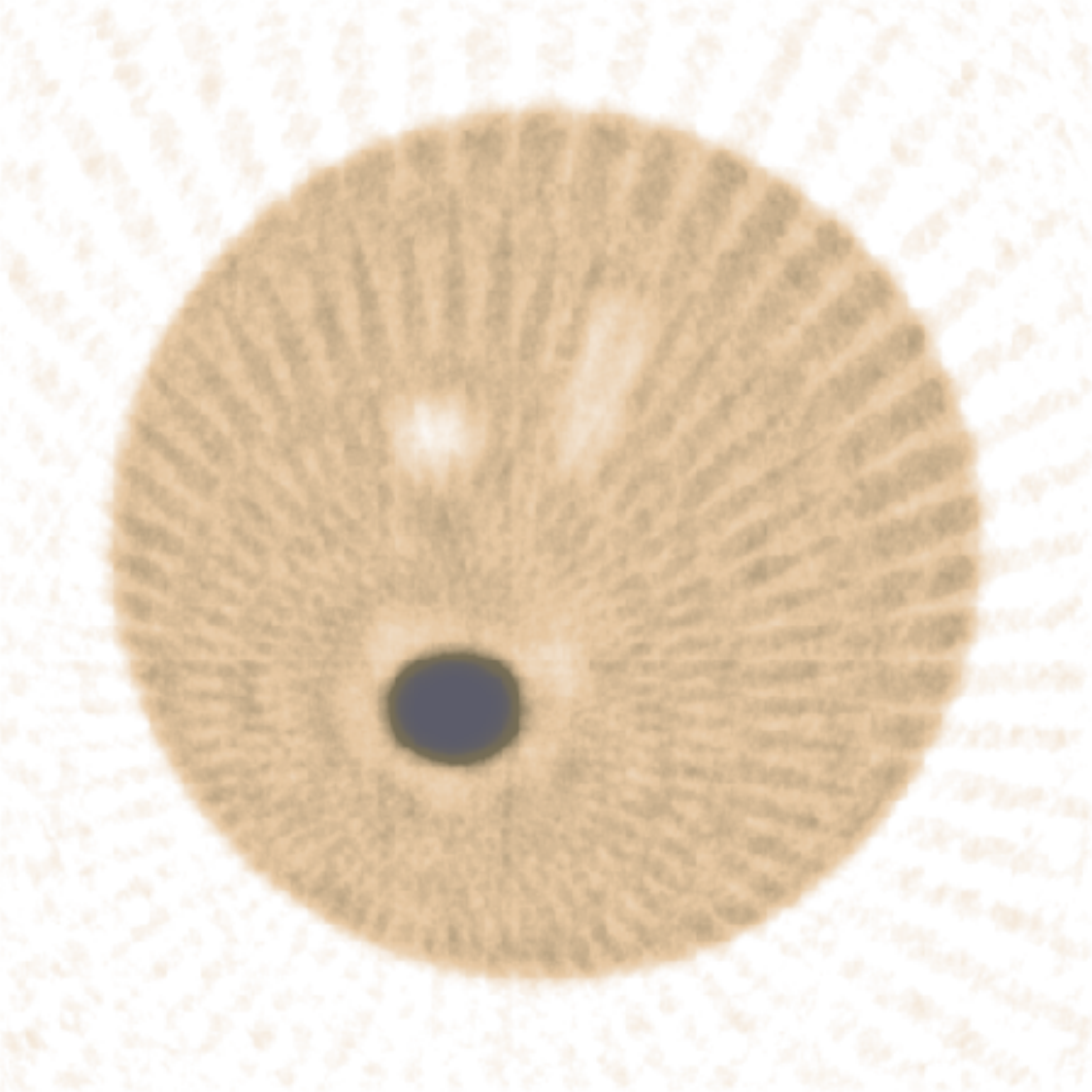}
    \caption{HMC, TV, 30  }\label{hmctv18030}\end{subfigure}

    \begin{subfigure}[b]{3.3cm}
    \includegraphics[height=3.3cm]{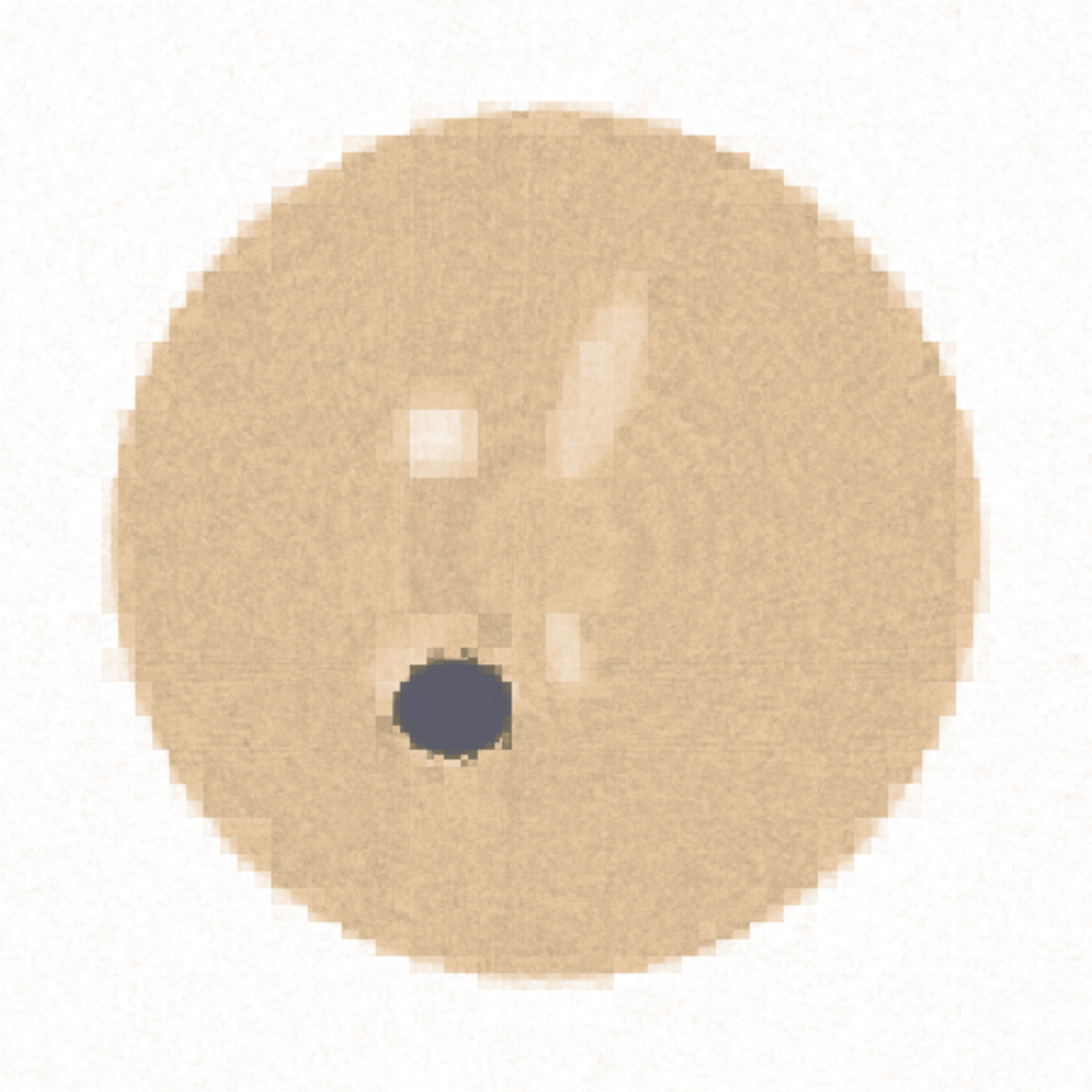}
    \caption{MwG, Besov, 90  }\label{mwghaar18090}\end{subfigure}     
     \begin{subfigure}[b]{3.3cm}
    \includegraphics[height=3.3cm]{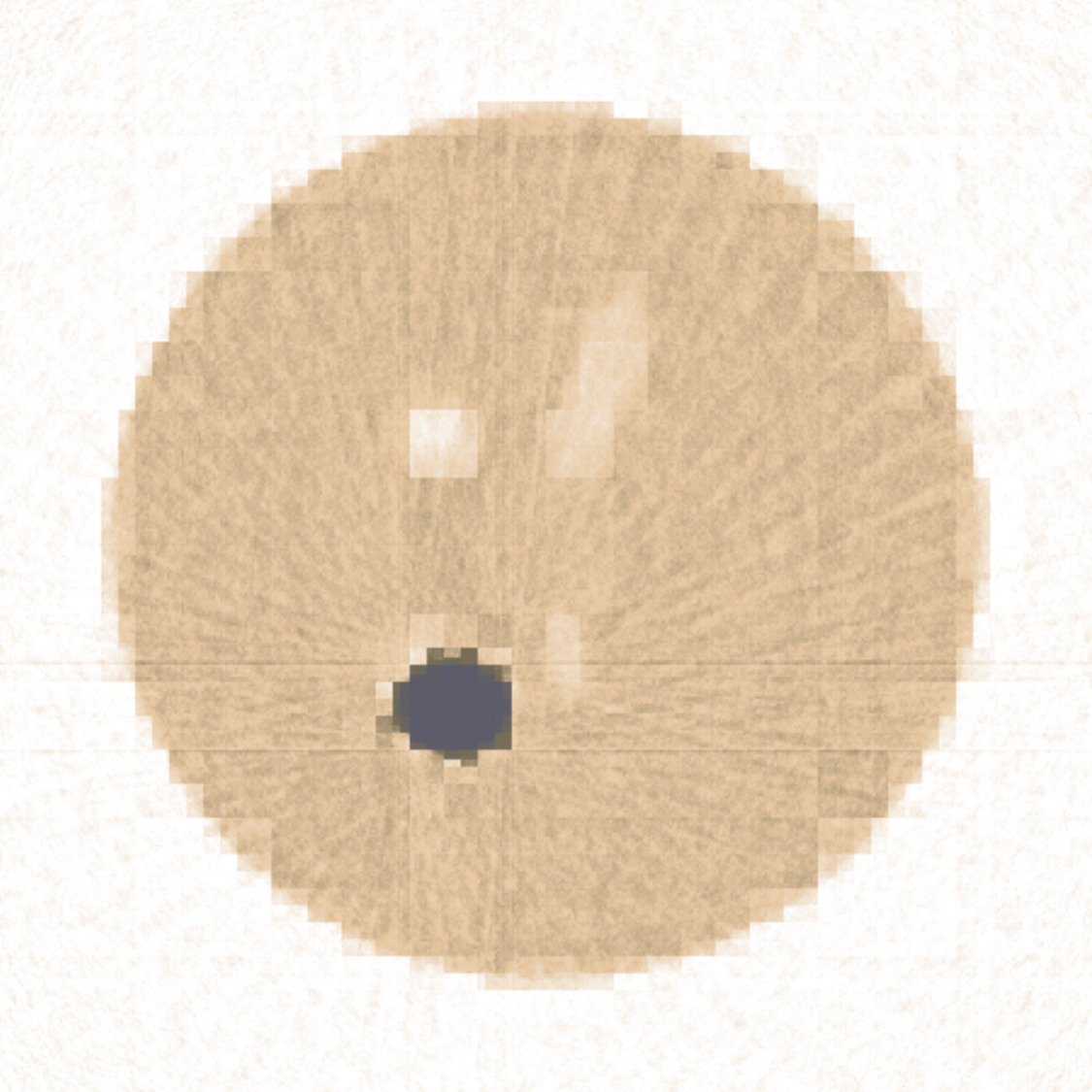}
    \caption{MwG, Besov,  30  }\label{mwghaar18030}\end{subfigure} 
     \begin{subfigure}[b]{3.3cm}
    \includegraphics[height=3.3cm]{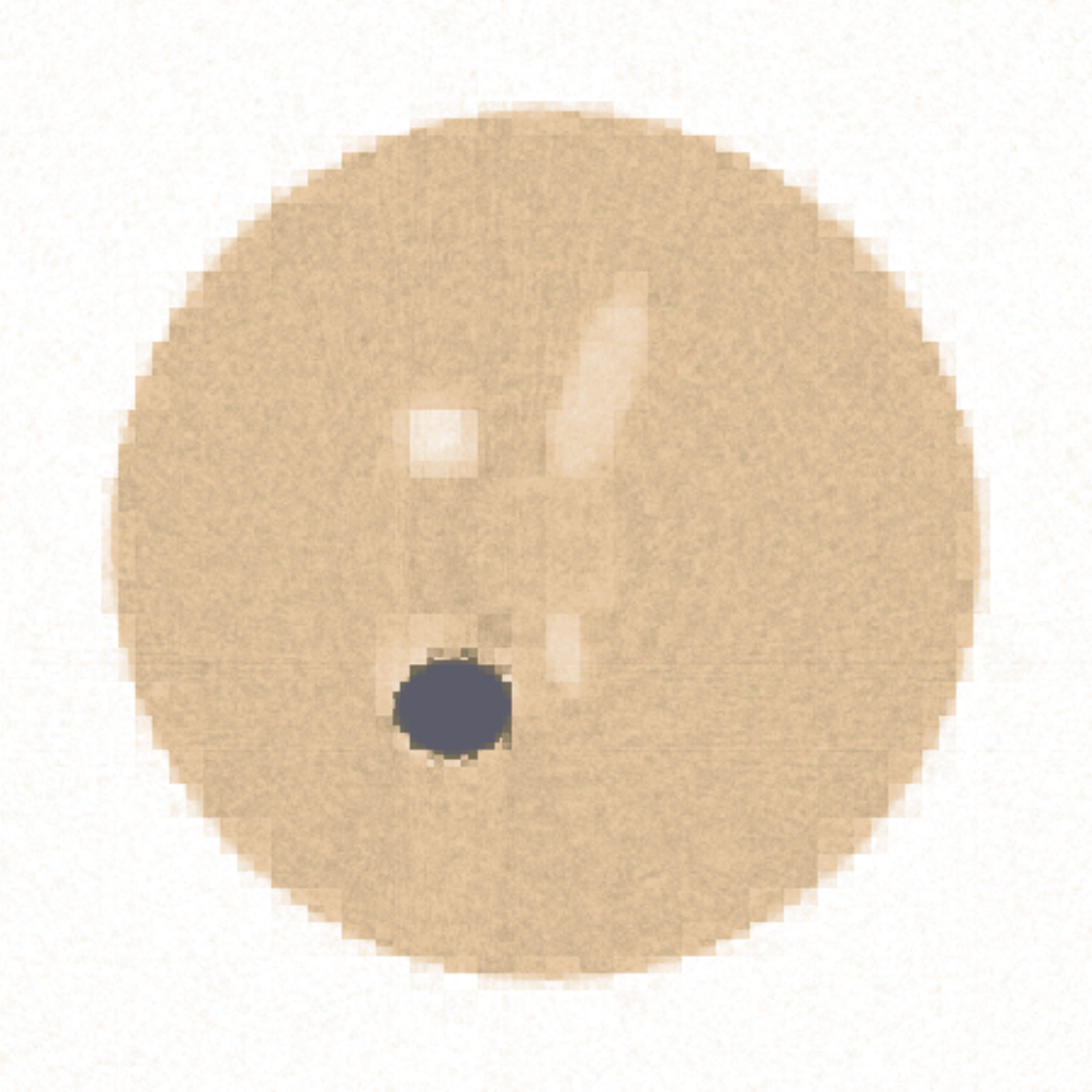}
    \caption{HMC, Besov,  90  }\label{hmchaar18090}\end{subfigure}
     \begin{subfigure}[b]{3.3cm}
    \includegraphics[height=3.3cm]{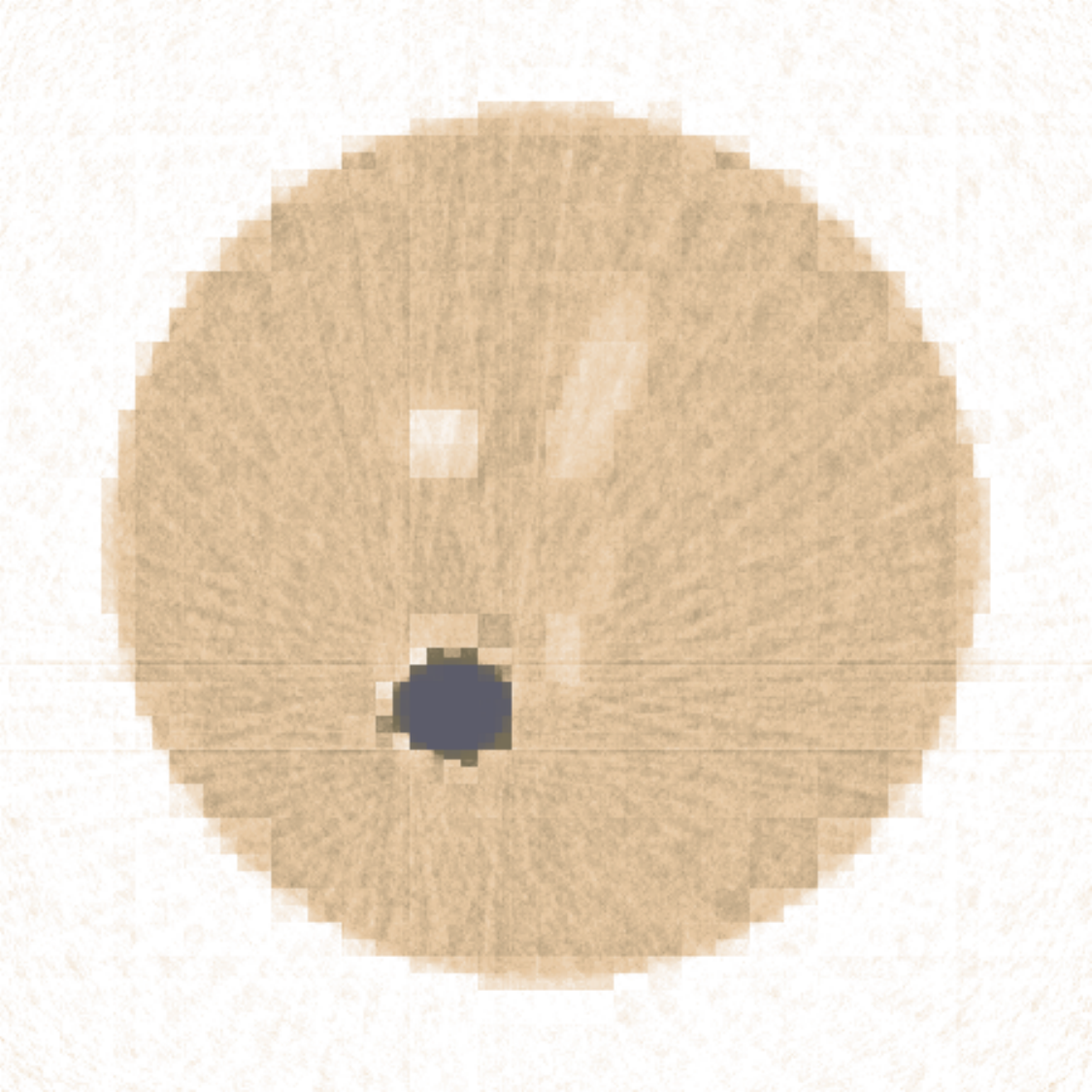}
    \caption{HMC, Besov,  30  }\label{hmchaar18030}\end{subfigure}
    \begin{subfigure}[b]{0.3cm}
    \includegraphics[height=4cm]{figs/wood/colorbar.pdf} 
    \end{subfigure}

    \begin{subfigure}[b]{3.3cm}
    \includegraphics[height=3.3cm]{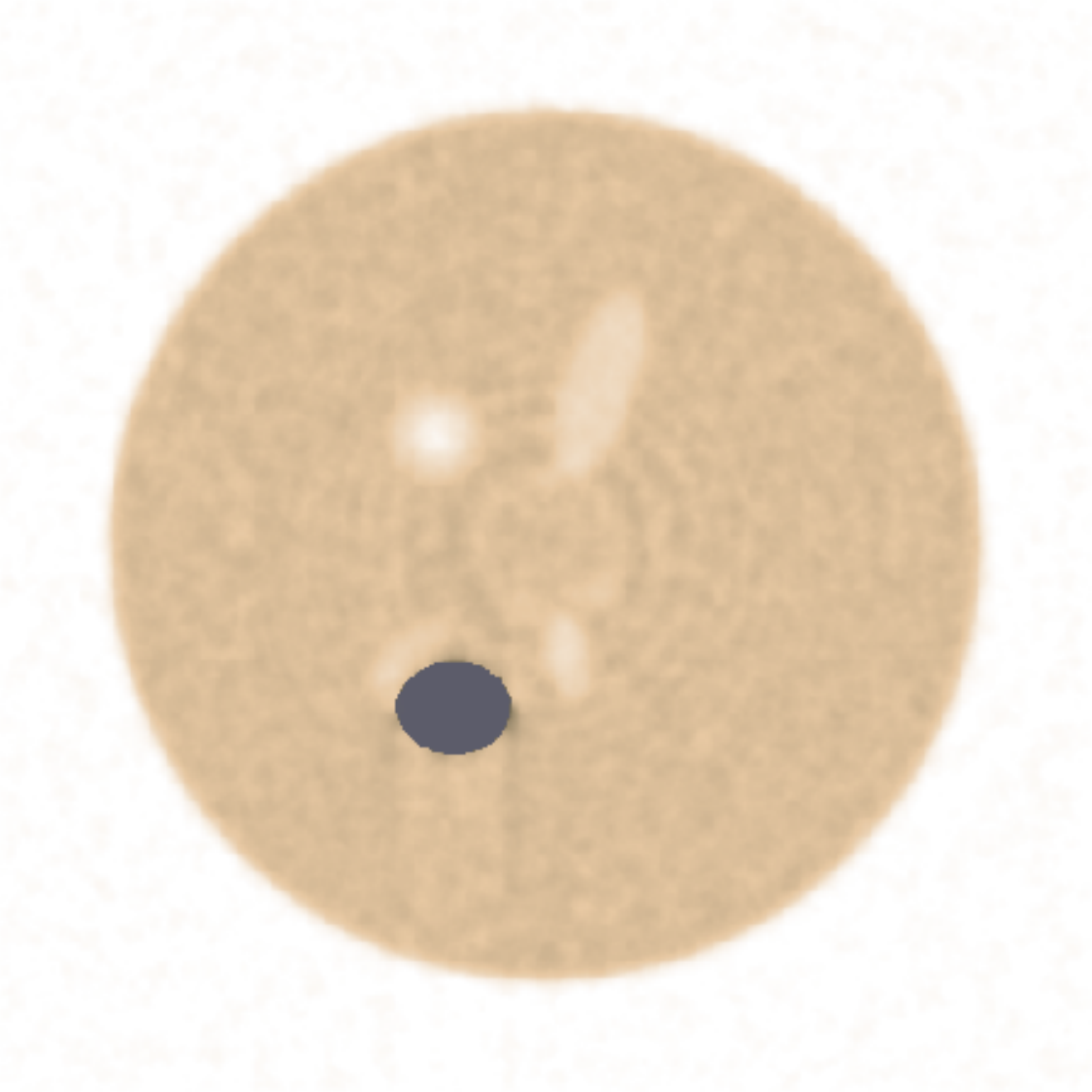}
    \caption{MwG, Cauchy, 90 }\label{mwgcauchy18090}\end{subfigure}
     \begin{subfigure}[b]{3.3cm}
    \includegraphics[height=3.3cm]{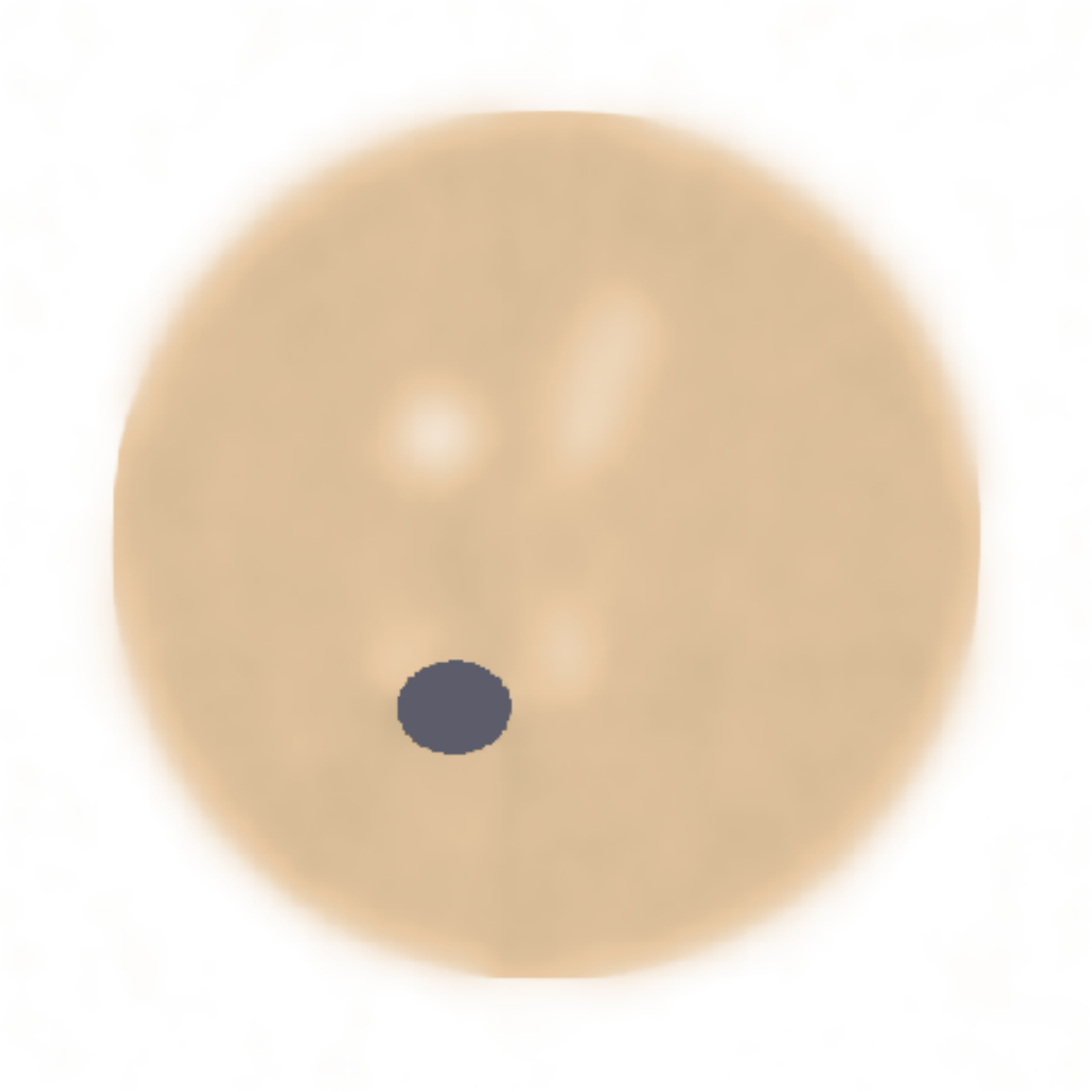}
    \caption{MwG, Cauchy, 30}\label{mwgcauchy18030}
    \end{subfigure}
     \begin{subfigure}[b]{3.3cm}
    \includegraphics[height=3.3cm]{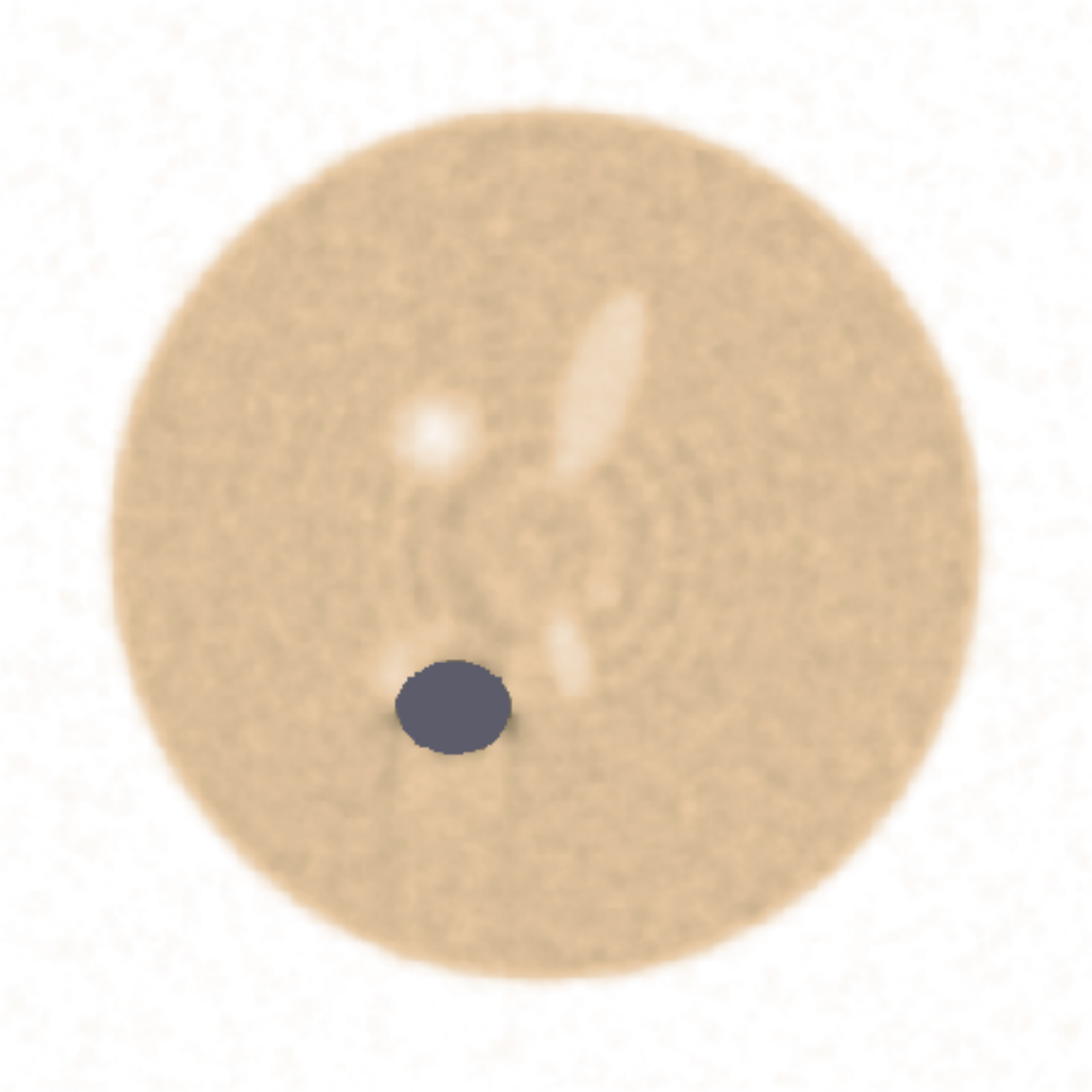}
    \caption{HMC, Cauchy, 90 }\label{hmccauchy18090}\end{subfigure}   
     \begin{subfigure}[b]{3.3cm}
    \includegraphics[height=3.3cm]{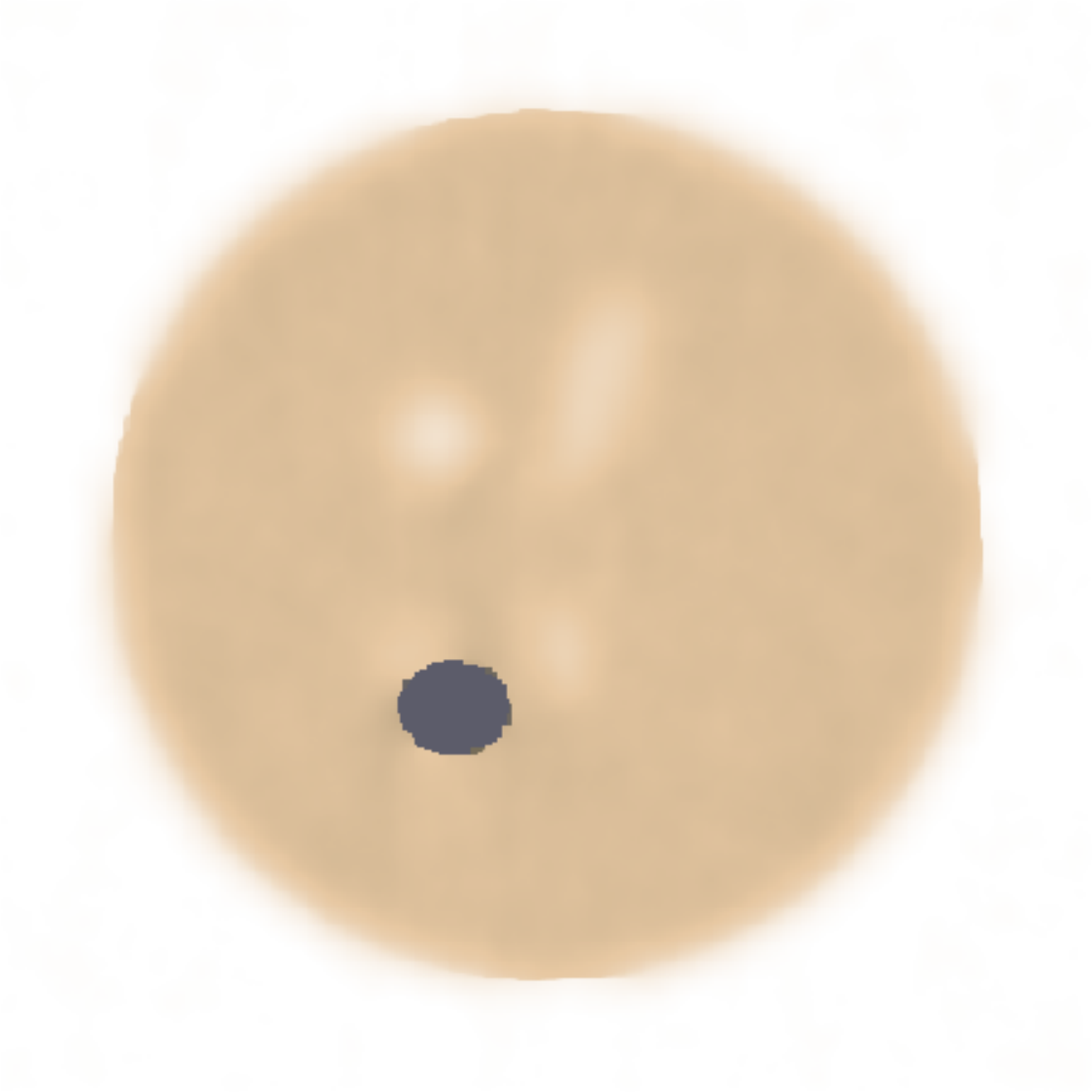}
    \caption{HMC, Cauchy, 30  }\label{hmccauchy18030}\end{subfigure}
    
    \caption{Two-dimensional CM estimates with different measurement angles, prior assumptions, and samplers}
    \label{logwidecm}
\end{figure}

In Figure \ref{logwidecm}, we have plotted the CM estimates computed with MwG and HMC. Here we skip Gaussian priors, as the CM and MAP estimates with Gaussian priors are the same.
It is notable that the CM estimate with TV prior differs noticeably from its MAP estimate (Figures \ref{slices} and \ref{logwide_map}). 
For high-dimensional inverse problems, it has been demonstrated by \citet[Figure 4]{lassas_siltanen:2004} that CM  estimates might be wildly oscillatory, even though the same prior  produces reasonable MAP estimates. 
Also, the great number of scanning artefacts  is noteable. 
On the other hand, CM estimates with Cauchy difference prior have less coordinate-axis related artefacts,  and CM estimates of Besov  prior are less blocky than their corresponding MAP estimates.

\begin{figure}
    \begin{subfigure}[b]{3.5cm}
    \includegraphics[height=2.6cm]{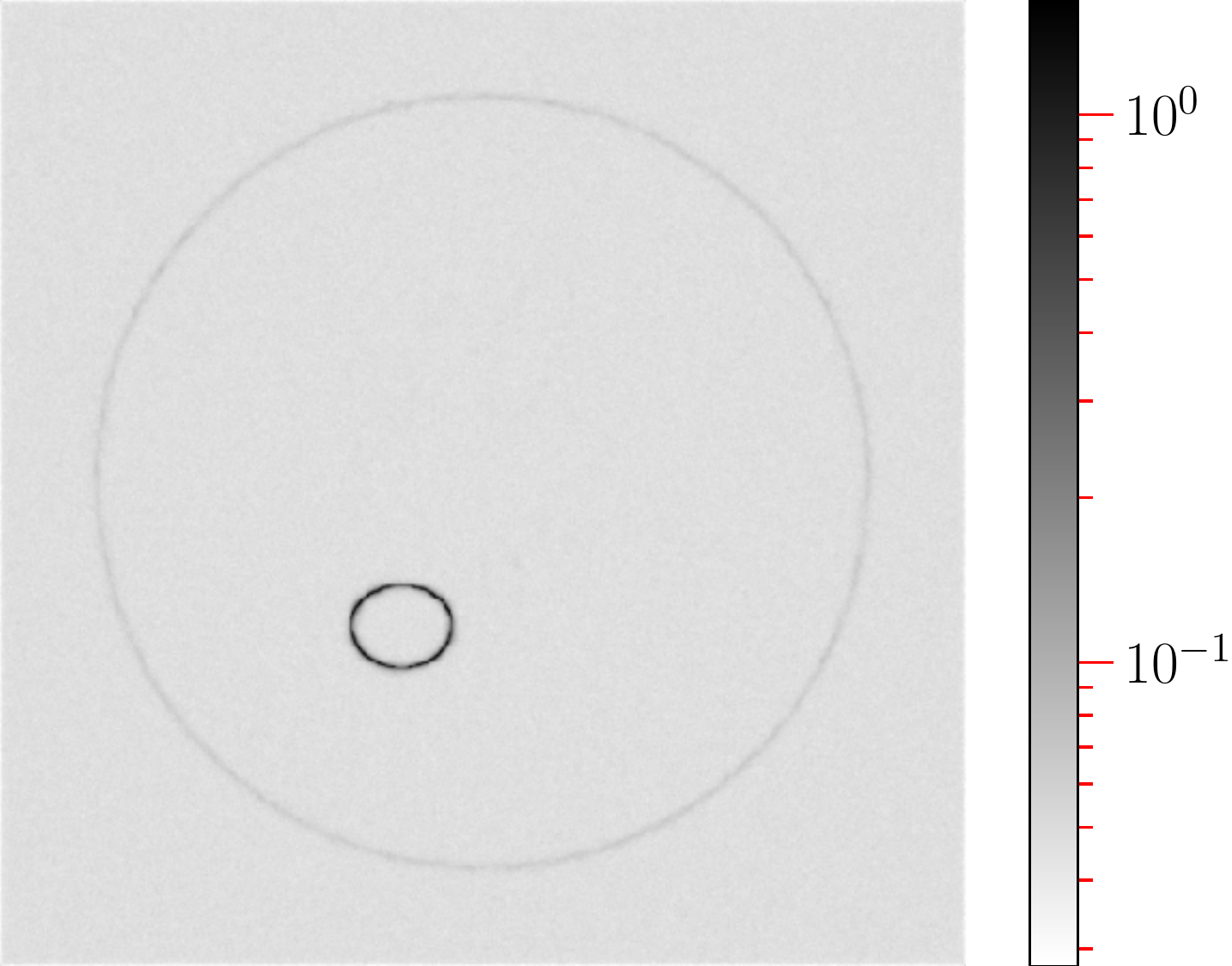}
    \caption{MwG, TV,  90  }\label{mwgtvvar90}\end{subfigure}
    \begin{subfigure}[b]{3.5cm}
    \includegraphics[height=2.6cm]{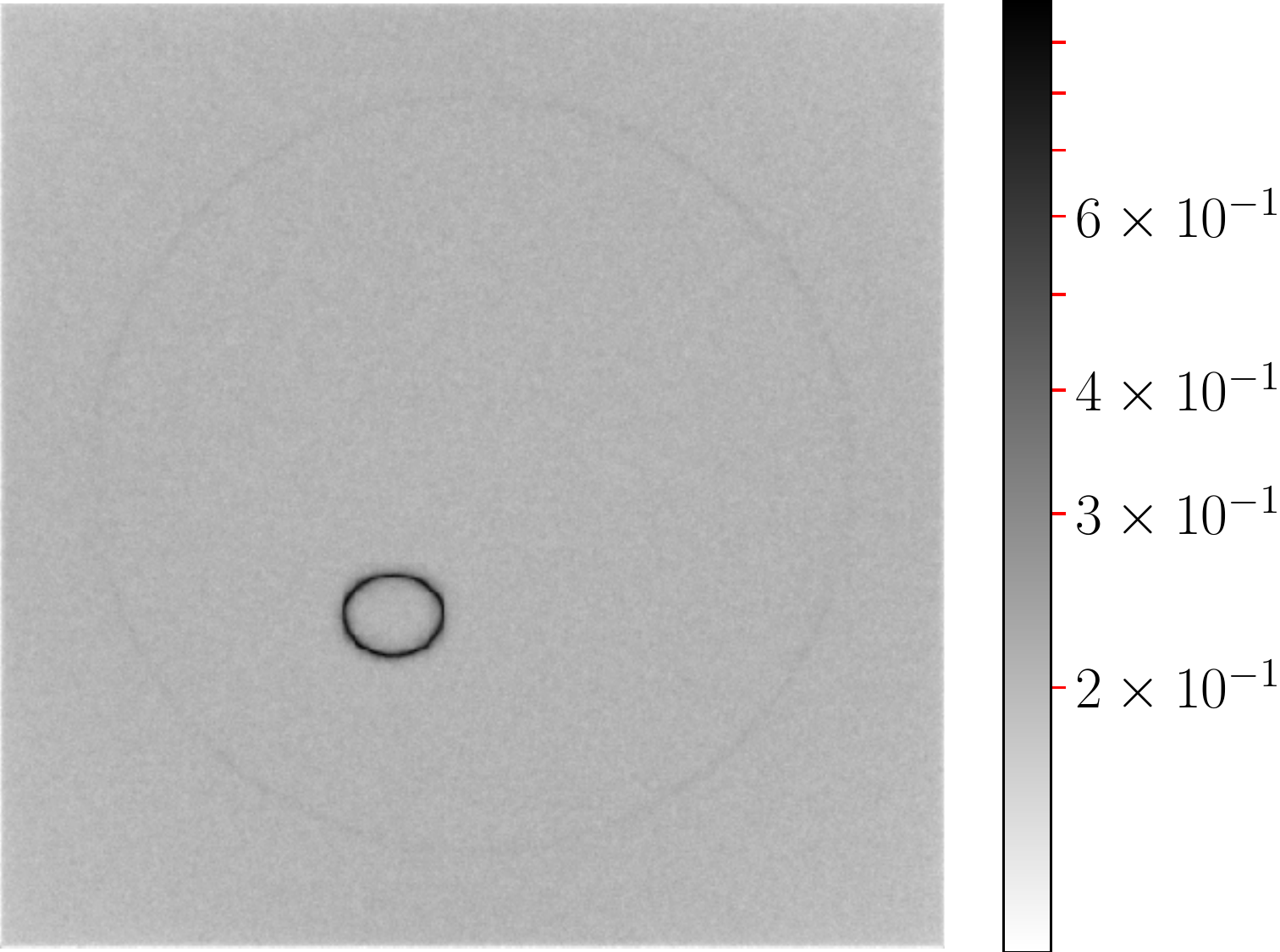}
    \caption{MwG, TV, 30  }\label{mwgtvvar30}\end{subfigure}
    \begin{subfigure}[b]{3.5cm}
    \includegraphics[height=2.6cm]{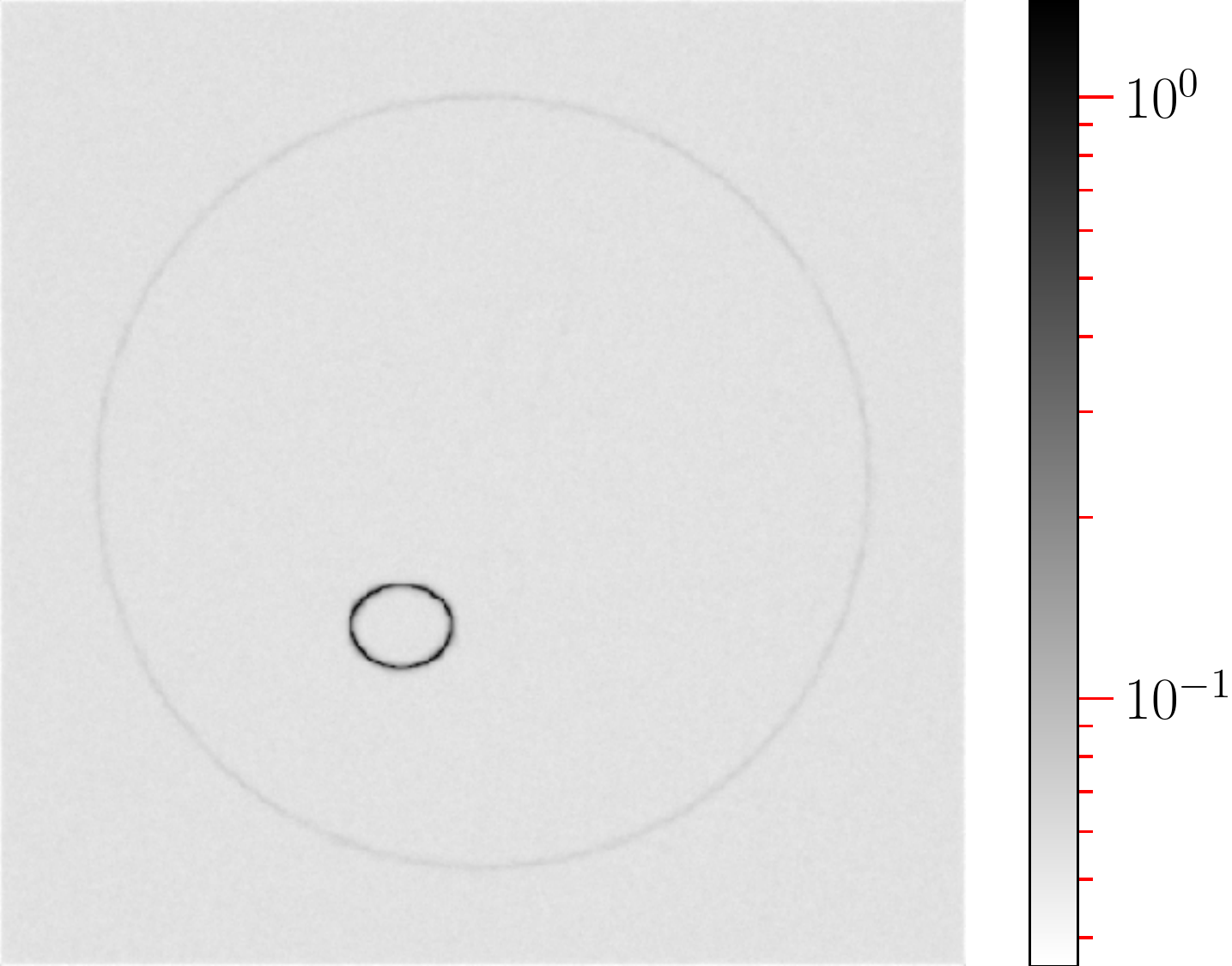}
    \caption{HMC, TV, 90}\label{hmctvvar90}\end{subfigure}
    \begin{subfigure}[b]{3.5cm}
    \includegraphics[height=2.6cm]{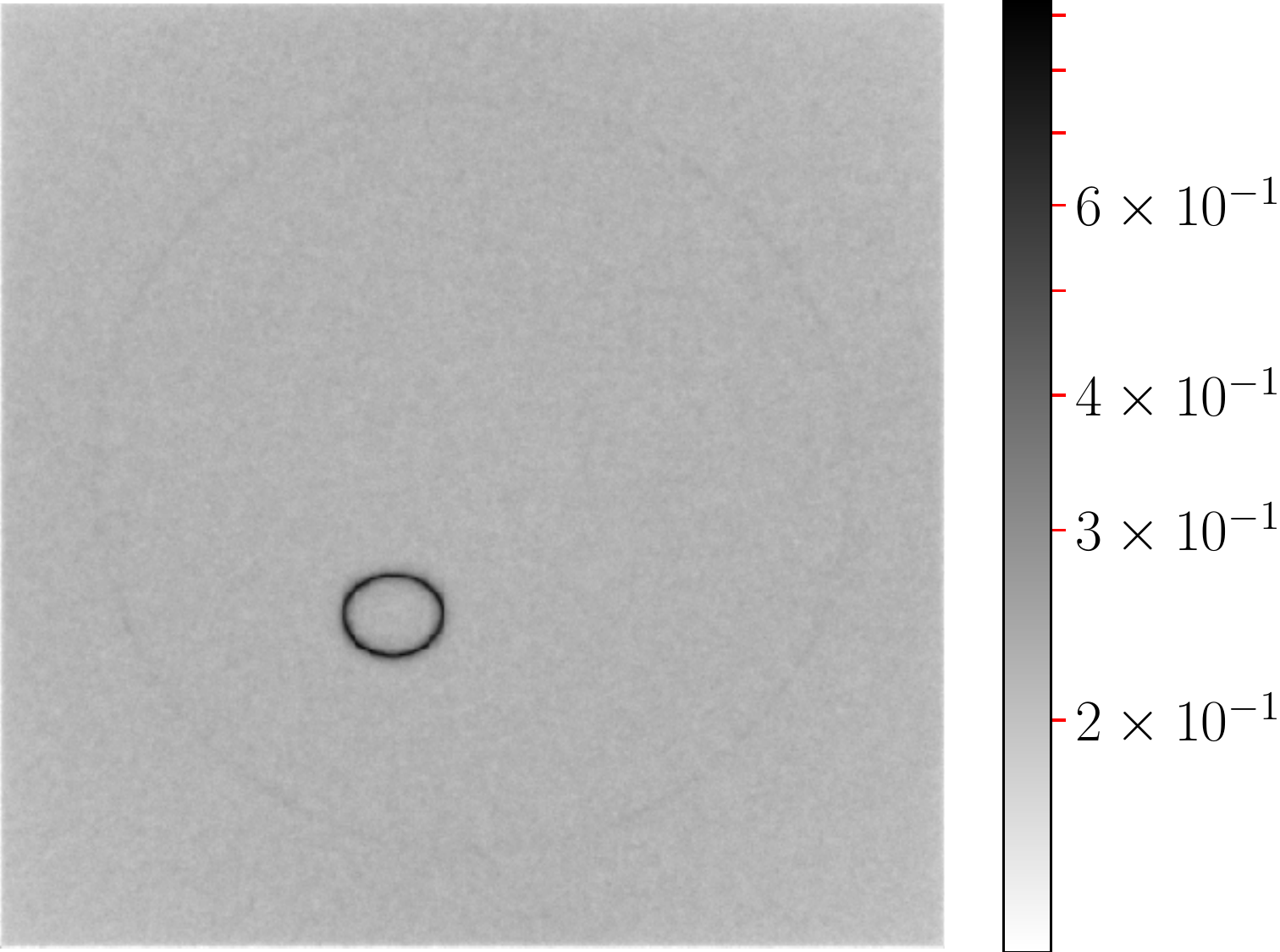}
    \caption{HMC, TV, 30  }\label{hmctvvar30}\end{subfigure}

    \begin{subfigure}[b]{3.5cm}
    \includegraphics[height=2.6cm]{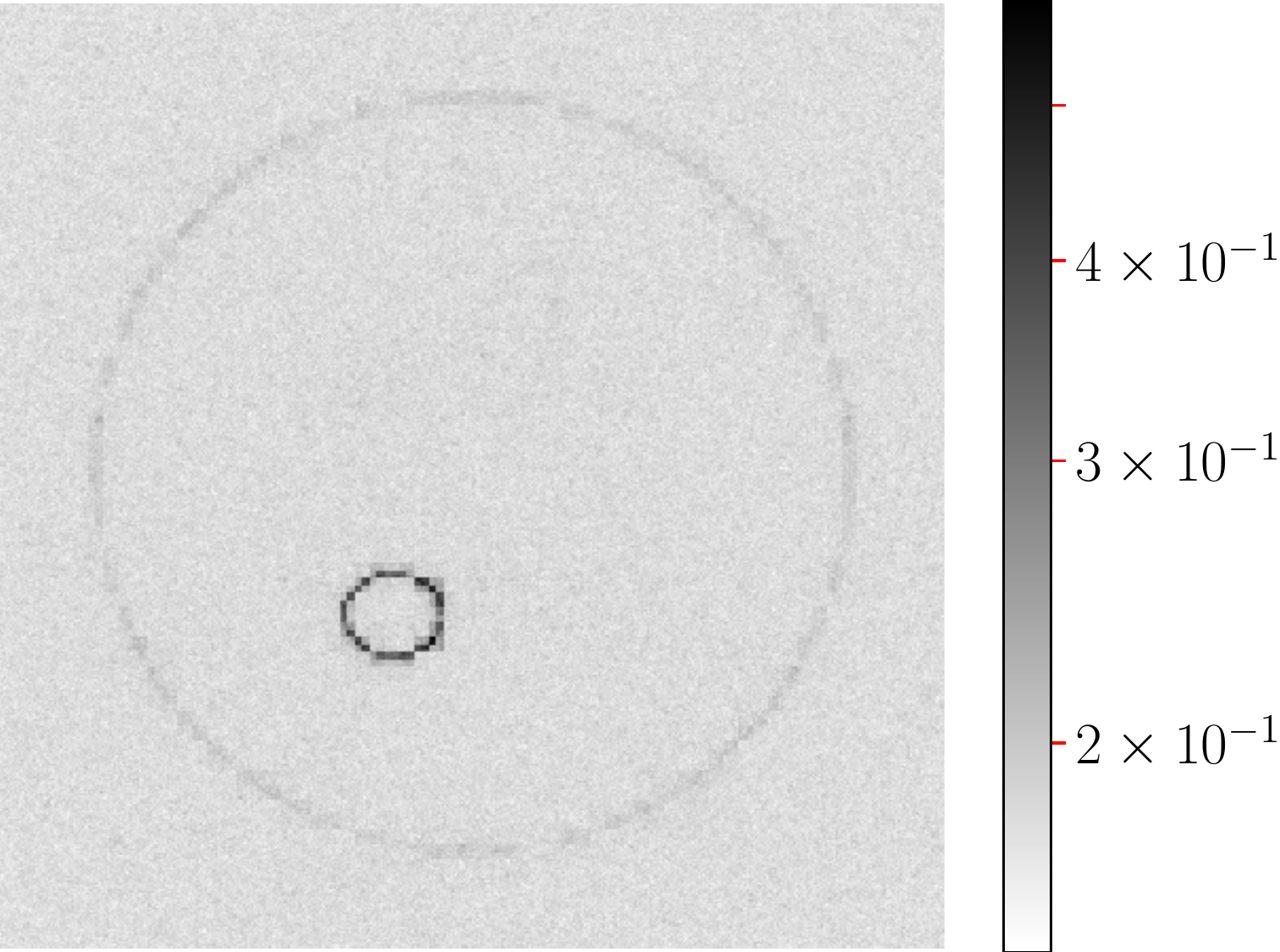}
    \caption{MwG, Besov, 90  }\label{mwgbesovvar90}\end{subfigure}     
     \begin{subfigure}[b]{3.5cm}
    \includegraphics[height=2.6cm]{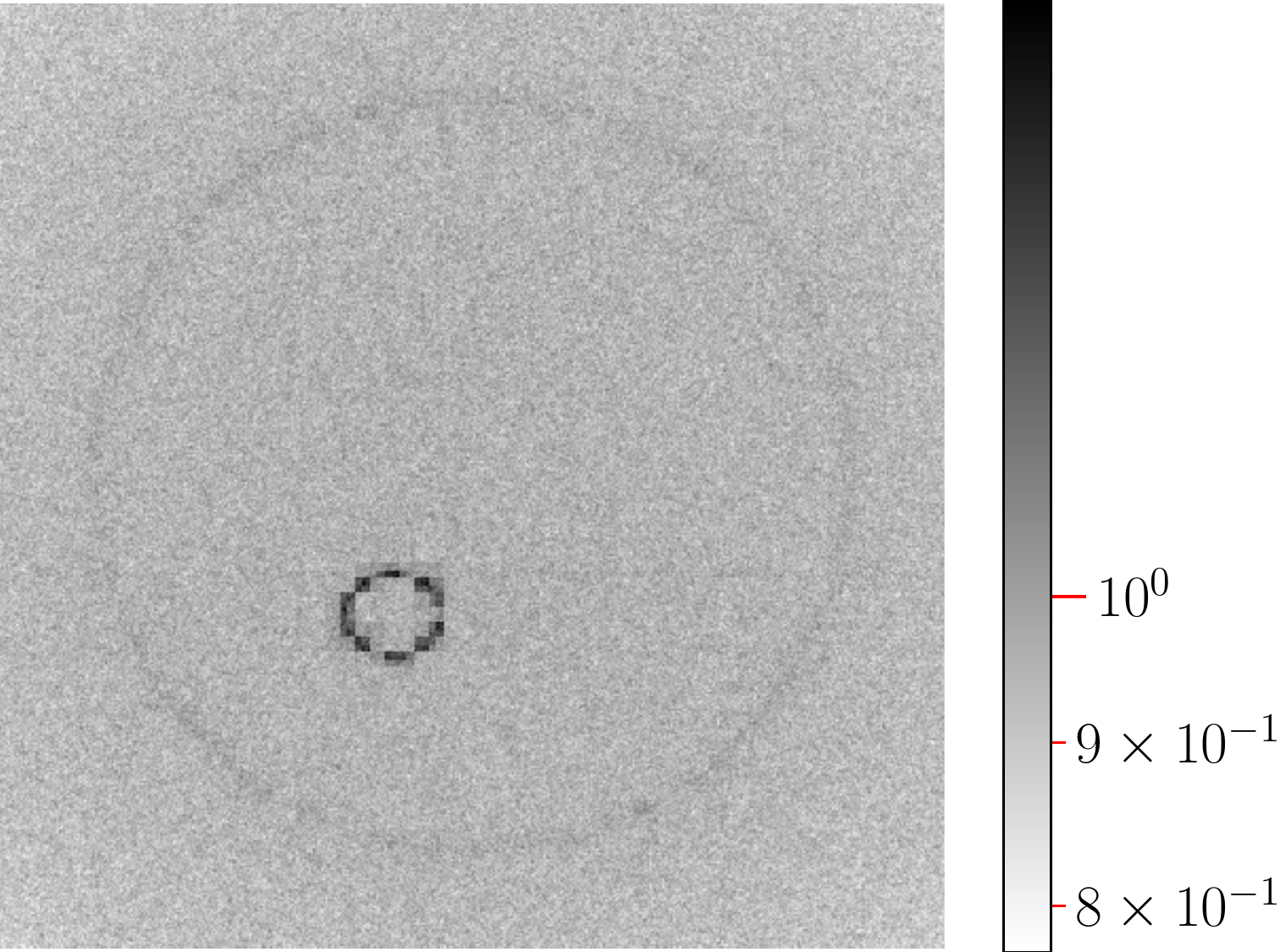}
    \caption{MwG, Besov,  30  }\label{mwgbesovvar30}\end{subfigure} 
     \begin{subfigure}[b]{3.5cm}
    \includegraphics[height=2.6cm]{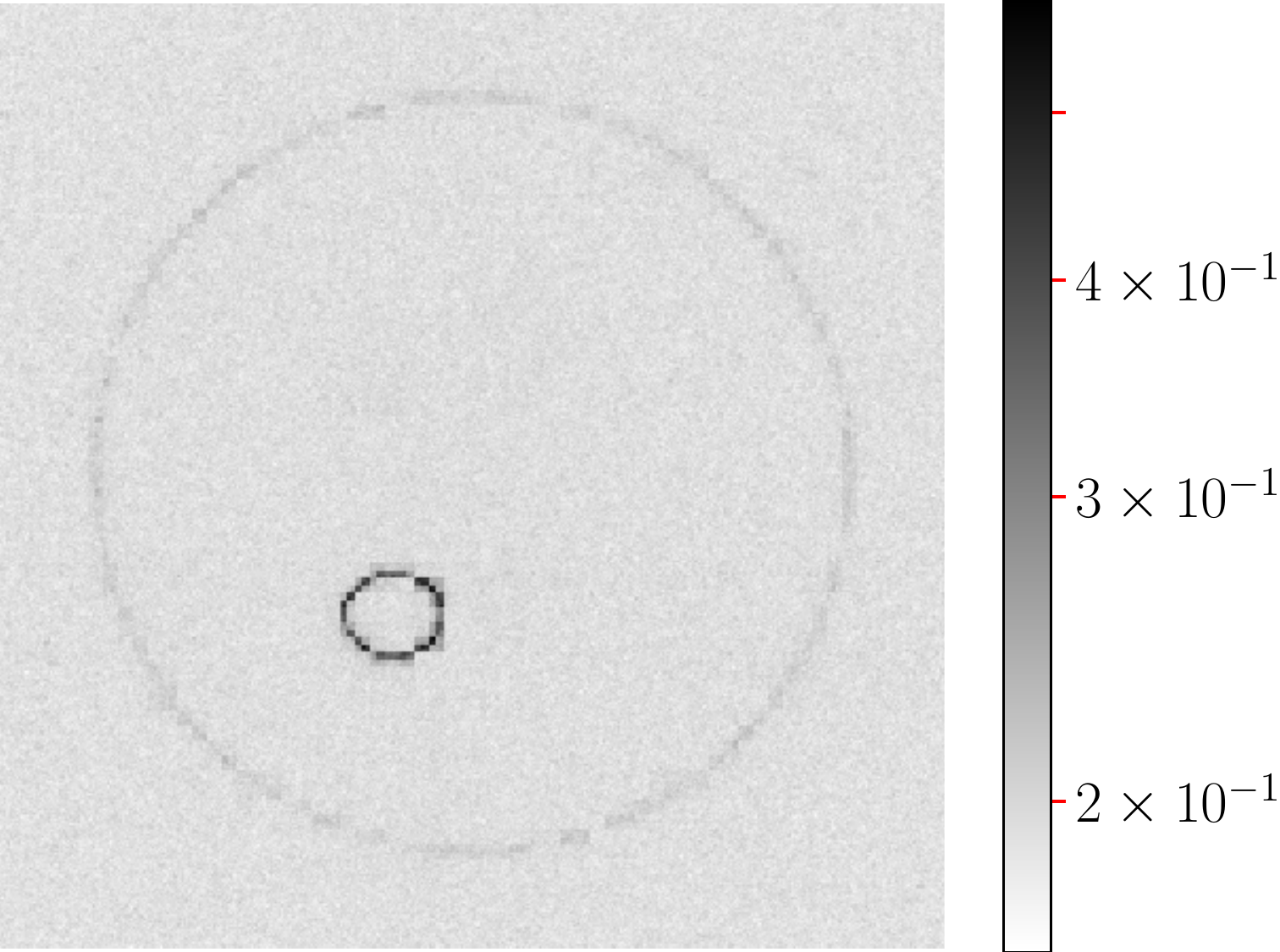}
    \caption{HMC, Besov,  90  }\label{hmcbesovvar90}\end{subfigure}
     \begin{subfigure}[b]{3.5cm}
    \includegraphics[height=2.6cm]{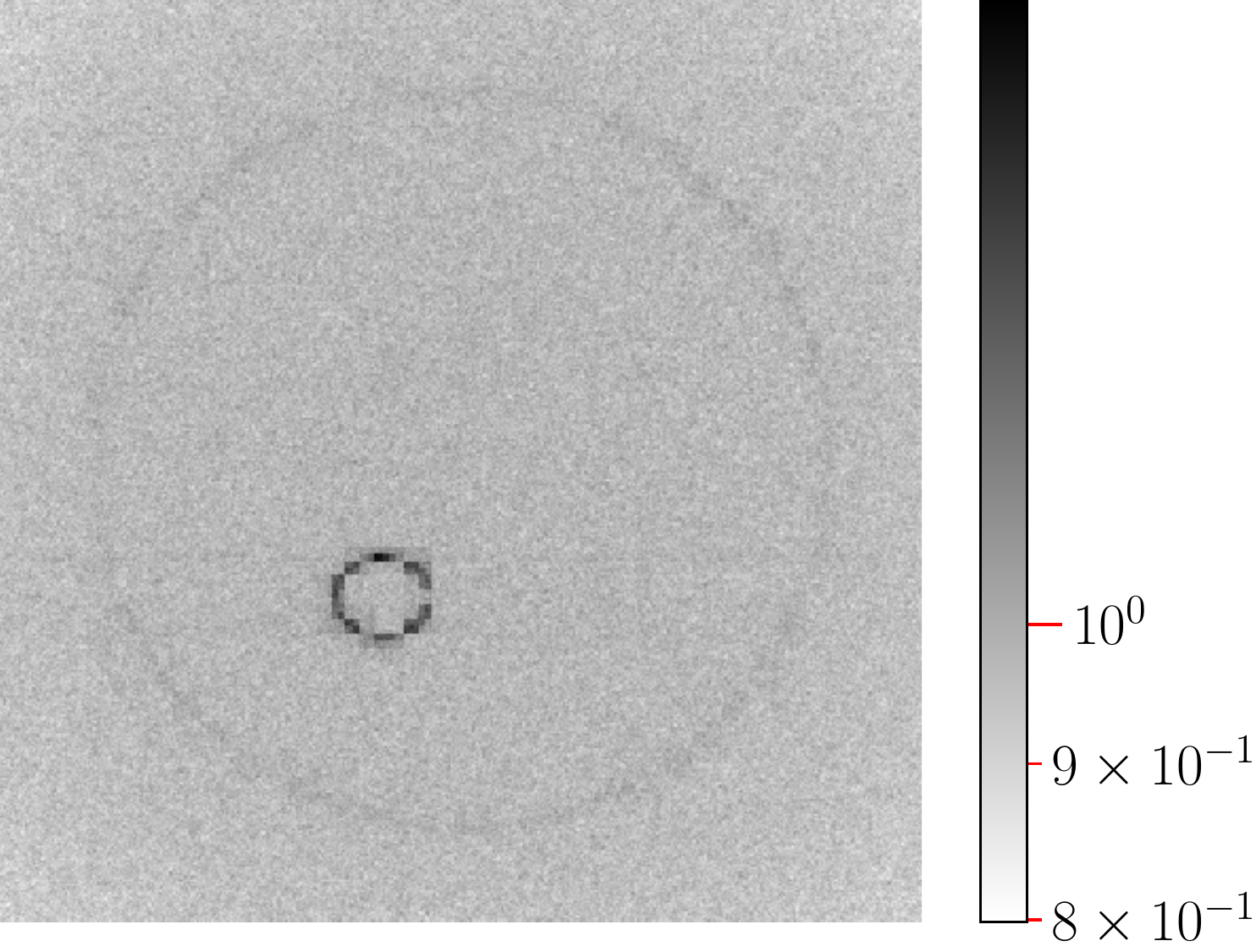}
    \caption{HMC, Besov,  30  }\label{hmcbesovvar30}\end{subfigure}

    \begin{subfigure}[b]{3.5cm}
    \includegraphics[height=2.6cm]{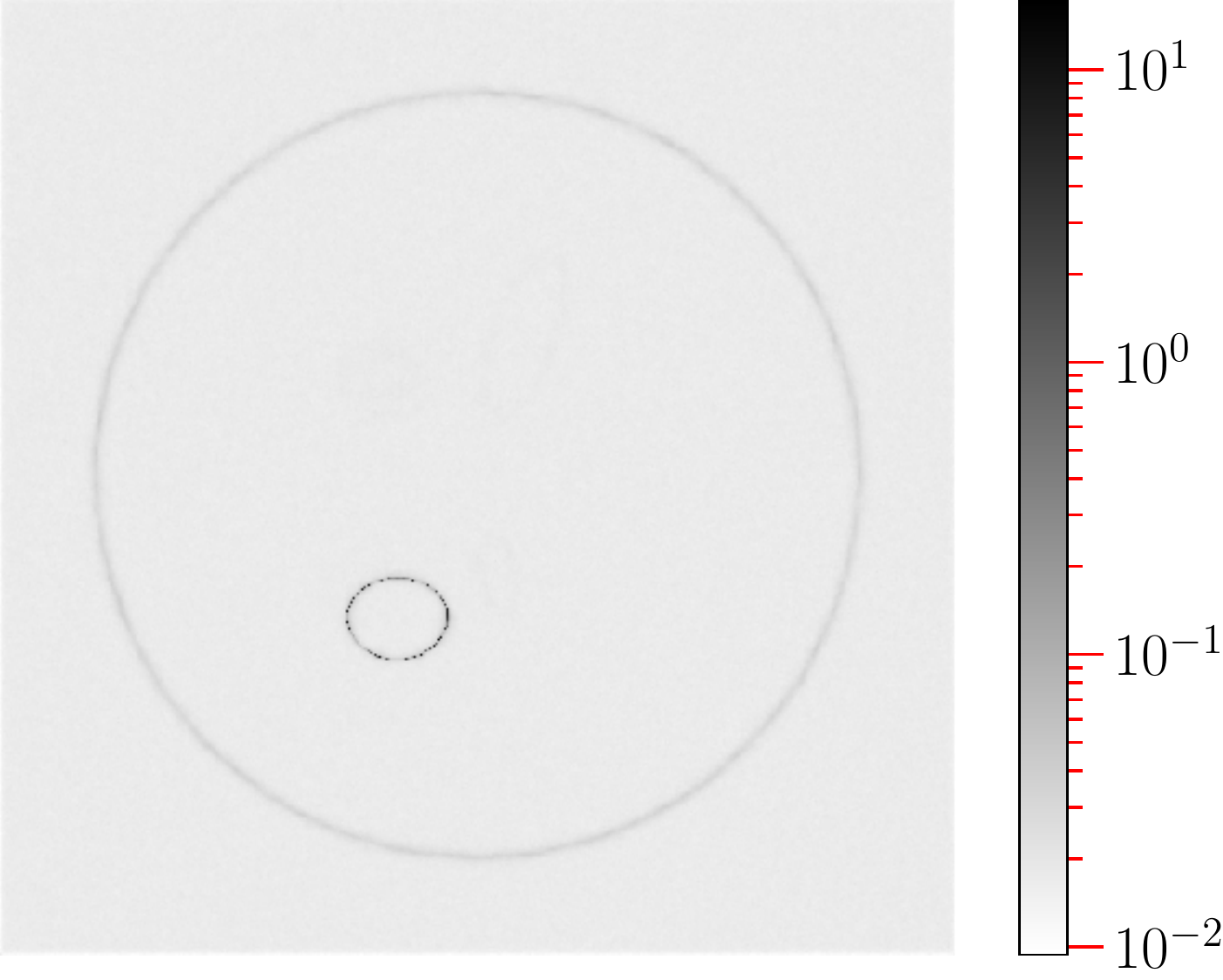}
    \caption{MwG, Cauchy, 90 }\label{mwgcauchyvar90}\end{subfigure}
     \begin{subfigure}[b]{3.5cm}
    \includegraphics[height=2.6cm]{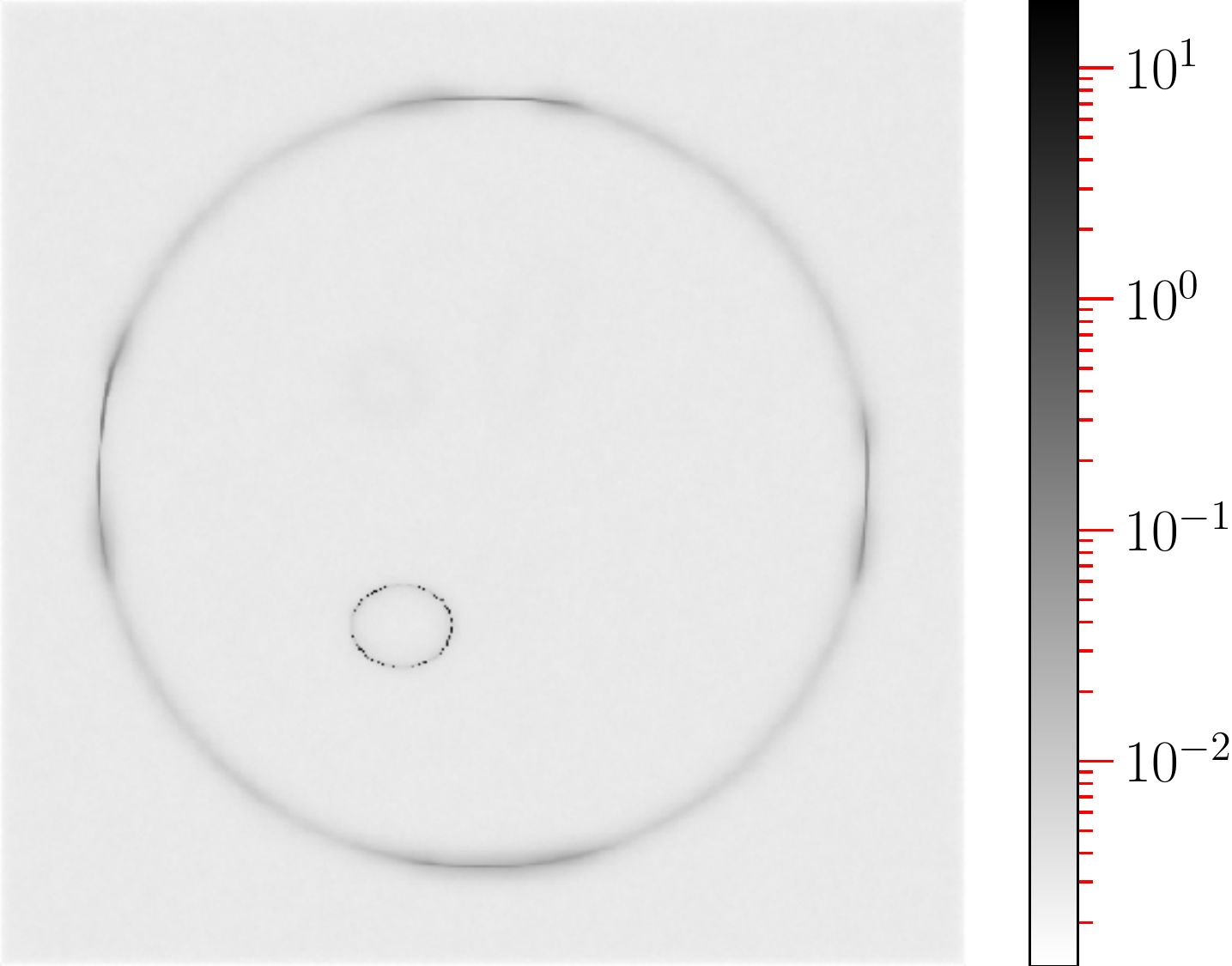}
    \caption{MwG, Cauchy, 30}\label{mwgcauchyvar30}
    \end{subfigure}
     \begin{subfigure}[b]{3.5cm}
    \includegraphics[height=2.6cm]{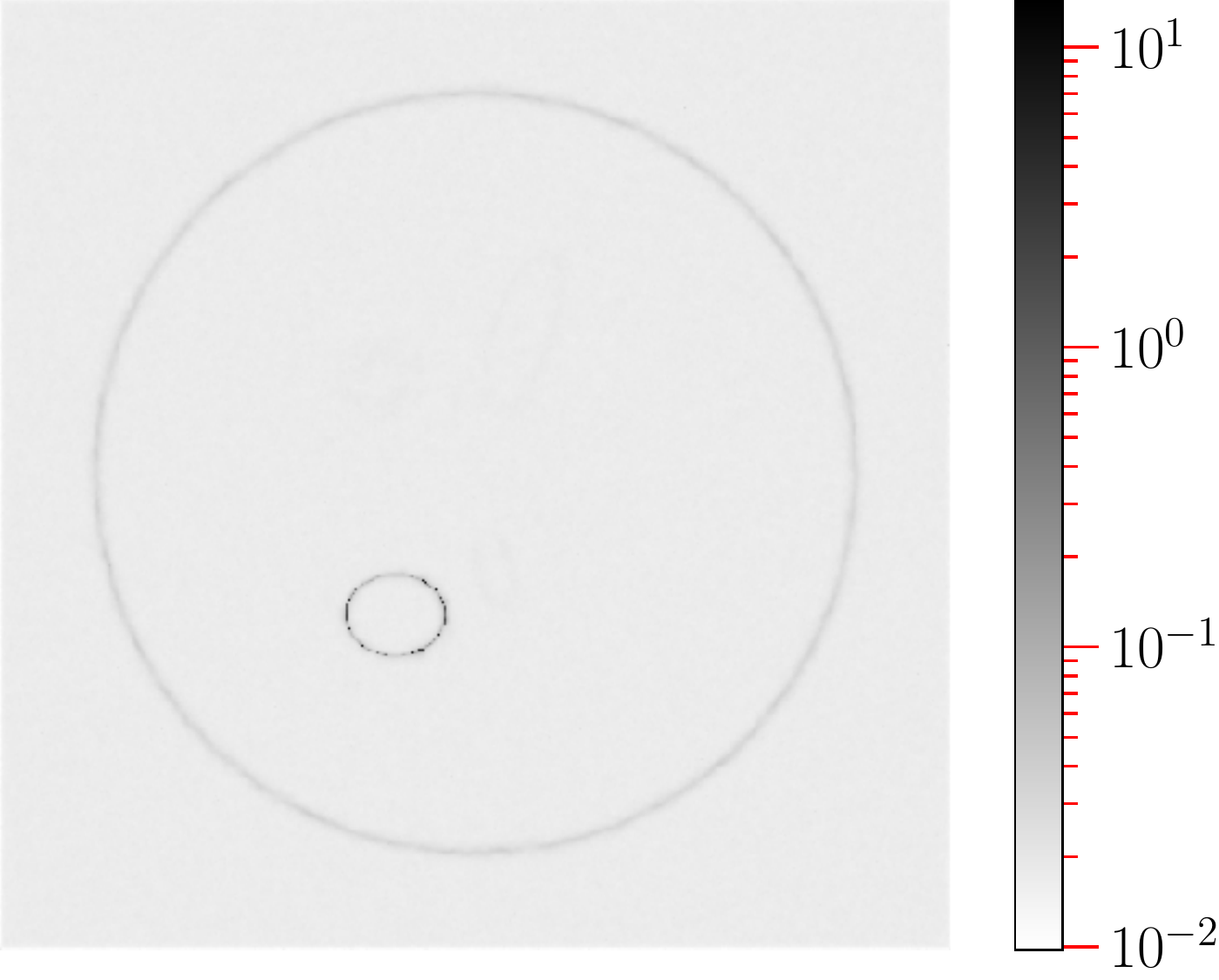}
    \caption{HMC, Cauchy, 90 }\label{hmccauchyvar90}\end{subfigure}   
     \begin{subfigure}[b]{3.5cm}
    \includegraphics[height=2.6cm]{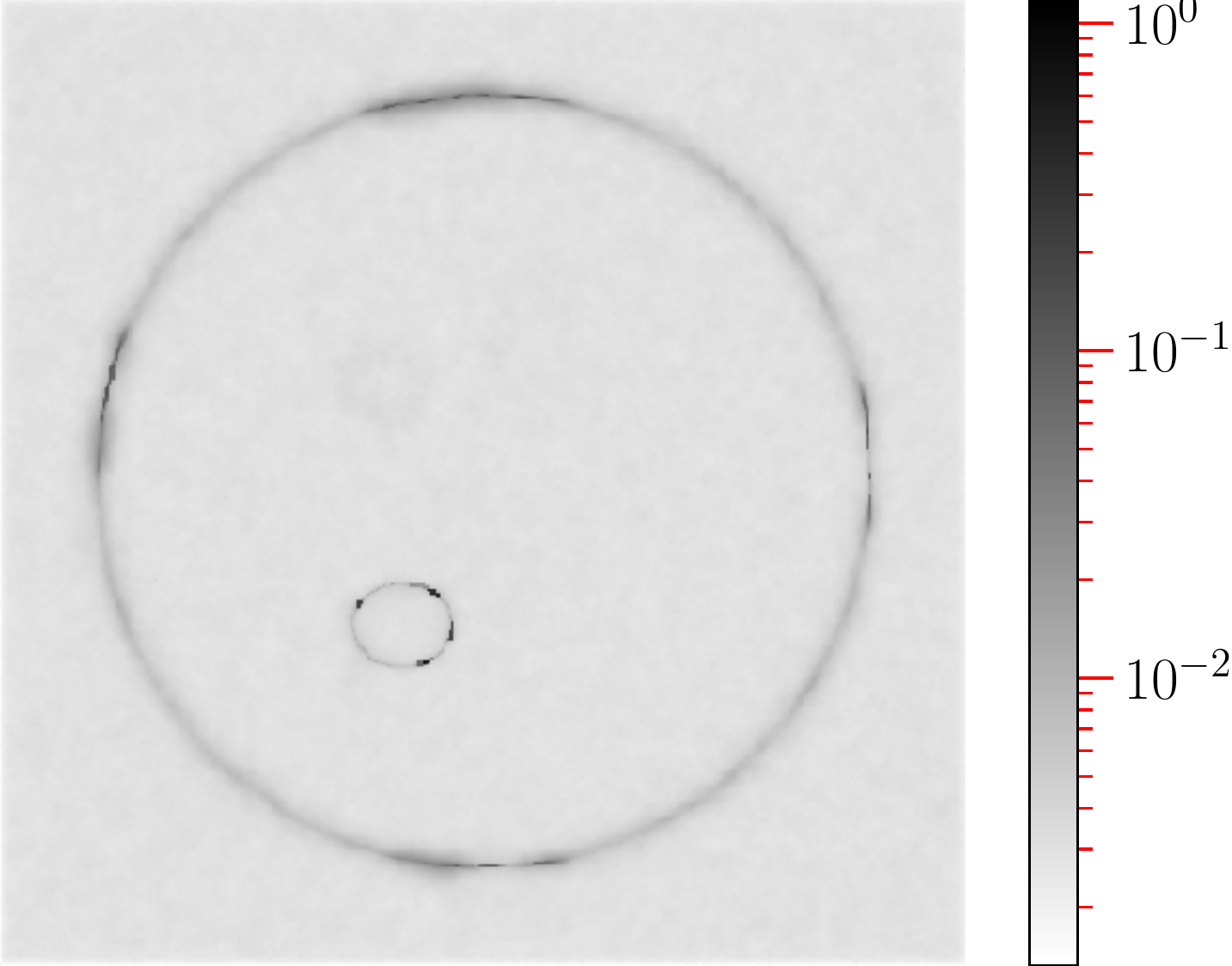}
    \caption{HMC, Cauchy, 30  }\label{hmccauchyvar30}\end{subfigure}

    \caption{Pixel-wise variance estimates for the CM estimates in Figure \ref{logwidecm}}
    \label{logwidevar}
\end{figure}

In Figure \ref{logwidevar}, we have logarithmic pixel-wise variances estimated from the MCMC chains with MwG and HMC-NUTS.
For TV and Besov priors, MwG and HMC produce similar results. 
However, they produce different variance estimates for Cauchy prior at the 30 angle case. 
This is because some of the  pixels at the boundary of the high density object have much higher variance than the other boundary pixels.  
This is a mixing issue, and we notice that with heavy tails the MwG and HMC require longer chains for Cauchy than for TV and Besov.
We could improve the sampling capabilities of NUTS by adapting a diagonal global mass matrix different than the identity one by running a preliminary run to estimate the variance of posterior. Adapting a full covariance matrix or using a local metric and therefore changing HMC-NUTS to Riemannian Manifold Monte Carlo would be computationally infeasible in this kind of a high-dimensional case.  On the other hand, there is not much we can do to improve the space exploration in MwG. We might change the adaptation strategy  to the component-wise robust adaptive Metropolis-Hastings, which should explore the heavy tails more efficiently.

\subsection{Drill-core tomography}

Here we take a simpler setting for a lower-contrast target, and we detect pores and mineralised soil within a  core sample. Unlike in the log tomography experiment, there are no  objects with drastically different absorption coefficient than the surrounding matter has in the domain of interest. 
However, the drill-core sample has lots of pores and soil cobs and their sizes also vary.  

\begin{figure}
    \begin{subfigure}[b]{3.3cm}
    \includegraphics[height=3.3cm]{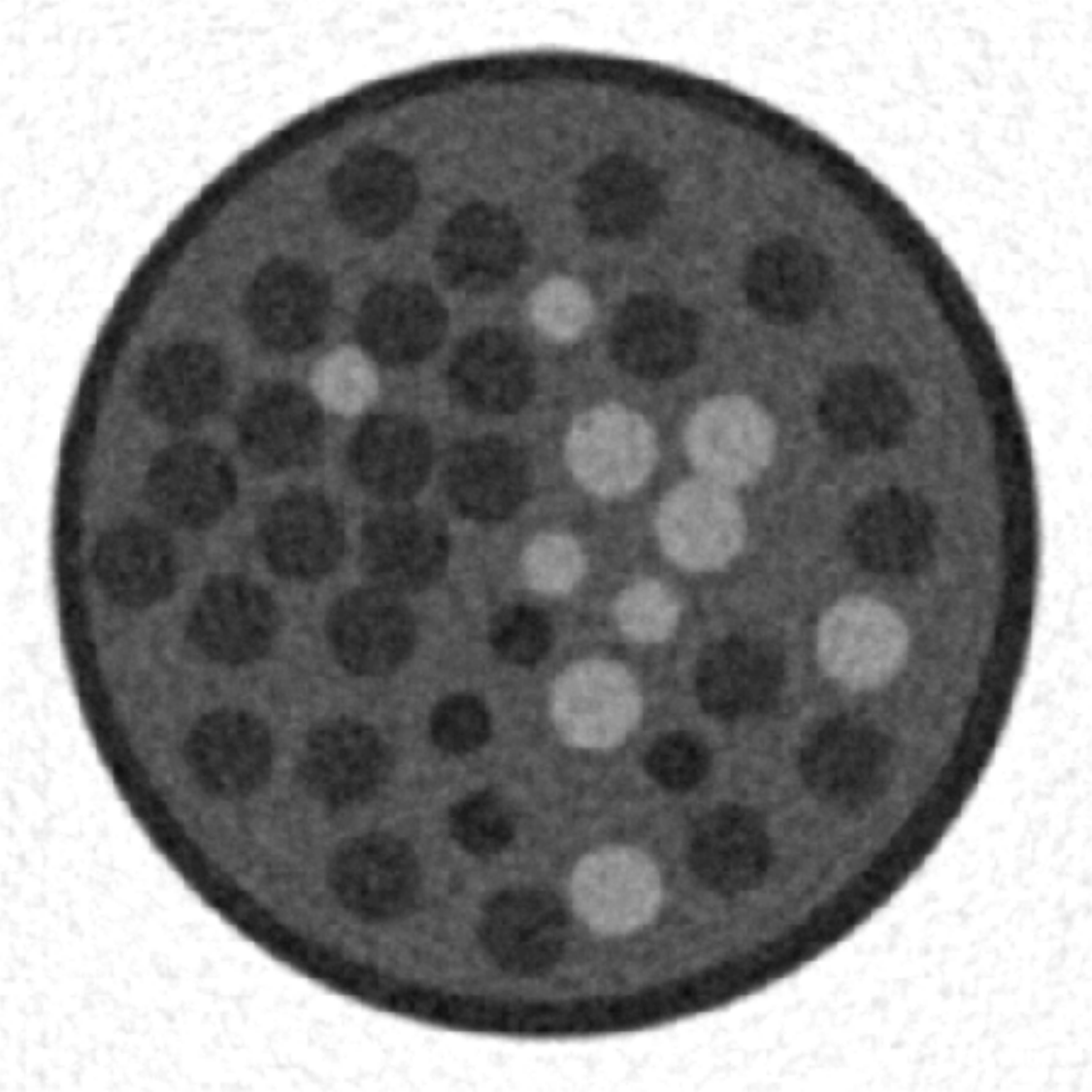}
    \caption{Gaussian, 90 }\label{kivitikhonov18090}
    \end{subfigure}
    \begin{subfigure}[b]{3.3cm}
    \includegraphics[height=3.3cm]{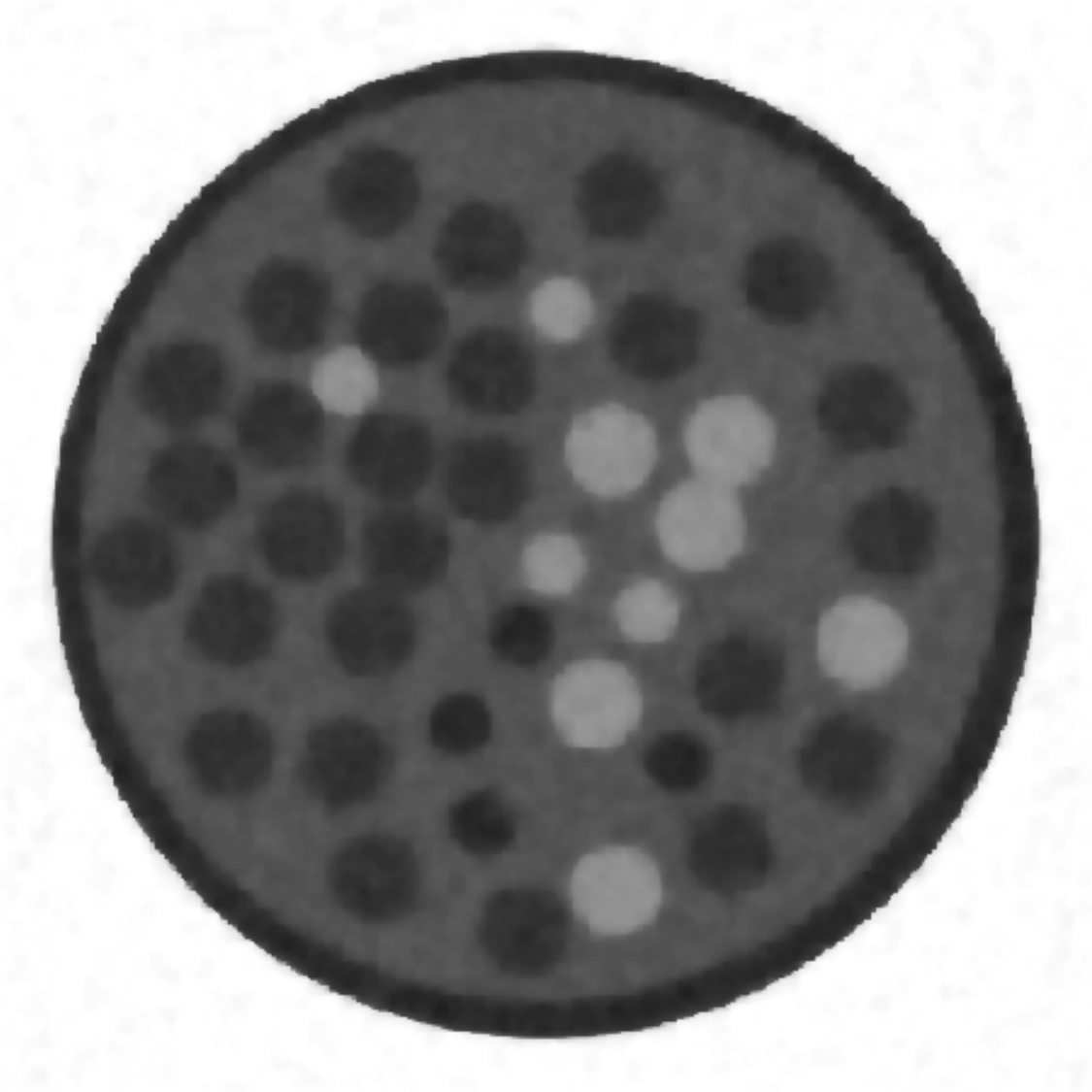}
    \caption{TV, 90 }\label{kivitv18090}
    \end{subfigure}
    \begin{subfigure}[b]{3.3cm}
    \includegraphics[height=3.3cm]{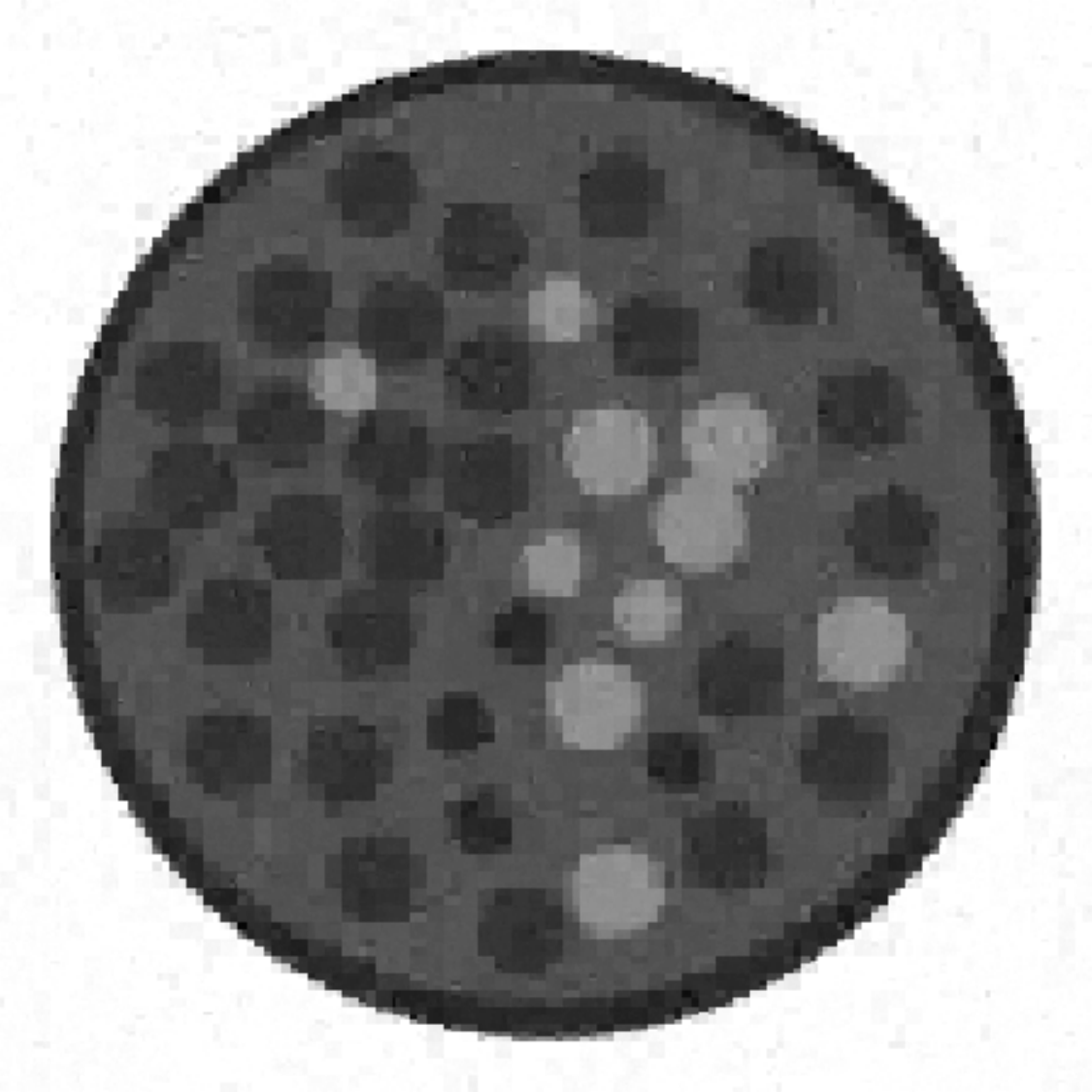}
    \caption{Besov, 90 }\label{kivihaar18090}
    \end{subfigure}
    \begin{subfigure}[b]{3.3cm}
    \includegraphics[height=3.3cm]{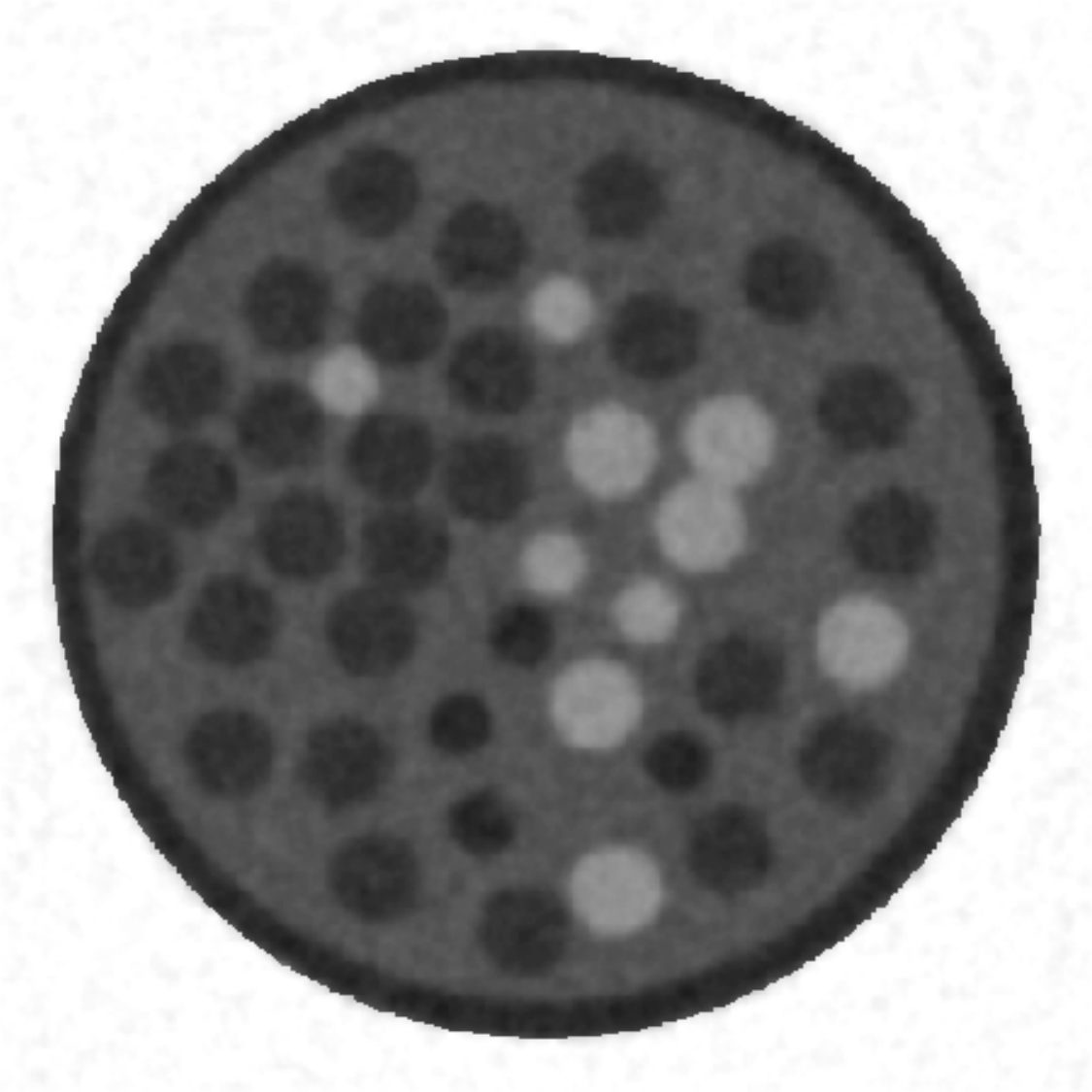}
    \caption{Cauchy, 90 }\label{kivicauchy18090}
    \end{subfigure}

    \begin{subfigure}[b]{3.3cm}
    \includegraphics[height=3.3cm]{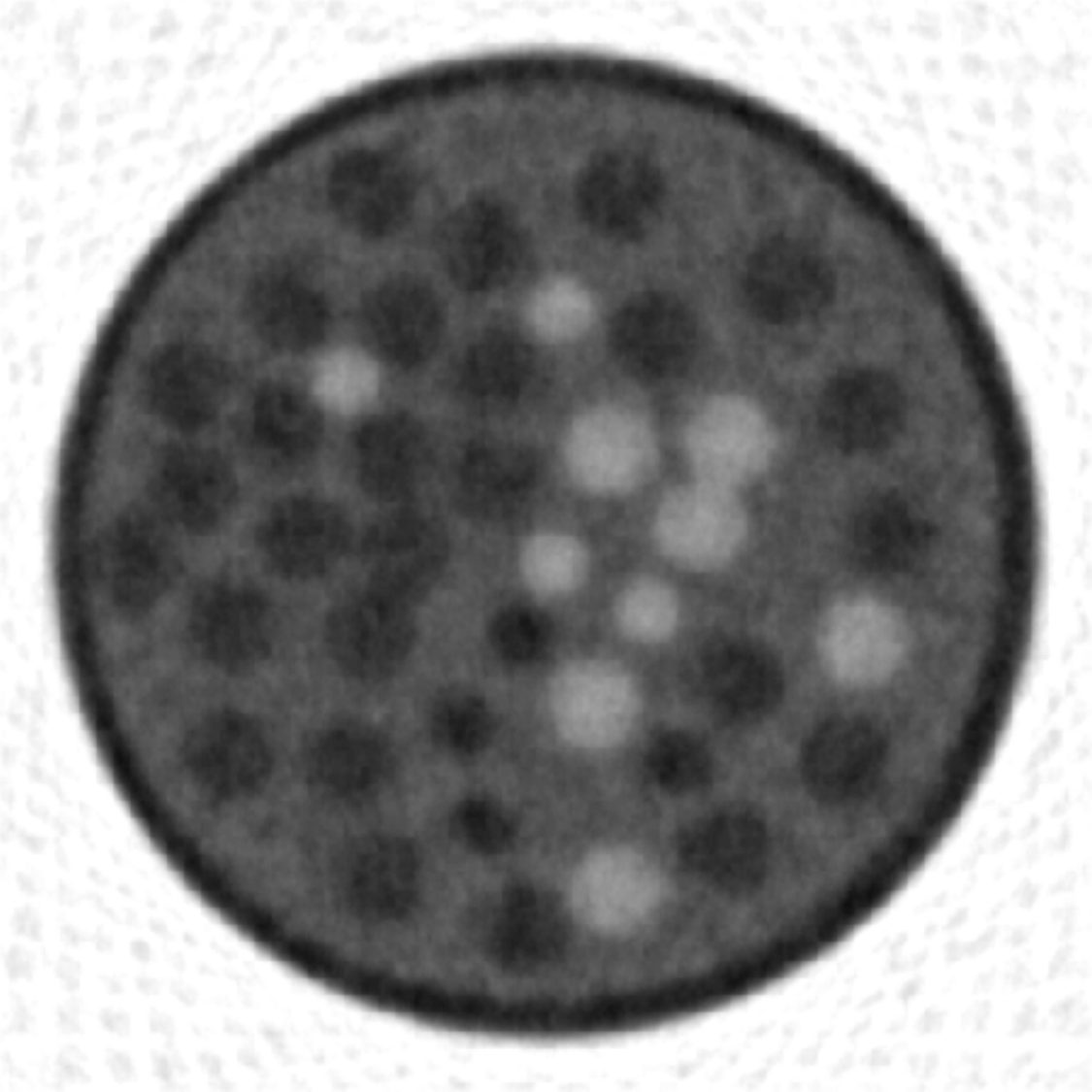}
    \caption{Gaussian, 30 }\label{kivitikhonov18030}
    \end{subfigure}    \begin{subfigure}[b]{3.3cm}
    \includegraphics[height=3.3cm]{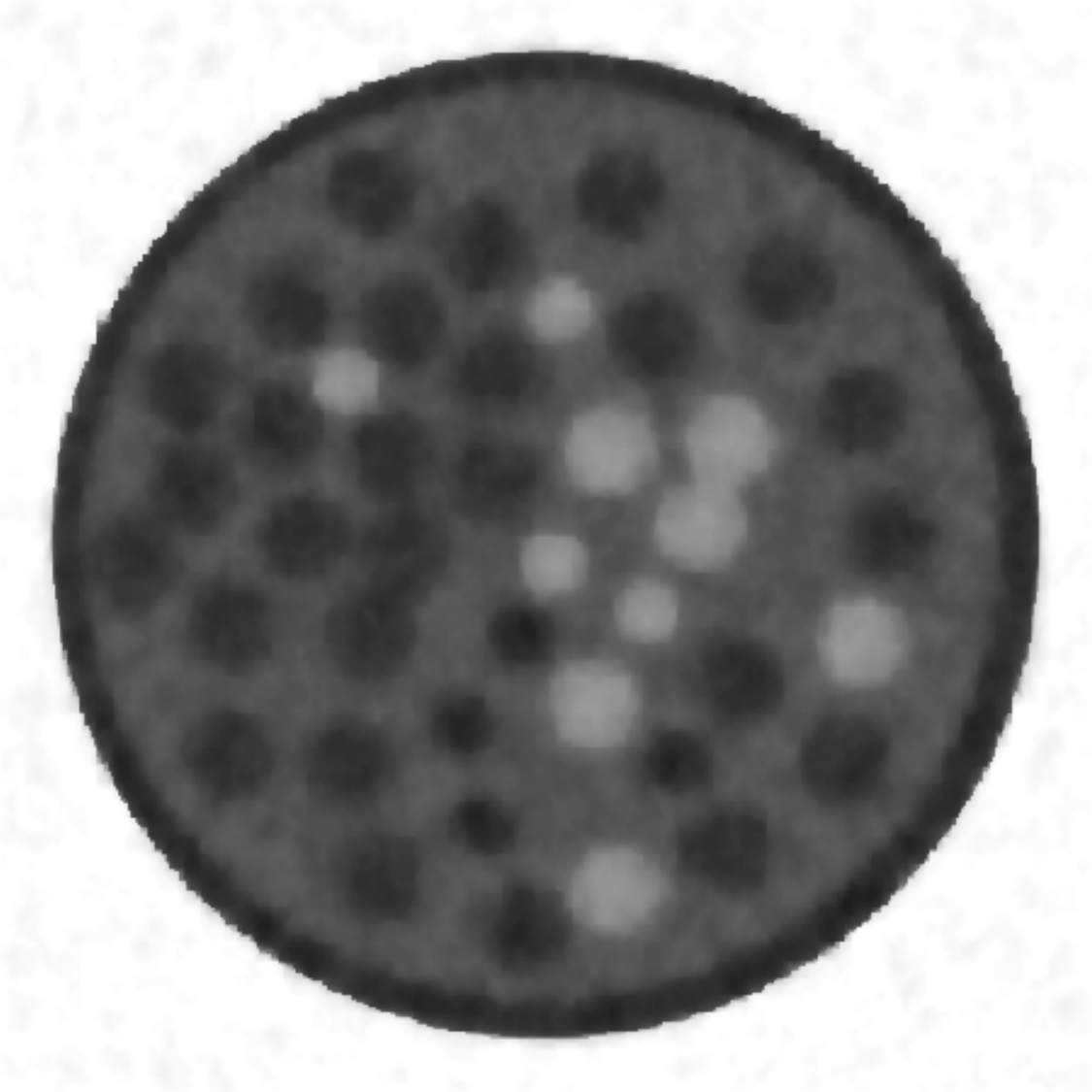}
    \caption{TV, 30}\label{kivitv18030}
    \end{subfigure}
    \begin{subfigure}[b]{3.3cm}
    \includegraphics[height=3.3cm]{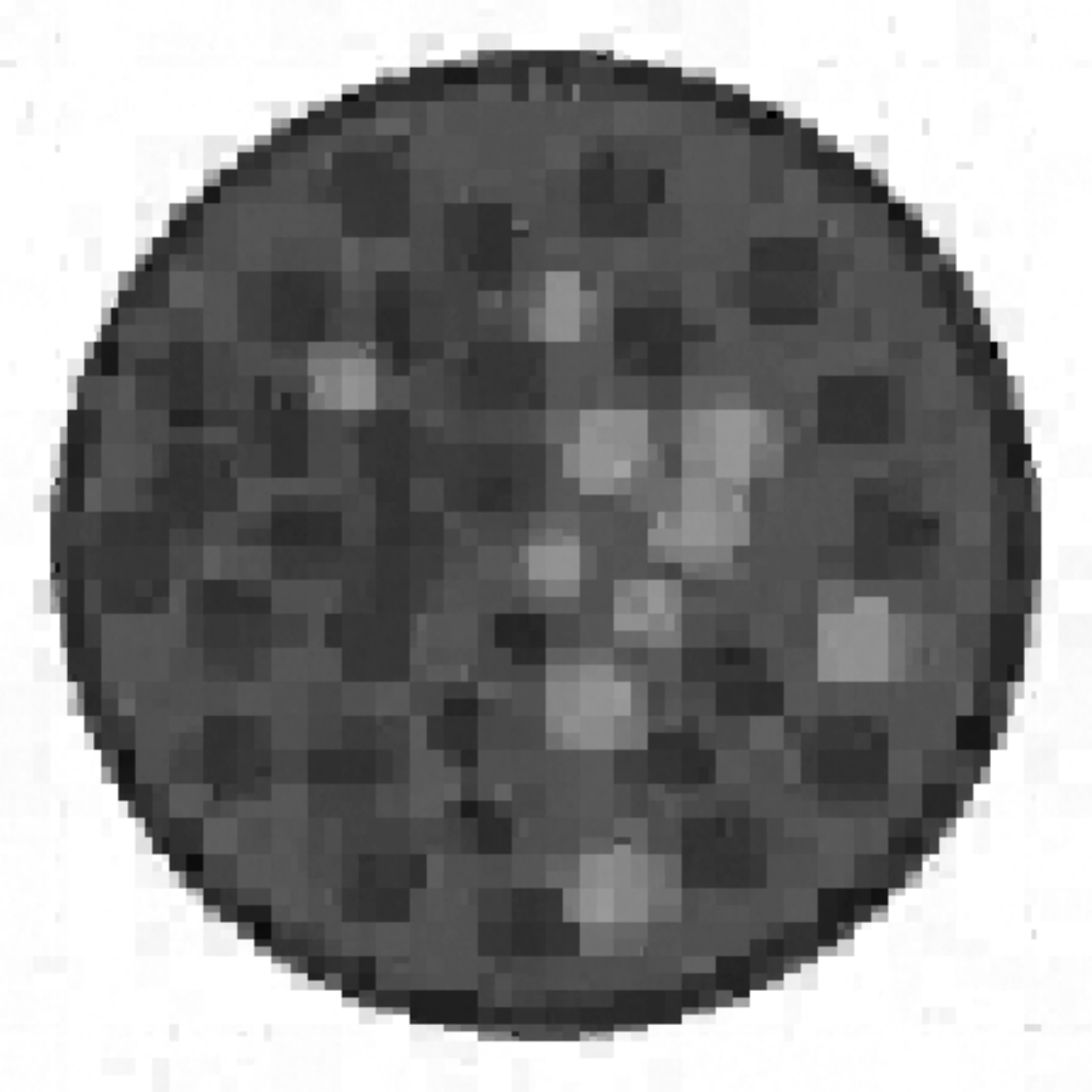}
    \caption{Besov, 30 }\label{kivihaar18030}
    \end{subfigure}
    \begin{subfigure}[b]{3.3cm}
    \includegraphics[height=3.3cm]{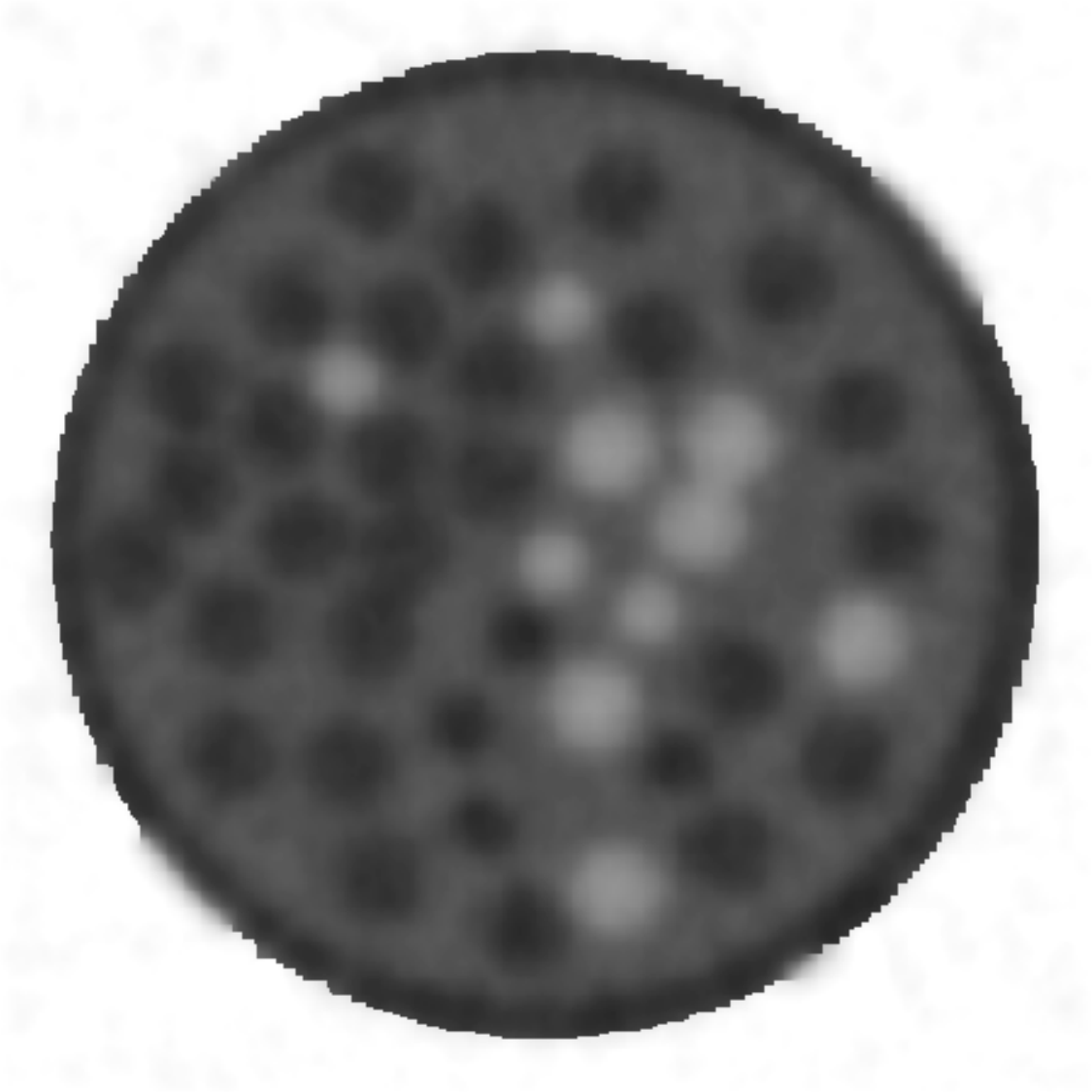}
    \caption{Cauchy, 30 }\label{kivicauchy18030}
    \end{subfigure}
    \begin{subfigure}[b]{0.3cm}
    \includegraphics[height=4cm]{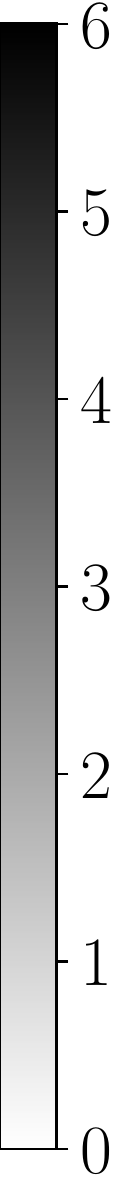} 
    \end{subfigure}    

    \begin{subfigure}[b]{3.3cm}
    \includegraphics[height=3.3cm]{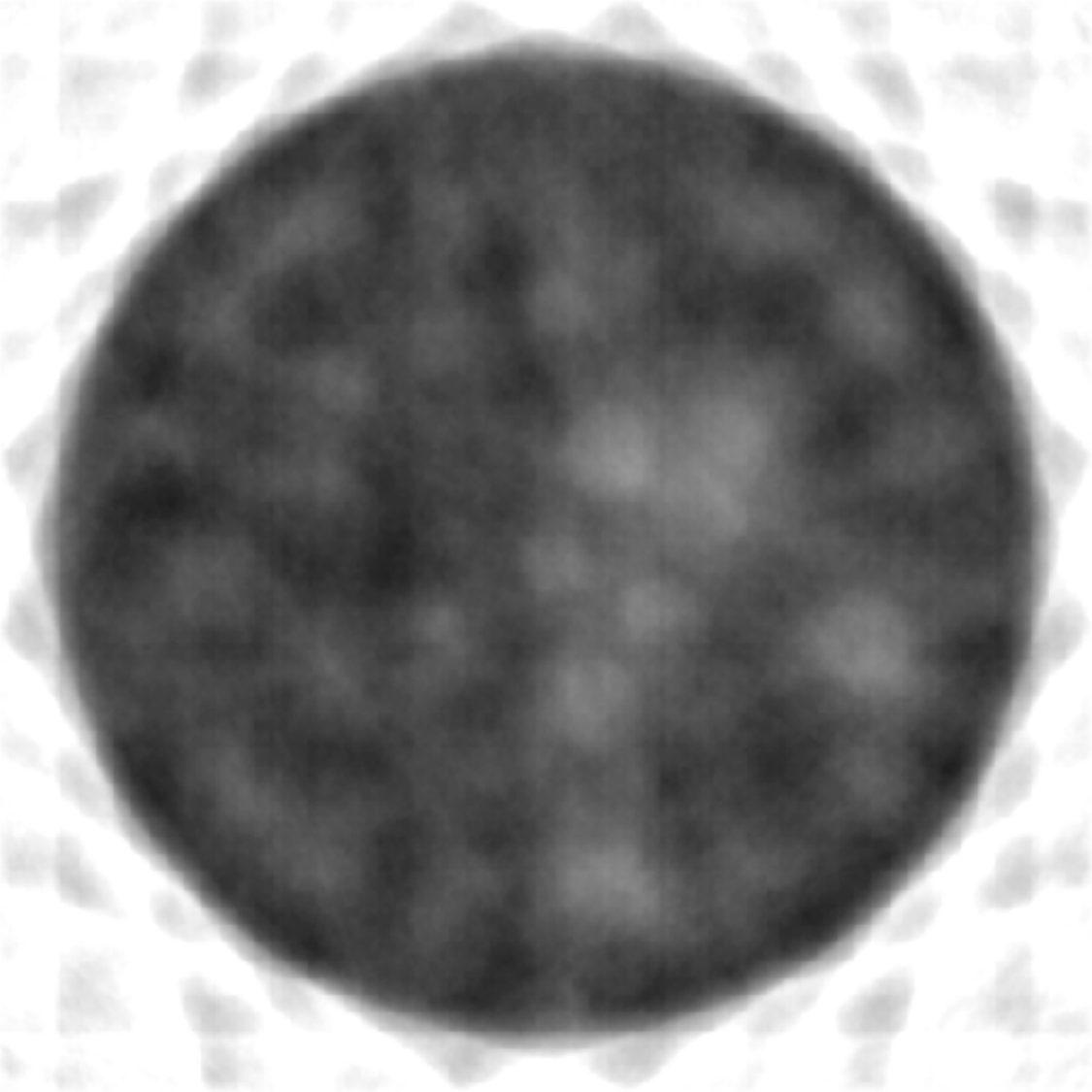}
    \caption{Gaussian, 10 }\label{kivitikhonov18010}
    \end{subfigure}
    \begin{subfigure}[b]{3.3cm}
    \includegraphics[height=3.3cm]{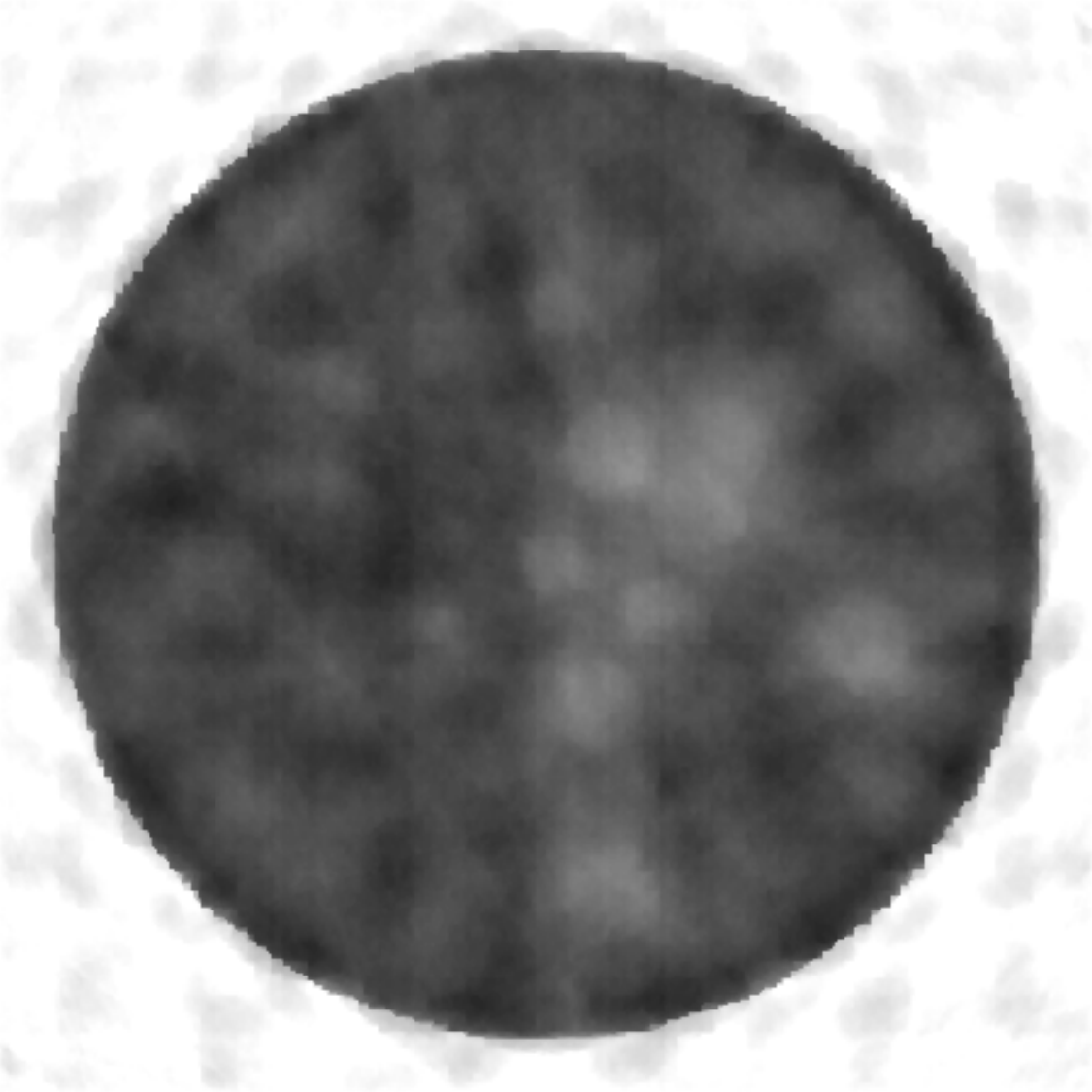}
    \caption{TV, 10 }\label{kivitv18010}
    \end{subfigure}
    \begin{subfigure}[b]{3.3cm}
    \includegraphics[height=3.3cm]{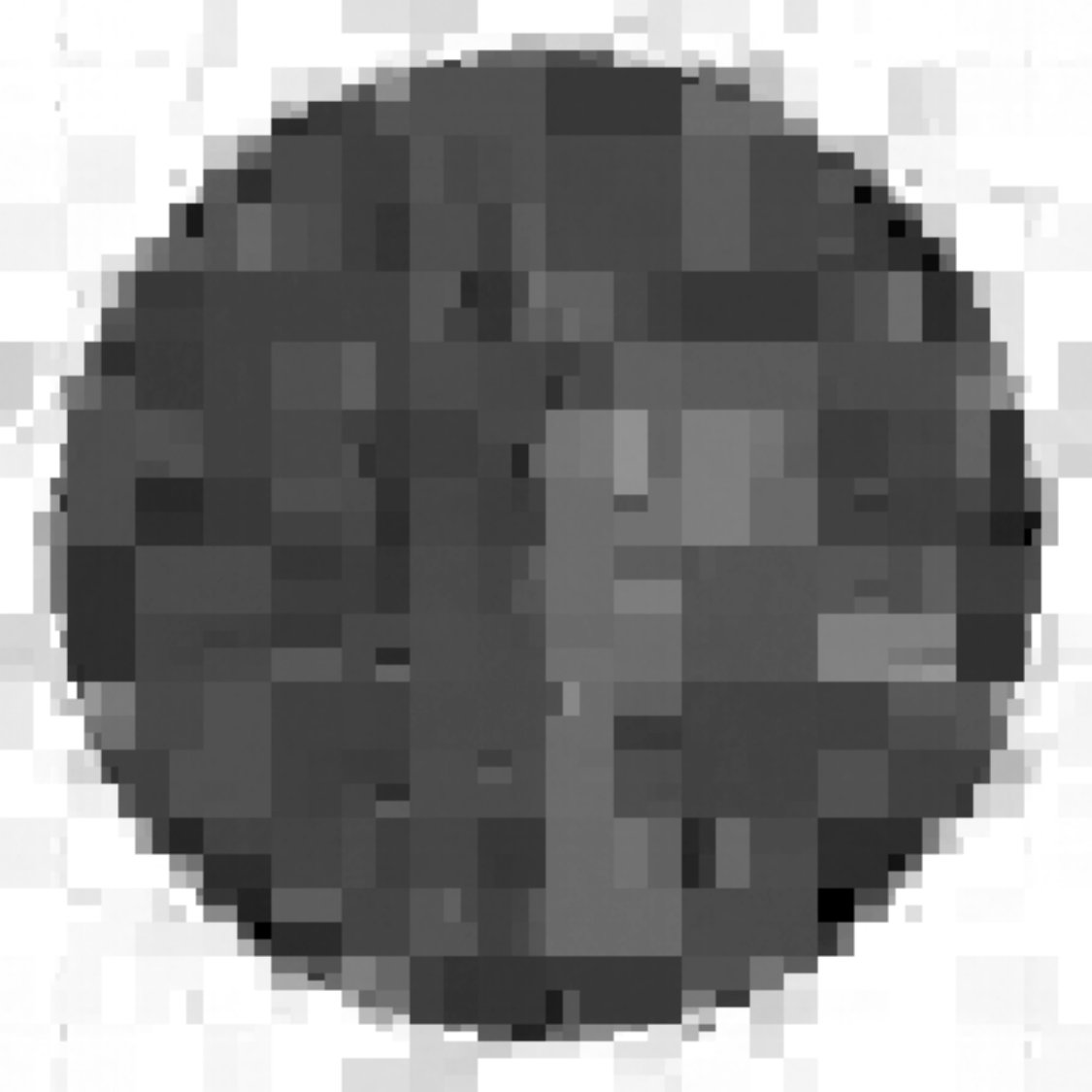}
    \caption{Besov, 10 }\label{kivihaar18010}
    \end{subfigure}
    \begin{subfigure}[b]{3.3cm}
    \includegraphics[height=3.3cm]{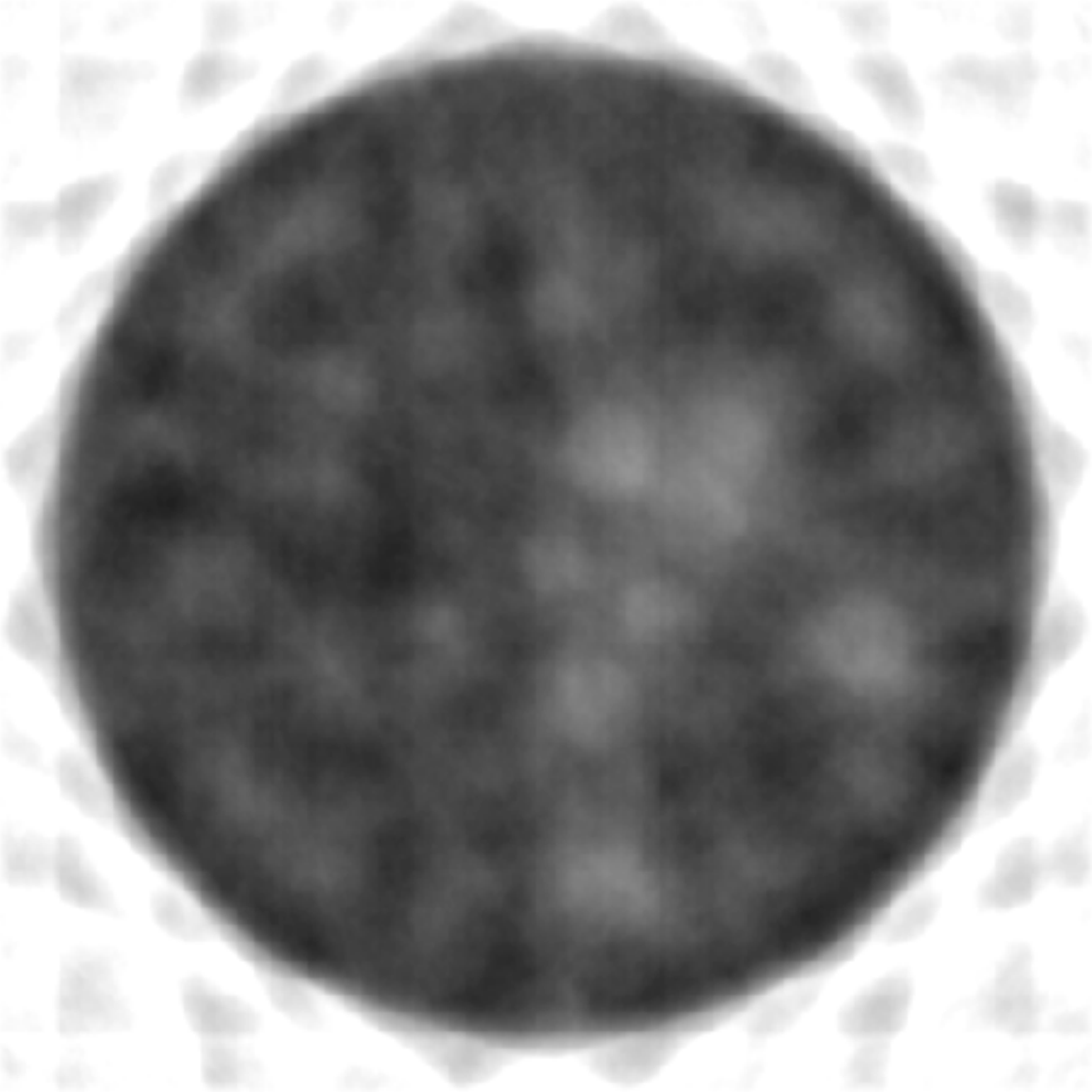}
    \caption{Cauchy, 10 }\label{kivicauchy18010}
    \end{subfigure}
    \caption{MAP estimates for the drill-core experiment with different measurement angles and prior assumptions} 
    \label{rockwide}
\end{figure}

We shall only demonstrate MAP estimation, and the results are in Figure \ref{rockwide}.
At the first glance, there are no significant differences within the MAP estimates if we compare TV, Cauchy  and  Gaussian difference priors to each other at 90 measurement angle scenario. 
Nevertheless, we consider the Cauchy and TV priors as the best ones, since they have the least amount of noise present and they still preserve the edges of the pores with low density and and denser regions of the core sample well. If we wanted to achieve the same level of noise with Gaussian difference prior, we would end up having clearly oversmoothed MAP estimates. 

At the 30 angle measurement angle scenario, MAP estimate with Besov  prior is  severely  ruined with block artefacts and in the 10 angle scenario, it is completely useless.  This is likely due to the fact that the region of interest consists of many spherical objects which are close to each other, but the wavelet coefficients of Haar family do not approximate well such details.   
Unlike in the log tomography case, there are not so many  artefacts present in the MAP estimates calculated with Gaussian prior.   In the 10 angle case, MAP estimates calculated with  Gaussian, TV and Cauchy priors are very close to each other and neither of them can be considered superior over the others. We could try to fix the coordinate-axis dependency of Cauchy prior by introducing isotropic Cauchy prior so that we set a bivariate Cauchy distribution for  differences of each pixel to their neighbours. 
 
\section{Conclusion}
\label{section:Conclusion}

We have made a Bayesian inversion industrial tomography comparison study with four different prior assumptions. 
The targets have different contrasts, that is, sharp edges were to be reconstructed.
This is a classical problem, where one needs either hierarchical or non-Gaussian models, and hence inherently computational complexity becomes a problem. 
The chosen methods, to use optimisation or MCMC, is a balance between fast computations and uncertainty quantification. It should be noted that for many industrial applications the computational cost is still too high for many dynamical systems. However, for static objects and case studies where computational time is not critical, the correctly chosen methods can significantly enhance the reconstruction results for difficult measurement geometries and targets.
It is notable that systematic choices of different parameters, including mesh refinement parameters, dictate the reconstruction accuracy. Here, we have used simple grid searches, in order to show the maximum possible reconstruction capabilities  of the priors  while paying attention to which kind of artefacts each prior had. In reality, choosing the parameters would require implementing some kind of a hierarchical model. This would naturally further increase computational complexity.

In all the scenarios we considered, MAP estimates with Cauchy and TV priors were considered the best -- their artefacts were mild in general and, while their edge-preserving performances were still excellent. The point estimates using Besov  prior with at least Haar wavelets underperformed remarkably compared to the other priors in most cases due to the major block artefacts  it tends to produce. We could reduce that behaviour by reducing the number of DWT levels and  using another wavelet family, but even that cannot improve the location-dependency of the prior. Likewise, we could try fixing the anisotropic of the Cauchy prior with a bivariate Cauchy distribution or using the alternative formulations of the prior.  
For further development,  acceptance-ratio-based adaptation in MwG and setting a non-identity mass matrix for HMC would improve sampling. 

Further improvements in addition to the aforementioned ones, we should consider using other prior distributions, like general $\alpha$-stable priors, or using hierarchical priors. For instance, instead of the Gaussian  prior we might test various hierarchical Gaussian priors with possible many layers to potentially  improve the edge-preserving properties of it.

\begin{Backmatter}

\paragraph{Acknowledgments}
Authors thank Prof.\ Heikki Haario and Prof.\ Marko Laine for useful discussions.

\paragraph{Funding statement}
This work has been funded by Academy of Finland (project numbers 326240, 326341, 314474, 321900, 313708) and by European Regional Development Fund (ARKS project A74305).

\paragraph{Competing interests}
None.

\paragraph{Data availability statement}
Codes and the synthetic data used are available openly in GitHub \url{https://github.com/suurj/tomo/}.

\paragraph{Ethical standards}

We have followed the ethical code of conduct for research and publication of the Academy of Finland.

\paragraph{Author contributions}
Conseptualisation: S.L; S.S; L.R.
Funding acquisition: S.S; L.R.
Methodology: J.S; M.E; S.L; S.S; L.R.
Data visualisation: J.S.
Software: J.S; M.E.
Writing original draft: J.S; M.E; S.L; L.R.
All authors approved the final submitted draft.

\paragraph{Supplementary material}

None.

\bibliographystyle{apalike}
\bibliography{sources}



\end{Backmatter}

\end{document}